\documentclass[pdftex]{pasj01}
\draft
\usepackage{natbib}
 
\Received{$\langle$reception date$\rangle$}
\Accepted{$\langle$acception date$\rangle$}
\Published{$\langle$publication date$\rangle$}
\begin{document}

\title{Nobeyama 45-m mapping observations toward nearby molecular clouds, Orion A, Aquila Rift, and M17:
Project overview}
\author{Fumitaka \textsc{Nakamura}\altaffilmark{1,2,3},
Shun \textsc{Ishii}\altaffilmark{1,4},
Kazuhito \textsc{Dobashi}\altaffilmark{5},
Tomomi \textsc{Shimoikura}\altaffilmark{5},
Yoshito \textsc{Shimajiri}\altaffilmark{6},
Ryohei \textsc{Kawabe}\altaffilmark{1,2,3},
Yoshihiro \textsc{Tanabe}\altaffilmark{7},
Asha \textsc{Hirose}\altaffilmark{5},
Shuri \textsc{Oyamada}\altaffilmark{1,8},
Yumiko \textsc{Urasawa}\altaffilmark{9},
Hideaki \textsc{Takemura}\altaffilmark{1,2},
Takashi \textsc{Tsukagoshi}\altaffilmark{1},
Munetake \textsc{Momose}\altaffilmark{7},
Koji \textsc{Sugitani}\altaffilmark{10},
Ryoichi \textsc{Nishi}\altaffilmark{9}
Sachiko \textsc{Okumura}\altaffilmark{8},
%Chihomi \textsc{Hara}\altaffilmark{1,3},
%Kazushige \textsc{Sasaki}\altaffilmark{9},
Patricio \textsc{Sanhueza}\altaffilmark{1},
Quang \textsc{Nygen-Luong}\altaffilmark{1},
Takayoshi \textsc{Kusune}\altaffilmark{1}
%Sho \textsc{Katakura}\altaffilmark{1},
%Akifumi \textsc{Yamabi}\altaffilmark{1},
}%

\altaffiltext{1}{National Astronomical Observatory of Japan, 2-21-1 Osawa, Mitaka, Tokyo 181-8588, Japan}
\altaffiltext{2}{The Graduate University for Advanced Studies
(SOKENDAI), 2-21-1 Osawa, Mitaka, Tokyo 181-0015, Japan}
\altaffiltext{3}{Department of Astronomy, The University of Tokyo, Hongo, Tokyo 113-0033, Japan}
\altaffiltext{4}{Joint ALMA Observatory, Alonso de Co\'rdova 3107 Vitacura, Santiago, Chile}
\altaffiltext{5}{Department of Astronomy and Earth Sciences, 
Tokyo Gakugei University, 4-1-1 Nukuikitamachi, Koganei, Tokyo 184-8501, Japan}
\altaffiltext{6}{Laboratoire AIM, CEA/DSM-CNRS-Universit\'e Paris Diderot, 
IRFU/Service \'dAstrophysique, CEA Saclay, F-91191 Gif-sur-Yvette, France}
\altaffiltext{7}{College of Science, Ibaraki University, 2-1-1 Bunkyo, Mito, Ibaraki 310-8512, Japan}
\altaffiltext{8}{Faculty of Science, Department of Mathematical and Physical Sciences, Japan Women’s
University, 2-8-1 Mejirodai, Bunkyo-ku, Tokyo 112-8681, Japan}
\altaffiltext{9}{Department of Physics, Niigata University, 8050 Ikarashi-2, Niigata 950-2181, Japan}
\altaffiltext{10}{Graduate School of Natural Sciences, Nagoya City University, Mizuho-ku, Nagoya, 
Aichi 467-8601, Japan}
\email{fumitaka.nakamura@nao.ac.jp}


\maketitle

\begin{abstract}
We carried out mapping observations toward three nearby molecular clouds,
Orion A, Aquila Rift, and M17, using a new 100 GHz receiver, FOREST, on the Nobeyama 45-m telescope.
In the present paper, we describe the details of the data obtained such as 
intensity calibration, data sensitivity, angular resolution, and velocity resolution.
Each target contains at least one high-mass star-forming region.  
The target molecular lines were $^{12}$CO ($J = 1 - 0$), 
$^{13}$CO ($J = 1 - 0$), C$^{18}$O ($J = 1 - 0$), 
N$_2$H$^+$ ($J=1-0$), and CCS ($J_N=8_7-7_6$), with which
we covered the density range 
of 10$^2$ cm$^{-3}$ to 10$^6$ cm$^{-3}$ 
with an angular resolution of $\sim 20\arcsec$ and a velocity resolution of $\sim$ 0.1 km s$^{-1}$.
Assuming the representative distances of 414 pc, 436 pc,  and 2.1 kpc, 
the maps of Orion A, Aquila Rift, and M17 cover most of the densest parts
with areas of about 7 pc $\times$ 15 pc, 7 pc $\times$ 7 pc, 
and 36 pc $\times$ 18 pc, respectively.
On the basis of the $^{13}$CO column density distribution,
the total molecular masses are derived to be
$3.86 \times 10^4 M_\odot$, $2.67 \times 10^4 M_\odot$, and $8.1\times 10^5 M_\odot$ for
Orion A, Aquila Rift, and M17, respectively.
For all the clouds, the H$_2$ column density exceeds  
the theoretical threshold for high-mass star formation of $\gtrsim$ 1 g cm$^{-2}$, 
only toward the regions which contain current high-mass star-forming sites.
For other areas, further mass accretion or dynamical compression would be necessary for future high-mass star formation.
This is consistent with the current star formation activity.
Using the $^{12}$CO data, we demonstrate that our data have enough capability to identify molecular outflows, 
and for Aquila Rift, we identify 4 new outflow candidates.
The scientific  results will be discussed in details in separate papers.
\end{abstract}

\section{Introduction}
\label{sec:intro}

Star formation not only determines the observed properties of galaxies, 
but also significantly influences galaxy evolution
\citep{maclow04,mckee07}. 
What drives and regulates star formation in galaxies?  
There is little consensus to this apparently simple and fundamental question. 
Theoretical studies have demonstrated that once self-gravitating objects (cloud cores) form, 
their gravitational collapses lead to star formation at high rates \citep{klessen98}. 
However, star formation is known to occur in a very low rate in galaxies \citep{zuckerman74}. 
For example, the total molecular gas mass in our Galaxy is 
estimated to be 10$^9$ M$_\odot$ from the CO observations.
%\citep{roman16}. 
If all the molecular clouds are converted to stars within 
a cloud free-fall time which is a few Myr 
at the typical cloud density of a few thousand cm$^{-3}$, 
the {\it free fall} rate of star formation is calculated 
to be about 10$^3$ M$_\odot$ yr$^{-1}$, 
which is about 10$^3$ times larger than the observed 
star formation rate \citep{mckee97,murray10,robitaille10}.
More accurate estimations of star formation rates toward individual molecular clouds
are basically consistent with the above rough calculation \citep{krumholz07}.
The small Galactic star formation rate thus implies that some physical processes make star formation  slow and inefficient.

However, it remains uncertain what processes make star formation slow and inefficient. 
There are several processes discussed to slow down and 
regulate star formation, e.g., stellar feedback, magnetic field, and 
cloud turbulence \citep{shu87,mckee07,krumholz14}. 
It is thus crucial to characterize internal cloud structure and physical properties of
nearby molecular clouds to understand the roles of the processes in star formation.
In our project, we carried out mapping observations toward three nearby 
molecular clouds using the Nobeyama 45-m telescope, and attempt to 
address the issues of the inefficient star formation. 

Processes of star formation are often influenced by large-scale events. 
Protostellar jets and outflows are often extended to 0.1--10 pc-scale \citep{bally16}.   
Expanding bubbles with 1--10 pc generated by stellar winds from intermediate-mass and
high-mass stars are  discovered in nearby clouds \citep{arce11,feddersen18,pabst19} .  
FUV radiation from high-mass stars sometimes affects the parent clouds in 10 pc scale \citep{shimajiri11,ishii19}.  
Much larger bubbles created by supernovae interact with entire molecular clouds \citep{frisch15}.
Galactic spiral density waves influence molecular cloud formation and subsequent star formation \citep{elmegreen79}.
One of the immediate objectives of the present project is to reveal clouds structure and dynamics of target clouds 
and attempt to elucidate how these events influence  structure and dynamics of the target clouds.
In summary, wide-field mapping observations are important to understand the effects of stellar feedback 
and external events like large-scale shocks because they potentially affect cloud properties and structures 
in a cloud scale, 1$-$10 pc.  
For comparison, we summarize several recent wide-field survey toward our target clouds in table \ref{tab:survey}.
For Orion A, there are many molecular-line mapping surveys done so far, but many of these surveys were observed only in the northern parts including OMC-1.
As Aquila Rift, many molecular-line mapping surveys are just focusing on the two prominent star-forming regions, W40 and Serpens South.
As for M17,  only a few wide-field surveys that cover at least about 1 square degree areas have been done so far.

The main objectives of the present paper are to show the project overview and to describe the details 
of the observations and calibration of the obtained data.
In section \ref{sec:target},
we describe how we select our target clouds and lines. 
The details of the observations are described in section \ref{sec:obs}.
In section \ref{sec:calib}, we describe how to produce final data cubes.
Then, we summarize the quality of our maps such as sensitivity, angular resolution, and
velocity resolution.
In sections \ref{sec:global}, we present the spatial distributions of the molecular line emission obtained
and in section \ref{sec:abundance} we derive the spatial distribution of the $^{13}$CO and 
C$^{18}$O abundances toward Orion A.
We briefly discuss the global molecular gas distributions of the target clouds mainly using $^{12}$CO and
its isotopologues.
In section \ref{sec:outflow}, we show the preliminary results of CO outflow survey toward
L1641N in Orion A and Serpens South in Aquila Rift, demonstrating that our data can be 
used to detect outflows.
Finally, we summarize the main results of the present paper in section \ref{sec:summary}.

The detailed analysis will be presented in separate papers.
For example, \citet{tanabe19}, \citet{ishii19}, and \citet{nakamura19} describe
the results of the outflow survey, cloud structure analysis, and multi-line observations of the OMC-2 FIR 4 region for Orion A, respectively.
\citet{shimoikura19} and \citet{kusune19} present the detailed cloud kinematic structure and the relationship between filamentary structure and
magnetic field, respectively, toward Aquila Rift.
For M17, Nguyen Luong et al. (2019), Shimoikura et al. (2019), and \citet{sugitani19}
 will report the global cloud kinematics, dense core survey, and relationship between cloud structure and magnetic field, respectively.
In addition to the three main regions, a few other star forming regions such as 
Northern Coal Sack \citep{dobashi19} and DR 21 \citep{dobashi19b} were studied during our project.
These data are obtained for intensity calibration of the CO lines taken 
with a new 100 GHz receiver, FOREST.

\section{Project Overview}
\label{sec:target}

\subsection{Molecular line data with a $10^{-2}$ pc resolution within distances of a few kpc}

In 2011, the state-of-art facility, ALMA, started science operations. 
Since then, we can easily conduct observations with much higher sensitivity and higher angular resolution than ever done before.  
In the ALMA era, one of the important objectives of our project is
to make useful datasets to compare with ALMA maps of distant molecular clouds. 
Even using one of the largest millimeter telescopes, the Nobeyama 45-m telescope, 
the achieved beam size at 115 GHz is at most $\sim$ 15\arcsec.
On the other hand, we can easily achieve $\sim$ 1\arcsec \, resolution using ALMA.
To overcome the disadvantage of the coarse angular resolution, we choose nearby molecular clouds as our targets of the Nobeyama 45-m telescope.
Large-scale ALMA mapping observations toward nearby molecular clouds are still limited and are impossible to do because of its small field of view
and extremely-long observation time.  On the other hand, the coarse beams of the single-dish telescopes allow us to conduct 
wide-field mapping observations.
In this sense, our Nobeyama mapping project is complementary to the ALMA observations toward distant molecular clouds.

Previous observational studies are often influenced by the effects of
different {\it spatial} resolutions when cloud structure and
physical properties of molecular clouds are compared. 
We would like to minimize the effects of 
the different spatial resolution to better understand the cloud structure and
the environmental effects. 
In our project, we therefore aim to obtain maps that can be 
compared directly with those obtained with ALMA 
in almost the same {\it spatial} resolution.

The number of pointings of the ALMA mosaic observations is 
limited below 150 for the 12-m Array,
which can cover the area with $\lesssim$ 5\arcmin $\times$ 5\arcmin \, with an angular resolution of 
1\arcsec \, at the ALMA band-3 ($\sim$ 100 GHz).
Assuming that we map about 1 square degree area with an angular resolution of 20 \arcsec \,  toward the regions whose distances are 20 times closer than the areas observed by ALMA with 1 \arcsec \, resolution, 
both maps obtained have comparable spatial coverage and spatial resolution.
As described below,  we thus chose Orion A, Aquila Rift, and M17 
as our targets of the Nobeyama 45-m mapping observations.
In table \ref{tab:resolution}, we compare the achieved spatial 
resolutions of several 
molecular clouds, mainly including our target regions. 
In our previous Nobeyama 45-m observations, we have mapped nearest molecular clouds
($\sim$ 140 pc) such as L1551 \citep{yoshida10,lin17} and $\rho$ Ophiuchus cloud 
(Maruta et al. 2010; Nakamura et al. 2011)
%\citep{maruta10,nakamura11}
using the Nobeyama 45-m telescope.
These maps have a spatial resolution of about 2000 $-$ 3000 au, 
which is comparable to those of the maps of Orion A (414 pc) combined with the CARMA 
data \citep{kong18}. 
For more distant clouds with distances of a few kpc, ALMA observations can
achieve the spatial resolution of a few thousand au at an angular resolution of $\sim$ 1\arcsec.
Thus, our Nobeyama maps of nearby clouds  can be directly compared 
with the maps of the molecular clouds 
within a few kpc 
in a comparable spatial resolution of a few thousand au.

We opened all the mapping data we took in this project to the public. The data are available through
the Japanese Virtual Observatory (JVO) archival system.\footnote{http://jvo.nao.ac.jp/index-e.html}

%\subsection{Selection of Targets}
%\label{sec:target}

\subsection{Target Lines}

We  chose the following five molecular lines: $^{12}$CO ($J=1-0$), 
$^{13}$CO ($J=1-0$), C$^{18}$O ($J=1-0$), N$_2$H$^+$ ($J=1-0$), 
and CCS ($J_N=8_7-7_6$). 
The $^{12}$CO molecule is the second most abundant after the hydrogen molecule 
in molecular clouds, and therefore it can trace the basic cloud structure reasonably well. 
Rarer isotopologues of $^{12}$CO such as $^{13}$CO and C$^{18}$O can trace
relatively denser gas with densities of 10$^3 - 10^4$ cm$^{-3}$. 
N$_2$H$^+$ has a critical density of $\sim$ 10$^5$ cm$^{-3}$ and 
traces cold dense gas, particularly in prestellar phase, very well \citep{caselli02,punanova16}. 
Thus, we can cover the densities 
from 10$^2$ cm$^{-3}$ to 10$^6$ cm$^{-3}$ using these 4 lines.
The main reason why we choose CCS is that CCS can be obtained simultaneously with 
$^{13}$CO ($J=1-0$), C$^{18}$O ($J=1-0$), and N$_2$H$^+$ ($J=1-0$)
as we mention below.
CCS is abundant in early prestellar phase and traces a similar density range to N$_2$H$^+$.
Thus, CCS can provide additional information of chemical evolution of prestellar dense gas
\citep{suzuki92,marka12,loison14}.
 
Another reason why we mainly chose the CO lines is that the main beam efficiencies 
of the 45-m telescope at 100 GHz band ($\lesssim$ 0.5) 
is not so superb as those of other telescopes such 
as the IRAM 30-m and LMT telescopes (see table \ref{tab:obs})
and thus observations in strong emission lines such as $^{12}$CO and its isotopologues 
can be done more efficiently within limited observation time.
Although N$_2$H$^+$ is not so strong as $^{12}$CO, $^{13}$CO, and C$^{18}$O, 
a new multi-beam SIS receiver, FOREST, allows us to observe N$_2$H$^+$ line 
simultaneously with $^{13}$CO and C$^{18}$O.
In addition, CCS ($J_N=8_7-7_6$) can also be obtained simultaneously with 
 N$_2$H$^+$, $^{13}$CO and C$^{18}$O using a spectral window mode 
 which has become available since the 2016-2017 season.
%We thus obtained CCS maps in the 2016-2017 season.

We also adopted a velocity resolution of $\sim$ 0.1 km s$^{-1}$, so that we can reasonably 
identify dense cores with typical internal velocity dispersion of a few tenths km s$^{-1}$. 
Our observed emission lines are summarized in table \ref{tab:obs}.

\subsection{Target Clouds}

We chose the target clouds taking into account the following conditions:
\begin{itemize}
\item[(1)] The main part of the target cloud can be covered by one square degree
with a few hundred hours 
using the new 100 GHz 4-beam SIS receiver FOREST, 
including overhead such as pointing observations and flux calibration.

\item[(2)] The target clouds are relatively well studied, and
 additional datasets are  available.

\item[(3)] A target cloud contains regions of ongoing high-mass star formation.
%at least one massive star-forming region.

\item[(4)] The spatial resolution of the Nobeyama 45-m data 
can be comparable to or smaller than the typical size of
dense cores ($\sim 0.1$ pc).

\end{itemize}

Taking the above conditions into account, we chose the following three regions as our targets: 
(1) Orion A ($d\sim 414$ pc, \citet{menten07,kim08}), (2) Aquila Rift  
($d\sim 436$ pc, \cite{ortez17b}), and 
(3) M17 ($d\sim 2000$ pc, \cite{xu11}).

%Table ? briefly summarize some characteristics of individual clouds.
In this survey, we chose nearby clouds which contains the formation sites of O-type and early B-type stars.
There are not so many such regions within $\sim$1 kpc (e.g., Orion A, Aquila Rift, California, Mon R2, and Orion B).  
%There are a few other regions such as the California molecular cloud, Mon R2, and Orion B. 
Among them, Orion A and Aquila Rift may be the nearest.
The evolutionary stages of the star-forming regions appear to be different in Aquila Rift and Orion A. For example, 
star formation lasts at least a few Myr in OMC-1 \citep{dario16}.  
On the other hand, in the Serpens South cluster in Aquila Rift, a fraction of the Class O/I protostars is extremely high, 
suggesting the age within 0.5 Myr.
In other words, Orion A contains more evolved star-forming regions 
than Aquila Rift.   The distances of Orion A and Aquila Rift are similar ($\sim$ 400 pc), 
and thus it is easy to directly compare the cloud structures
with the same spatial resolutions for molecular clouds in different evolutionary stages. 
These are the main reason why we chose these two clouds.
The star formation efficiencies of Orion A and Aquila Rift appear to be similar at a few \% to each other
\citep{evans09, maury11}, which is typical of nearby star-forming regions \citep{krumholz07}.
Thus, we believe that the two regions are representative of nearby star-forming molecular clouds.

The beam size of the Nobeyama 45-m telescope at 100 GHz is $\sim$ 15\arcsec, 
corresponding to $\sim$ 6200 au
and 30000 au at the distances of 414 pc and 2000 pc, respectively.
The spatial resolution of M17 is not satisfied with the last condition, but we 
chose it by the following two reasons. (1) the densest part of an infrared dark 
cloud in M17, M17 SWex, 
was observed  by ALMA and the N$_2$H$^+$ data are available (see e.g.,  \cite{ohashi16,chen19}),
and we can combine the 45-m data with the ALMA data to fill the zero spacing
so that we can make maps whose spatial resolution is comparable to
those of the maps of other two targets obtained 
with the Nobeyama 45-m telescope ($\sim$ 8000 au).
(2) Star formation activity in M17 appears to be triggered 
by a Galactic spiral wave passage \citep{elmegreen79}
and thus the M17 data are expected to 
provide us a clue to better understand the Galactic star formation process.
In addition, M17 is the nearest high-mass star forming region located to the Sagittarius arm. 
The stellar density at NGC 6811 is highest at $> 10^3$ star pc$^{-2}$ after the Carina cluster
among the regions in the MYStIX survey of massive star forming regions in X-Ray (Kuhn et al. 2015).
M17 is closer to us than the Carina region. \citet{povich10} proposed that high mass star formation may be delayed at the IRDC region and
the mass function of young stars in IRDC appears to be different from the Salpeter initial mass function.  M17 is expected to provide us 
key information to understand how high-mass stars form. 
Therefore, we chose the distant high-mass star-forming region M17.

As for Orion A, we combine our Nobeyama 45-m data with the CARMA interferometric 
data to obtain maps with $\sim$ 8'' resolution.
The combined maps have {\it spatial} resolution comparable to those of the Taurus and $\rho$ Ophiuchus molecular clouds previously obtained with the Nobeyama 45-m telescope, 
with a spatial resolution of about 3000 au ($\sim$ 20'') 
at a distance of 140 pc. 
Thus, using the Nobeyama data and ALMA data, 
we can directly compare internal structure and physical properties of several clouds located 
at the different distances in the range from 140 pc to a few kpc with the same {\it spatial} resolutions (see table \ref{tab:resolution}).  
We expect that our maps obtained  are useful as templates of
nearby molecular clouds which can be compared with maps 
toward more distant molecular clouds, which can be obtained with ALMA, 
e.g., Infrared Dark Clouds at a few kpc such as the Nessie nebula and G11.11-0.12 (Snake).

We determine the mapping areas by reference to  the 2MASS $A_V$ maps 
\citep{dobashi05, dobashi11}, {\it Herschel} column density maps, and {\it Spitzer} images. 
Taking the observation time into account, we planned to map the target areas with
$A_V$ $\gtrsim$ a few corresponding to a few tenth g cm$^{-2}$, 
so that we can detect almost all the self-gravitating cores in our target
clouds (see the $A_V$ threshold derived by \cite{andre10}).

\begin{figure}
 \begin{center}
 \includegraphics[width=12cm, bb = 0 0 1200 1600]{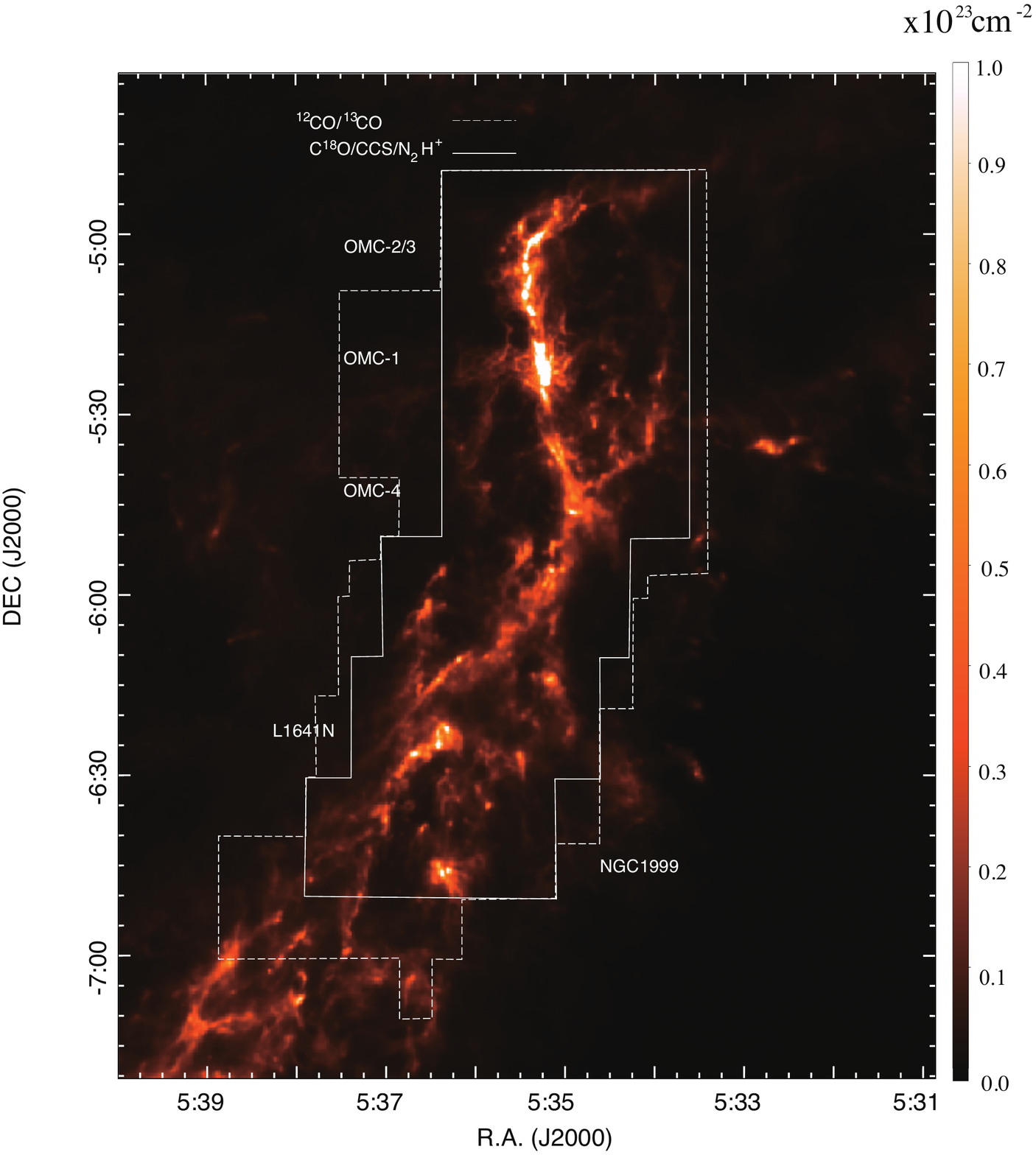}
 \end{center}
\caption{Observation areas overlaid on the H$_2$ column density maps of Orion A.
The solid and dashed lines indicate the observation boxes for $^{12}$CO+$^{13}$CO set and C$^{18}$O+N$_2$H$^+$+CCS set, respectively.
The H$_2$ column density map has about an effective angular resolution of  $\sim$36$"$ (see Lombardi et al. 2016).
%, (b) Aquila Rift, and (c) M17, respectively.
\label{fig:obsarea}}
\end{figure}

\begin{figure}
 \begin{center}
 \includegraphics[width=16cm,bb=0 0 900 900]{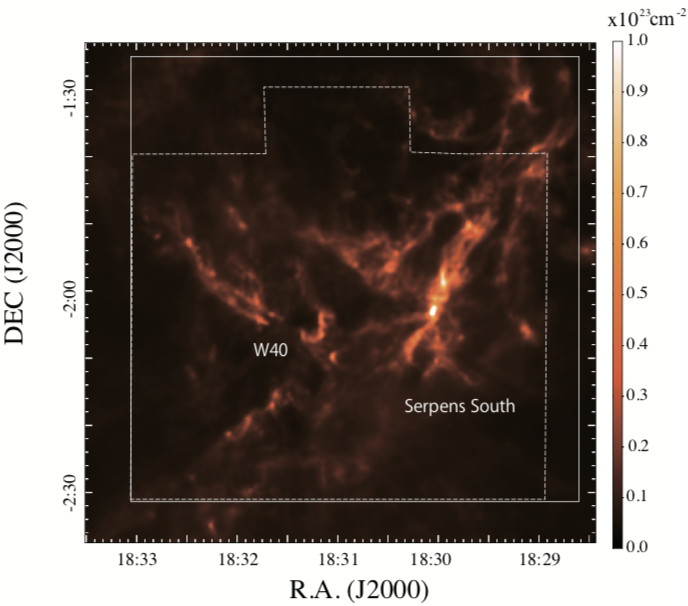}
 \end{center}
\caption{Same as figure \ref{fig:obsarea} but for Aquila Rift.
The color image shows  the H$_2$ column density map whose fits data were downloaded via the {\it Herschel} Gould Belt Survey Archive system.
The solid and dashed lines are the same as those in figure \ref{fig:obsarea}.
\label{fig:obsarea_aquila}}
\end{figure}

\begin{figure}
 \begin{center}
 \includegraphics[angle=90, width=10cm,bb=0 0 1088 583]{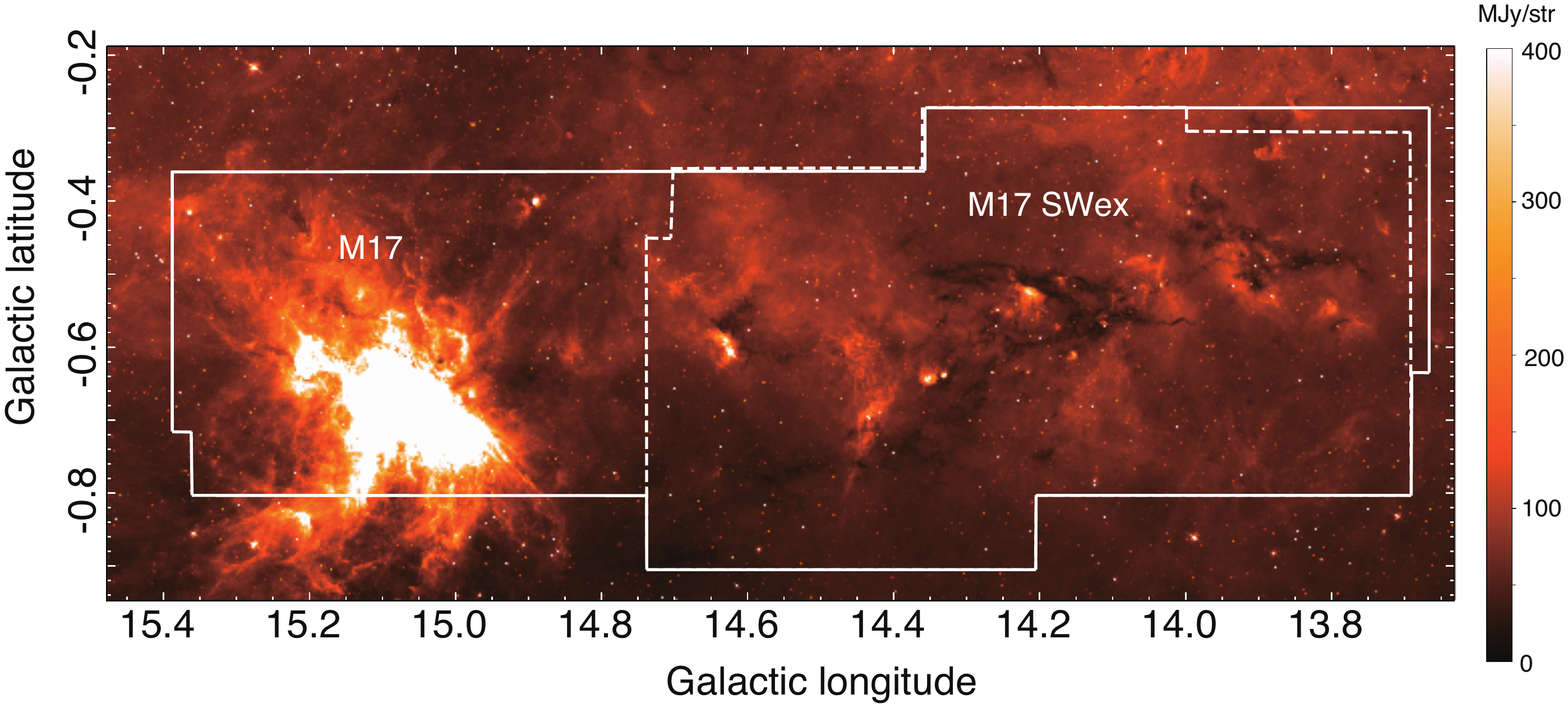}
 \end{center}
\caption{Observation areas overlaid on the {\it Spitzer} 8$\mu$m image of M17.
The color image shows  the  {\it Spitzer} 8$\mu$m image of M17 downloded from
the Glimpse Archival system.
The solid and dashed lines are the same as those in figure \ref{fig:obsarea}.
\label{fig:obsarea_m17}}
\end{figure}

Figures \ref{fig:obsarea},  \ref{fig:obsarea_aquila}, and \ref{fig:obsarea_m17} present the mapping areas of 
Orion A, Aquila Rift, and M17, respectively, overlaid on the {\it Herschel} or {\it Spitzer} images.

\section{Observations}
\label{sec:obs}

As described in the next section, we used the molecular line data taken with  BEARS for intensity calibration.
In this section, 
we first describe the details of the FOREST observations, and then 
describe the details of the BEARS observations.
BEARS is the SIS 25-element focal plane receiver with the frequency coverage of 82 -- 116 GHz, 
and has been used for many mapping observations since 2000. 
Therefore, the intensity calibration scheme is well established compared to the new receiver, FOREST, which was
also demounted  for repair after the first season and re-installed in the second season.  
In addition, the surface accuracy of the telescope dish was significantly improved for the first two seasons 
by applying the holograph to the dish surface.  A careful data reduction of the obtained data is crucial
to verify the absolute intensity scale of the FOREST data. Thus, we used the data taken with BEARS 
for the intensity calibration.
For both observations, we adopted the OTF mapping mode 
with a position switching method using the emission-free positions areas summarized in table \ref{tab:off}.

For FOREST, we adopted two frequency set to observe the target lines. 
Set 1 is for $^{12}$CO and $^{13}$CO.
Set 2 is for C$^{18}$O, N$_2$H$^+$, $^{13}$CO and CCS with a spectral window mode. 
For BEARS, we observed only single line for the individual observations.

%%%%%%%%%%%%%%%%%%%%%%%%%%%%%%%%%

\subsection{FOREST}
\label{subsec:forest}

%The procedure of the FOREST observations is basically similar to that of BEARS.
FOREST is a 4-beam dual-polarization sideband-separating SIS receiver 
\citep{minamidani16} and has 16 intermediate frequency (IF) outputs in total, 
then 8 IFs in the upper-sideband and other 8 IFs in the lower-sideband were used 
for the molecular line observations.
As backends, we used a digital spectrometer based on an FX-type correlator, SAM45, that is 16 sets of 4096 channel array.
We divided the mapping area into smaller sub-areas whose sizes are summarized in table \ref{tab:box}.
Then, we carried out the OTF observations toward each sub-area.
The parameters of the OTF observations are summarized in table \ref{tab:otf}.
Scans of the OTF observation are separated in the interval of \timeform{5''.17}. 
Thus, individual scans by 4 beams of FOREST are overlapped. 
%The observation parameters are given in table \ref{tab:otf}.
The positions of the emission-free areas used for the observations are summarized in table \ref{tab:off}.
We note that the coordinates of the emission-free areas are the ones for the first beam of FOREST.

Calibration of the observations was done by the chopper wheel technique
to convert the output signal into the antenna temperature $T_{A}^*$, 
corrected for the atmospheric attenuation.
Some details of the observations are also summarized in table \ref{tab:forest}.
The telescope pointing was checked every 1 hour by observing the SiO maser lines from 
the objects presented in table \ref{tab:off}. 
The pointing accuracy was better than $\sim$ 3$''$ throughout the entire observation.

In order to minimize the scanning effects, the data with orthogonal scanning directions along the R.A. 
and decl. axes were combined into a single map.
We adopted the same gridding convolution function (spheroidal function) as the BEARS data to calculate the intensity at each grid point of the final cube data 
with a spatial grid size of \timeform{7.5"}.

The FOREST receiver and telescope conditions have not been stable for the first two seasons.
For example, the receiver was demounted 
on June, 2016 (right after the first season) to improve several internal components. 
Second, the surface accuracy of the 45-m dish was significantly 
improved for the first two seasons by applying the holograph to the dish surface.
Therefore, the observation conditions were different from season to season. 
To minimize the effects of these factors in determining the intensity scale
of observed lines, we scaled  the data obtained with FOREST by those of BEARS
whose  intensity calibration method is well established.
The detail of the flux calibration of the FOREST data is described in the next section.

In the last season (2016-2017), a new observational capability called a spectral window mode was available, which allows
us to obtain more lines simultaneously.  We therefore observed $^{13}$CO ($J=1-0$), C$^{18}$O ($J=1-0$), 
N$_2$H$^+$ ($J=1-0$), and CCS ($J_N=7_6-6_5$) simultaneously.
In the spectral window mode, we equally divided the whole bandwidth into two, so that the bandwidth and frequency resolution of 
the spectrometer arrays were set to 31.25 MHz and 15.26 kHz, respectively.

\subsection{BEARS}
\label{sec:bears}

The procedure of the BEARS observations is basically similar to that of FOREST.
The details of the BEARS observations are summarized in table \ref{tab:bears}. 
Some of the results of the BEARS observations are described in the references listed in the last column of table \ref{tab:bears}.
In brief, we divided the mapping area into many 15$' \times 15'$ or $20' \times 20'$ rectangle sub-areas.
The sizes of these sub-areas are determined so as to complete an OTF scan within 1 or 1.5 hours.
We carried out mapping observations toward each sub-region in an OTF mode \citep{sawada08}
using BEARS and 25 sets of 1024 channel Auto Correlators 
(ACs) which have the bandwidth of 32 MHz and the frequency 
resolution of 37.8 kHz.  
The velocity resolutions of the observations for individual lines are listed in table \ref{tab:bears}.

Calibration of the observations was done by the chopper wheel technique
to convert the output signal into the antenna temperature $T_{A}^*$, 
corrected for the atmospheric attenuation.
%During the observations, the system noise temperatures 
%were in the range between 210 and 400 K in a double sideband. 
At 110 GHz, the half-power beam width was about 15$''$, which  
corresponds to about 0.03 pc at a distance of 414 pc.
The main beam efficiency ($\eta_{\rm 45m}$) was about 0.5 at 110 GHz for the corresponding observation season. 
The telescope pointing was checked every 1 or 1.5 hours by observing 
SiO maser sources, Orion KL and IRC+00363 for the Orion A and Serpens South observations, respectively, 
and was better than 3$''$ during the whole observing period.

We obtained a map by combining scans along the two axes 
that run at right angles to each other.
We adopted a convolutional scheme with a spheroidal function 
to calculate the intensity at each grid point 
of the final cube data with a grid size of 7.5$''$. 
This convolutional scheme is the same as that of the FOREST data.
The spheroidal function is given in \citet{sawada08}. % and is described by \citet{schwab84} in detail.  
The resultant effective resolution was about 21$''$ at 110 GHz.
Finally, we converted the intensities in the antenna temperature scale ($T_{\rm A}^*$) 
into those in the brightness temperature scale ($T_{\rm mb}$) 
by dividing with the main beam efficiencies ($\eta$), $T_{\rm mb} =T_{\rm A}^* /\eta$.
The typical rms noise levels of the final maps are  listed in table \ref{tab:bears}.
We note that for Orion A, the coverages of the $^{12}$CO and $^{13}$CO maps taken with BEARS 
are from Decl. $\sim$ --5$^\circ$20$\arcmin$ to $\sim$ --6$^\circ$30$\arcmin$, 
which is slightly smaller than the actual mapping area of the FOREST observations.
The coverage of the C$^{18}$O map were much smaller than those of $^{12}$CO and $^{13}$CO.
For the other lines(N$_2$H$^+$ and CCS), no OTF data are available.
The typical noise level achieved for the BEARS observations were $\sim$ 0.4 K and 0.7 K at 0.1 km s$^{-1}$ 
for $^{12}$CO and $^{13}$CO, respectively.

\section{Flux Calibration}
\label{sec:calib}

The procedure of the flux calibration is complicated as we describe in this section.  
First, we briefly summarize our flux calibration procedure, and then we describe the details of the procedure in the subsequent subsections.
\begin{itemize}
\item[0.]  All the data were converted the intensity data in $T_A^*$ using the standard NRO data reduction tool, NOSTAR.
\item[1.] For the $^{12}$CO and $^{13}$CO of Orion A, the intensities of each emission line and each sub-box were 
multiplied by the scaling factor $SF_i^{\rm BEARS, \it l}$, where the scaling factor $SF_i^{\rm BEARS, \it l}$ is the ratio of the integrated intensity 
of the FOREST observations
and that obtained with BEARS for the corresponding sub-box $i$ and the superscript $l$ indicates the corresponding 
line ($^{12}$CO or $^{13}$CO).  
For the sub-boxes where BEARS data were not available, 
the average scaling factor of the corresponding season is adopted. 
Note that the BEARS data are in the $T_{\rm mb}$ scale.
\item[2.]  For the other data (in $T_A^*$), 
the intensities corrected for the daily variation were obtained by multiplying the intensity data of the corresponding sub-box 
by the scaling factor of the daily variation $SF_{\rm ref}$, where the scaling factor $SF_{\rm ref}$ is the ratio of
 the sum of the integrated intensity of the reference area obtained at the the corresponding observation session 
 ($S_{\rm ref}=\Sigma_i I_{i, \rm ref}$) and that obtained at the reference date  ($S_{\rm ref}^*= \Sigma_i I_{i, \rm ref}^*$), 
 $SF_{\rm ref}=S_{\rm ref}^*/S_{\rm ref}$). 
Note that we observed a small reference area at least once during each observation session.
\item[3.] Then, the intensities corrected for the daily variation were divided by the  beam efficiency at the corresponding frequency 
and converted the intensities in the $T_{\rm mb}$ scale.
\end{itemize}
We describe the details  of the procedure below.

%The procedure of the data calibration is somewhat different from Orion A and the other regions.
%For $^{12}$CO and $^{13}$CO of Orion A,  we summed up the BEARS and FOREST data to improve the data quality.
%For other line data and the data of the other regions, we applied a standard procedure.

\subsection{Orion A}

\subsubsection{$^{12}$CO and $^{13}$CO}

The $^{12}$CO and $^{13}$CO maps obtained with BEARS cover the northern part of the FOREST observation areas shown in figure \ref{fig:obsarea}.
The intensity scales of the $^{12}$CO and $^{13}$CO  data were thus calibrated by comparing with the previous survey data taken by the BEARS 
receiver \citep{shimajiri11, nakamura12, shimajiri14}, which were already corrected into the main beam 
temperature ($T_{\mathrm{mb}}$) scale.
We obtained the intensities by dividing the FOREST intensities obtained in $T_A^*$ by the scaling factor SF$_i$ determined in  each box. 
We determined a scale factor for the data of each observation sub-box and each line using the following equation,
\begin{equation}
{\rm SF}_i^{\rm BEARS,\it l} =  {  \displaystyle \sum_{k=1}^{N_i} {W_{k, \rm BEARS,\it l} / W_{k, \rm FOREST,\it l}} \over N_i}
\end{equation}
where $W_{k, \rm FOREST, \it l}$ is the integrated intensity of a given box in the antenna temperature scale
at the $k$-th grid, and
 $W_{k, \rm BEARS, \it l}$ is the integrated intensity of the same box in the brightness temperature scale 
 at the $k$-th grid. $l$ is the corresponding line, and 
 $N_i$ is the number of grids in the $i$-th box.
We measure the intensity ratio at every grid and took an average in each sub-box.  
For almost all the boxes,  the dispersions of the scaling factors were within 3 $\sigma$ of the mean value 
of each observation season.  For the boxes where the SF values are larger than 3 $\sigma$ of the mean value or the boxes where
the emission is too weak to measure the SF value, we adopted the value of SF measured immediately before or 
after the corresponding observation.   
However, we note that such sub-boxes with large dispersions were rare.
The scaling factors so derived are consistent with the results of the measurements of the telescope main beam efficiencies.
For example, in the 2015-2016 season, the mean value of the scaling factor was measured to  about 2.1, 
which corresponds to the main beam efficiency of 0.48 at 115 GHz. 
The actual main beam efficiency of the telescope was measured to be about 0.45.

Finally, we combined the FOREST $^{12}$CO and $^{13}$CO data with the BEARS data to reduce the noise levels.
The data combination enable us to typically reduce the noise levels by a factor of 1.5--2, which was crucial for making the 
CARMA and NRO combined Orion images, so that the noise levels in the uv space were well matched.

\subsubsection{C$^{18}$O, CCS, and N$_2$H$^+$}

Since the map coverage of the C$^{18}$O data obtained with BEARS is not so large and we do not 
have the corresponding N$_2$H$^+$ and CCS data,  we applied a different intensity calibration procedure.
For the C$^{18}$O and N$_2$H$^+$ data, we derived  the daily scaling factors or daily intensity variations 
by observing a small area containing FIR 4, which was observed at every observation session once or twice.
To determine the daily scaling factors, we observed a small area which contains FIR 3/4/5 at the date when the
wind speed at the telescope site was almost zero and $T_{\rm sys}$ was close to the lowest value, 
and we adopted the integrated intensity of the FIR 3/4/5 area at that date as a reference. Immediately after this measurement, 
we also observed a small area of B213 in Taurus to compare the data with those obtained with the IRAM 30-m telescope 
\citep{tafalla15} to check the intensity accuracy. 
To compare the intensities taken with different telescopes, we  
smoothed the Nobeyama data to match the map effective angular resolution.
The details of the FIR 3/4/5 area are given in \citet{nakamura19}.
For C$^{18}$O, we also checked the absolute intensity by using the C$^{18}$O map obtained with BEARS toward the northern part of
Orion A \citep{shimajiri14,shimajiri15} and we confirmed that our FOREST C$^{18}$O intensities agree with the BEARS data within an error less than 10 \%.
For N$_2$H$^+$, we checked the absolute intensity by using the N$_2$H$^+$ ($J=1-0$) fits data obtained 
with the IRAM 30-m telescope toward B213 in Taurus \citep{tafalla15}.  
We confirmed that the intensity of N$_2$H$^+$ ($J=1-0$) in B213 obtained with FOREST 
was only about 5 \% larger than that of the IRAM value, where we divided the intensity in the antenna temperature scale
by the main beam efficiency at 94 GHz.  Thus, we consider that the intensity scale of the N$_2$H$^+$ data is reasonably accurate. 

For CCS, we simply divided the intensities in $T_A^*$ by the main beam efficiency.
The CCS emission is extremely weak for all the targets.

\subsection{Aquila Rift and M17}

The BEARS maps of Aquila Rift were only toward the Serpens South region and the map coverages are very small compared to 
the FOREST mapping areas.  Therefore, we did not combine the BEARS and FOREST data.
Thus, we applied the same procedure as the Orion A C$^{18}$O data calibration to make the $^{12}$CO, $^{13}$CO,
C$^{18}$O, and N$_2$H$^+$ data of Aquila Rift.  After correcting the daily intensity variations of the FOREST data, 
we compared the FOREST intensities to the BEARS intensities toward
Serpens South to determine the scaling factors.  The scaling factors computed agreed with those determined 
with the main beam efficiencies within an error of $\sim$ 5 $-$ 10 \%.  
%Therefore, we simply divided the intensities by the main beam efficiencies to
%make the cube data in the brightness temperature scale.
For M17, we mapped small areas to measure the daily intensity variations and followed the same procedure as that fo Aquila Rift.

\subsection{Main Beam Efficiencies of the Nobeyama 45-m telescope}

The Nobeyama Radio Observatory measures the main beam efficiencies 
of the telescope at several frequencies every season, except the 2016-2017 season
that was our last (third) season.   The planets such as Mars and Jupiter are often 
used for the measurements.
See the web page of the observatory for details of the measurements.
Here, we obtain the main beam efficiencies of the observed lines by fitting the values
measured with several receivers on the 45-m telescope.

\begin{figure}
 \begin{center}
  \includegraphics[width=12cm, bb=-100 0 1200 600]{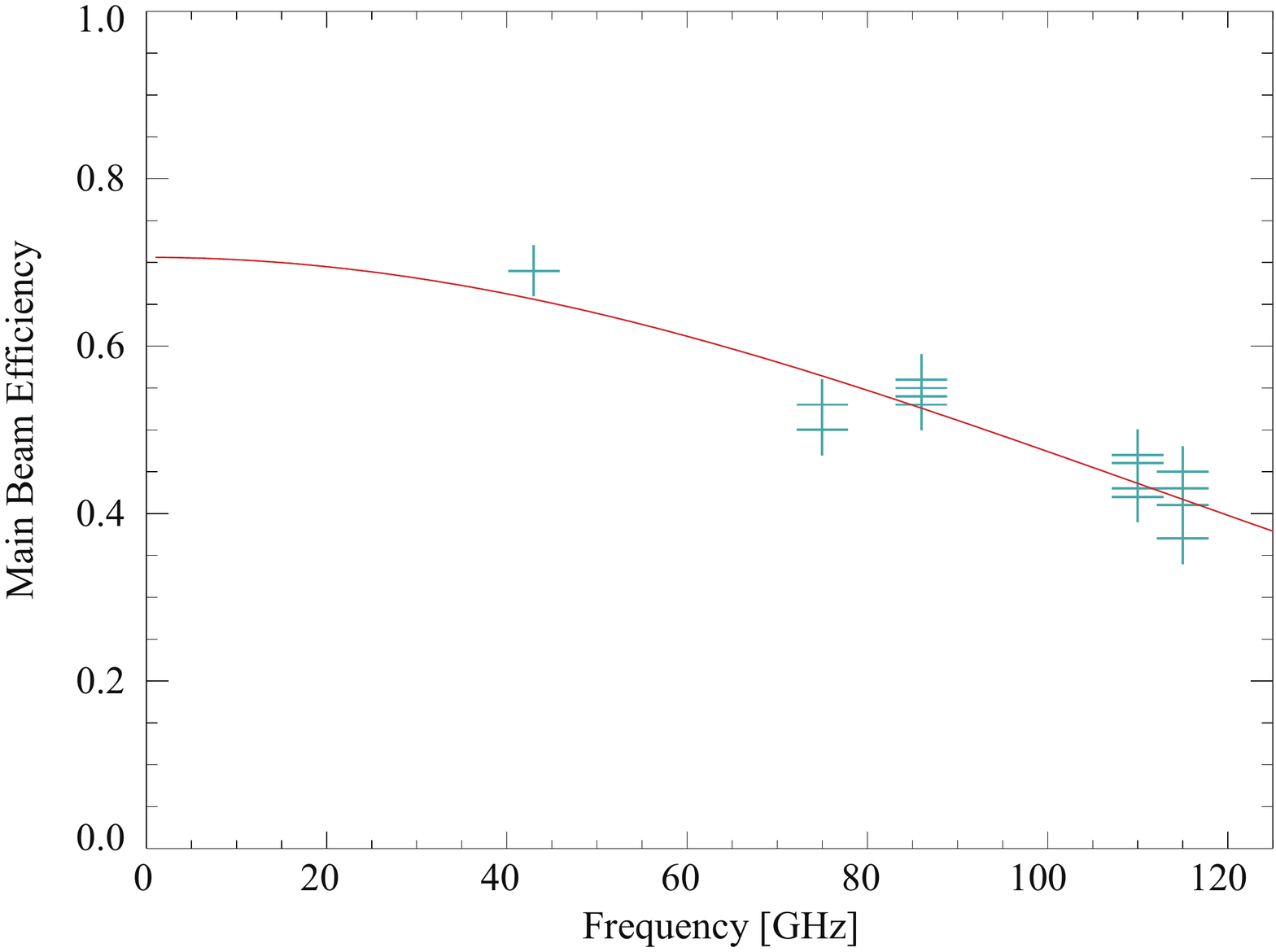}
 \end{center}
\caption{Main beam efficiency of the Nobeyama 45-m telescope as a function of
frequency.  The crosses indicate the efficiencies measured with the receivers installed 
on the 45-m telescope. The red line shows the line fitted with Equation [\ref{eq:eta}].
\label{fig:beam}}
\end{figure}

In figure \ref{fig:beam}, we show the telescope main beam efficiency measured in the 2015-2016 season. We plotted all the values of the main beam efficiencies measured 
with available receivers installed on the 45-m telescope.
By fitting the data points with the following function, 
\begin{equation}
\eta_A=\eta_{A0} \times \exp \left\{-\Bigl({4 \pi \varepsilon \over \lambda} \Bigr)^2 \right\}
\, ,
\label{eq:eta}
\end{equation}
we derive the main beam efficiencies at the frequencies
of our target molecular lines using the least-square fit,
and we obtain $\eta_{A0}=0.539$ and 
the accuracy of the parabola surface is given as $\varepsilon = 0.151$ mm.  
The derived main beam efficiencies are summarized in the last column of table \ref{tab:obs}. 
This procedure is basically the same as the recommended one by the observatory.

We use the obtained main beam efficiencies to convert the intensities measured in the antenna
temperature scale into those in the brightness temperature scale except for 
$^{12}$CO and $^{13}$CO of Orion A.

\section{Global Molecular Gas Distribution}
\label{sec:global}

In this section, we  present the global molecular gas distribution
toward the three target clouds. 
%The details of the individual targets will be discussed in separate papers.
The detailed characteristics, particularly the velocity structures, of the individual clouds will be described in separate papers.

\subsection{Average spectra of the three clouds}

\begin{figure}
 \begin{center}
\includegraphics[width=4cm, bb=-200 0 454 800]{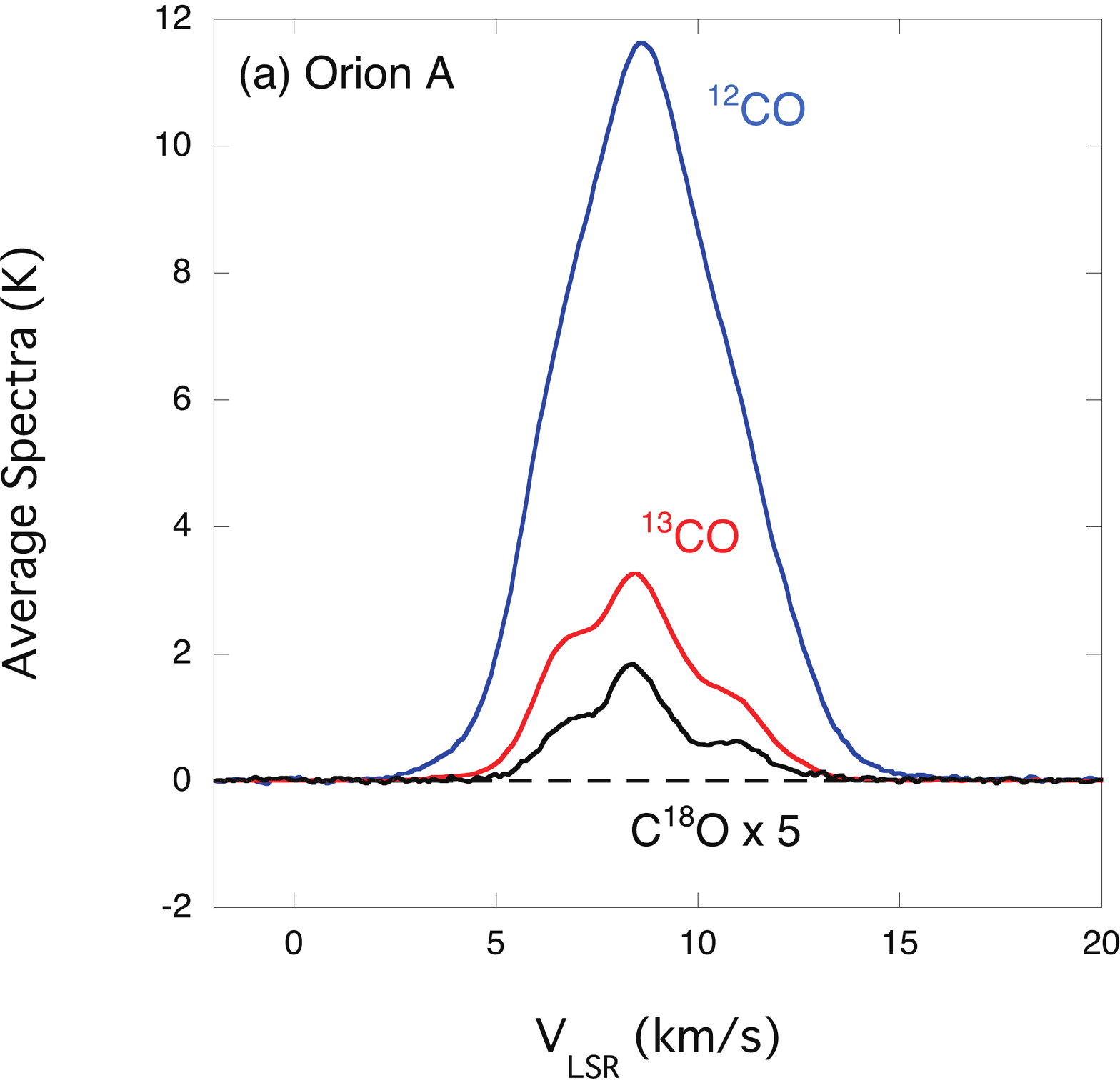}
\includegraphics[width=4cm, bb=-200 0 454 436]{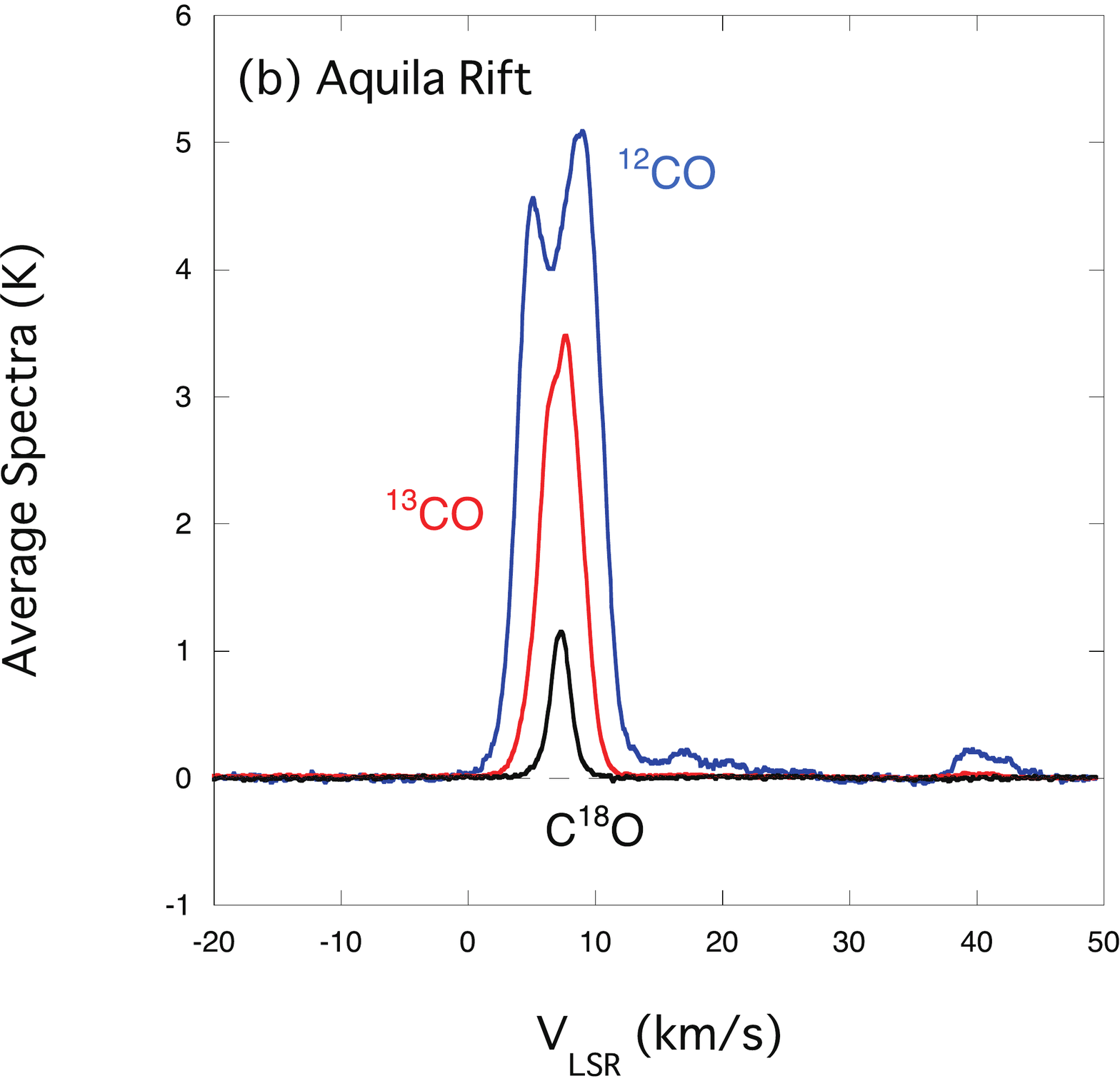}
\includegraphics[width=4cm, bb=-200 0 454 436]{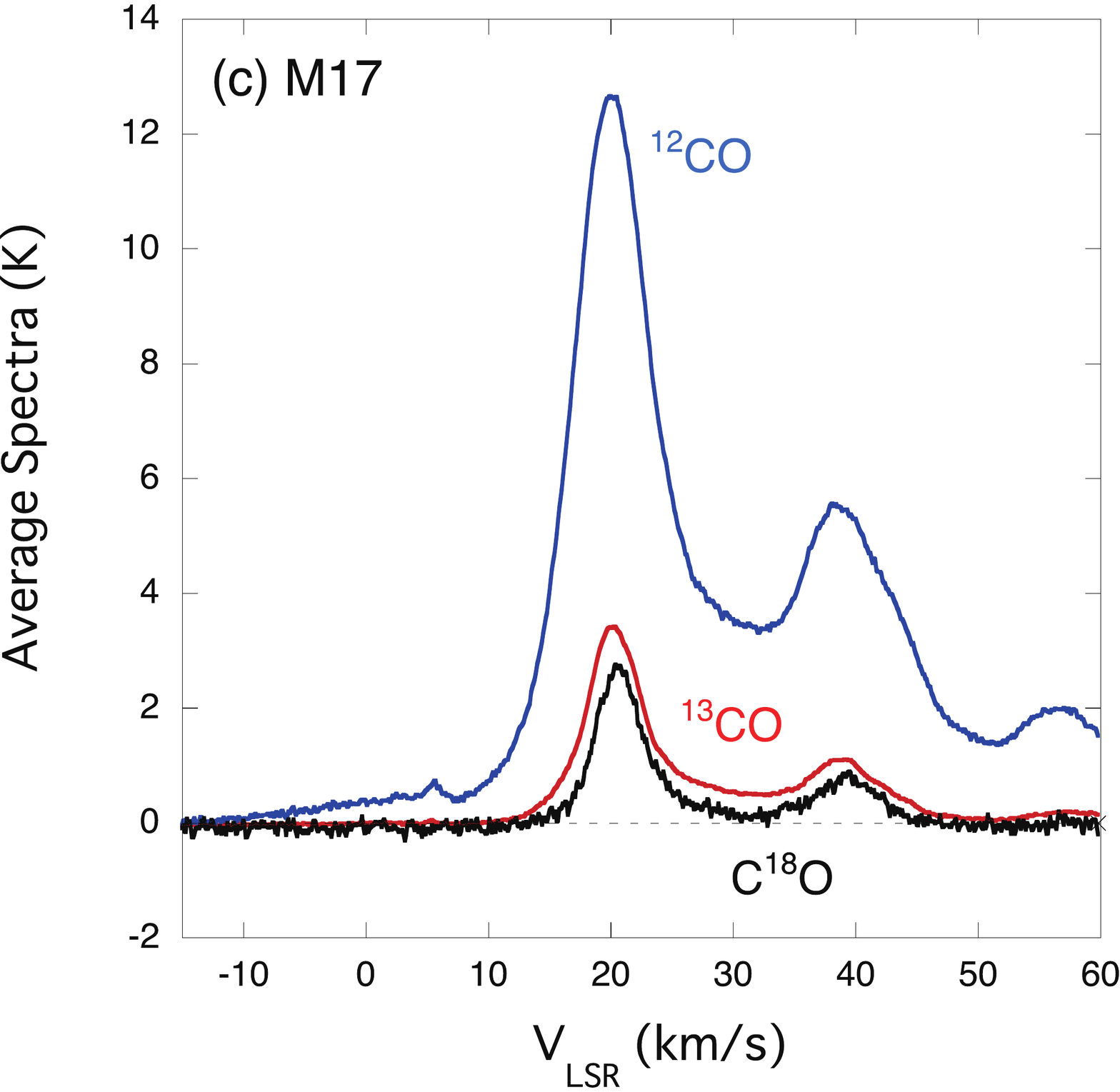}
 \end{center}
\caption{Avrage Spectra of $^{12}$CO, $^{13}$CO, and C$^{18}$O toward (a) Orion A, (b) Aquila Rift, and (c) M17.
The blue, red, and black lines indicate the $^{12}$CO, $^{13}$CO, and C$^{18}$O spectra, respectively.
\label{fig:coprofile}}
\end{figure}

In figure \ref{fig:coprofile}, we present the average spectra of the CO lines for the three clouds.
For Orion A, the average spectrum of the $^{12}$CO emission line has a single peak.
There are three peaks or shoulders in $^{13}$CO and C$^{18}$O at around 7 km s$^{-1}$, 9 km s$^{-1}$, and 11 km s$^{-1}$.
The Orion A filament has a large velocity gradient toward the northern part, and these components
are affected by the global velocity gradient along the filament. However, roughly speaking, 
the 11 km s$^{-1}$ component is dominant toward the northern part, while the 7 km s$^{-1}$ component
appears in the southern part.  The 9 km s$^{-1}$ component is strong in the main filamentary structure.

For the other two regions, the spectra have multiple peaks. For Aquila Rift and
M17, $^{12}$CO has at least three major peaks.  
For Aquila Rift, the C$^{18}$O has a single peak at 7.3 km s$^{-1}$ and thus the dips seen in
$^{12}$CO and $^{13}$CO are expected to be due to the self-absorption \citep{shimoikura18}.
The weak but distinct component at around 40 km s$^{-1}$ may be the molecular gas influenced by 
a supperbubble created by star formation in Scorpius-Centaurus Association \citep{breitschwerdt96,frisch15}.
This component is seen in $^{13}$CO but it is difficult to be recognized in C$^{18}$O.
The $^{12}$CO profile has a tail between 10 km s$^{-1}$ to 35 km s$^{-1}$. This comes from
several different components which reside in this region.  These components are more prominent in the
average spectra of smaller areas and we will discuss them later in a separate paper.
These complicated molecular gas distribution may be due to the interaction of molecular clouds with superbubbles 
\citep{frisch98,frisch15,nakamura17}.

For M17, the three components of $\sim$ 20 km s$^{-1}$, $\sim$ 40 km s$^{-1}$, and $\sim$ 55 km s$^{-1}$
are the molecular gas components which belong to the Sagittarius, Scutum, and Norma arms, respectively, \citep{zucker15}
as discussed in Nguyen Luong et al. (2019), and the main component is the one with 20 km s$^{-1}$
 where NGC 6811 and M17 SWex are located. 

\subsection{Orion A}

%\subsubsection{$^{12}$CO, $^{13}$CO, and C$^{18}$O}

\begin{figure}
 \begin{center}
 \includegraphics[width=4cm, bb=-100 0 442 610]{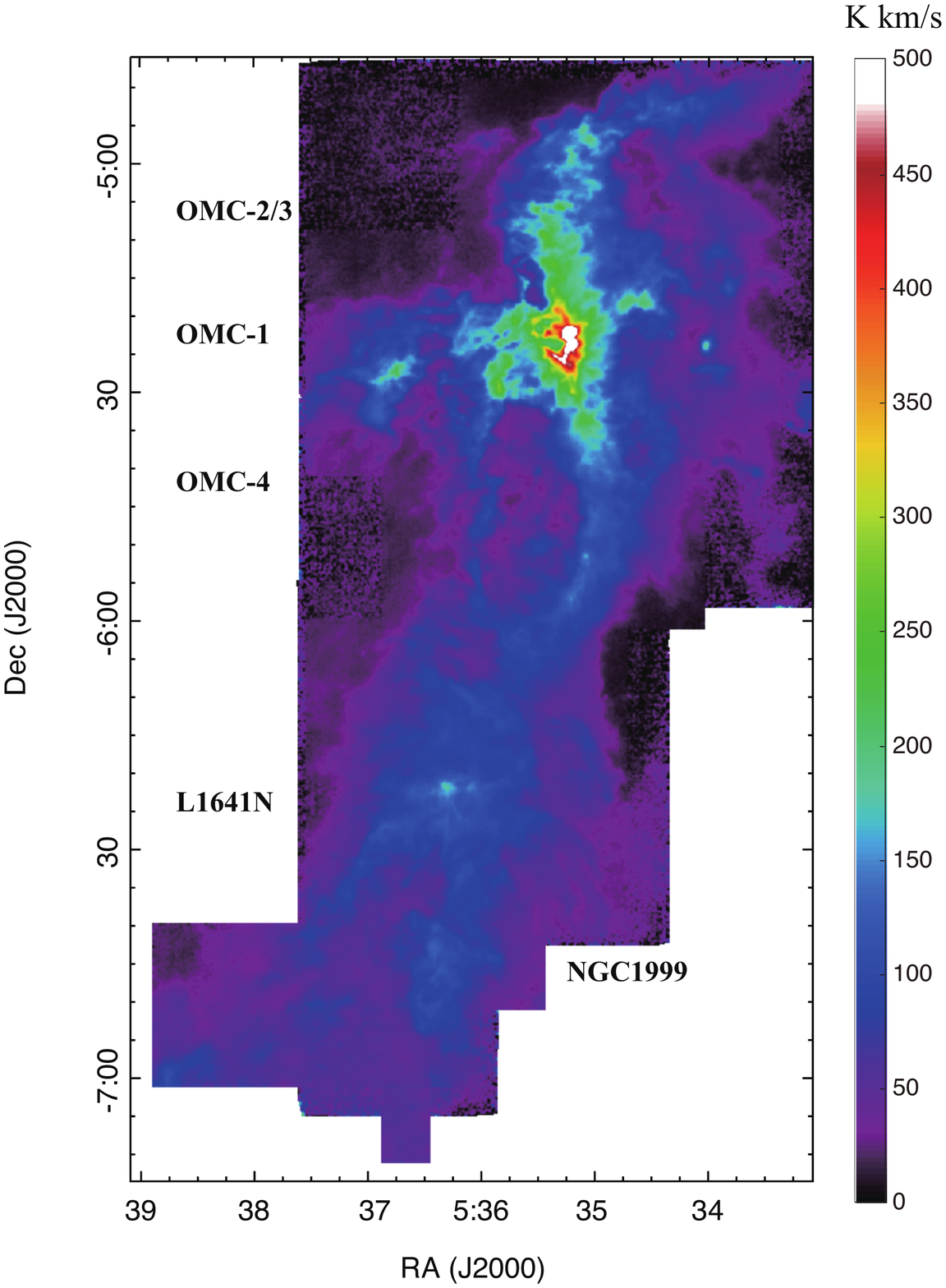}
 \includegraphics[width=8cm, bb=-100 0 442 610]{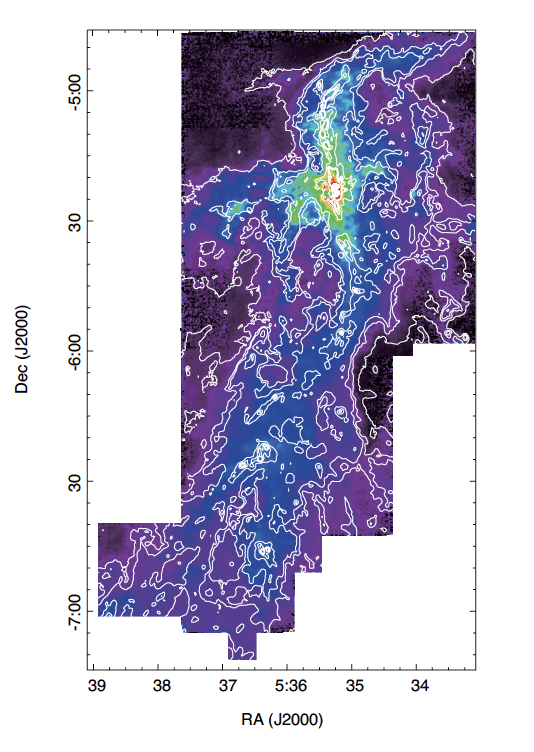}
 \end{center}
\caption{(a) $^{12}$CO ($J=1-0$) moment-0 map of Orion A, velocity-integrated from 2 km s$^{-1}$ to 20 km s$^{-1}$. (b) Same as panel (a) but the contours of the H$_2$ column density
are overlaid on the image.  The contour levels are drawn at $2.5\times 10^{21}$ cm$^{-2}$,
$5.0\times 10^{21}$ cm$^{-2}$, $7.5\times 10^{21}$ cm$^{-2}$, $2.5\times 10^{22}$ cm$^{-2}$,
$5.0\times 10^{22}$ cm$^{-2}$, $7.5\times 10^{22}$ cm$^{-2}$, $\cdots$. 
The effective angular resolution of the $^{12}$CO map is 21$"$.7. 
\label{fig:orion_12co}}
\end{figure}

\begin{figure}
 \begin{center}
 \includegraphics[width=4cm, bb=-100 0 442 610]{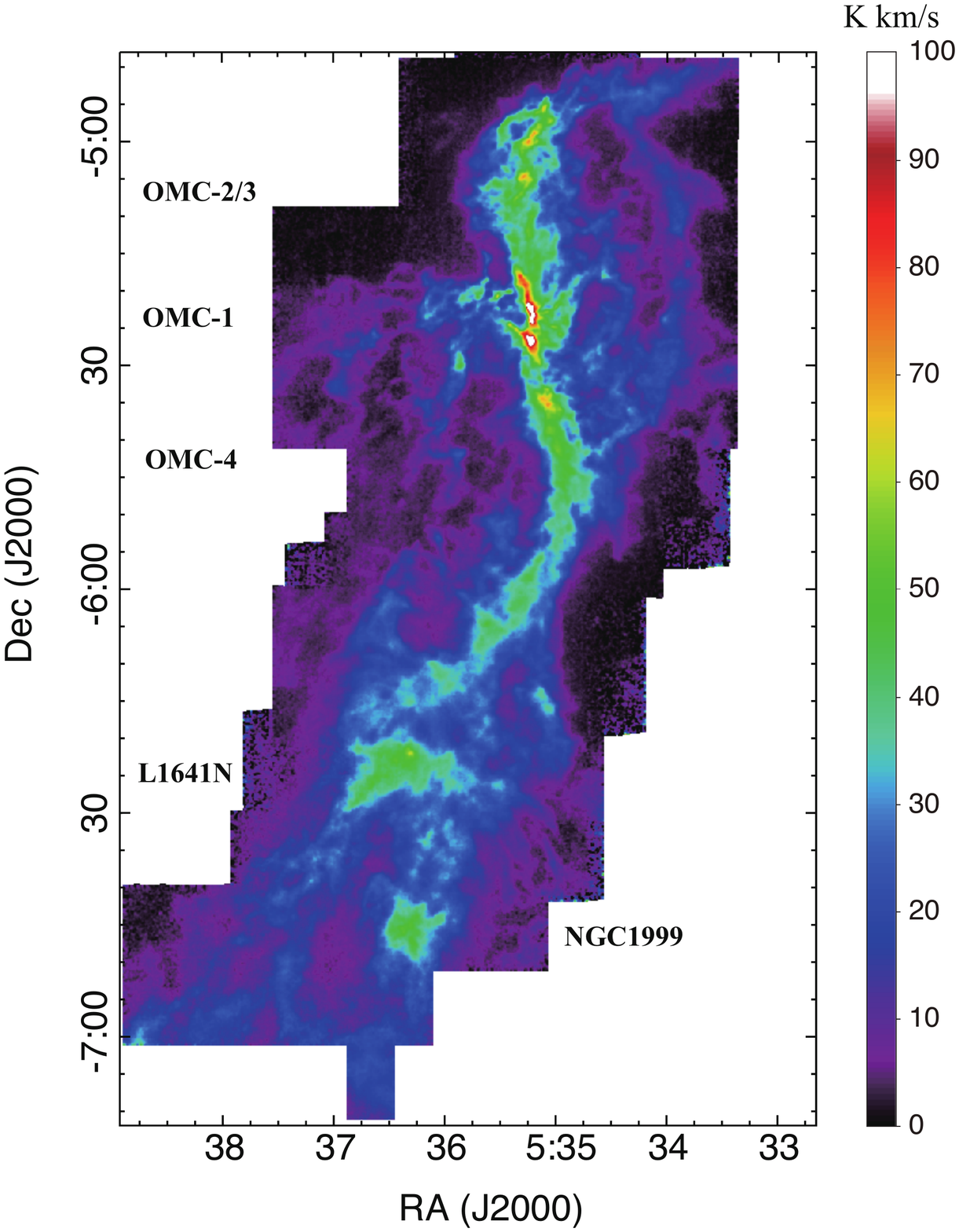}
 \includegraphics[width=12cm, bb=-100 0 442 610]{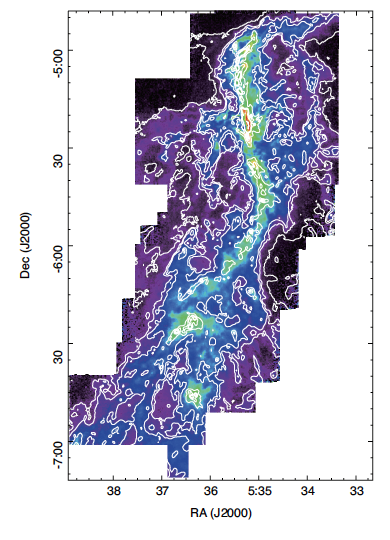}
 \end{center}
\caption{(a) $^{13}$CO ($J=1-0$) moment-0 map of Orion A, velocity-integrated from 2 km s$^{-1}$ to 20 km s$^{-1}$. (b) Same as panel (a) but the contours of the H$_2$ column density
are overlaid on the image.  The contour levels are the same as those of figure \ref{fig:orion_12co}.
The effective angular resolution of the $^{13}$CO map is 22$"$.1. 
\label{fig:orion_13co}
}
\end{figure}

\begin{figure}
 \begin{center}
 \includegraphics[width=8cm, bb=-100 0 442 610]{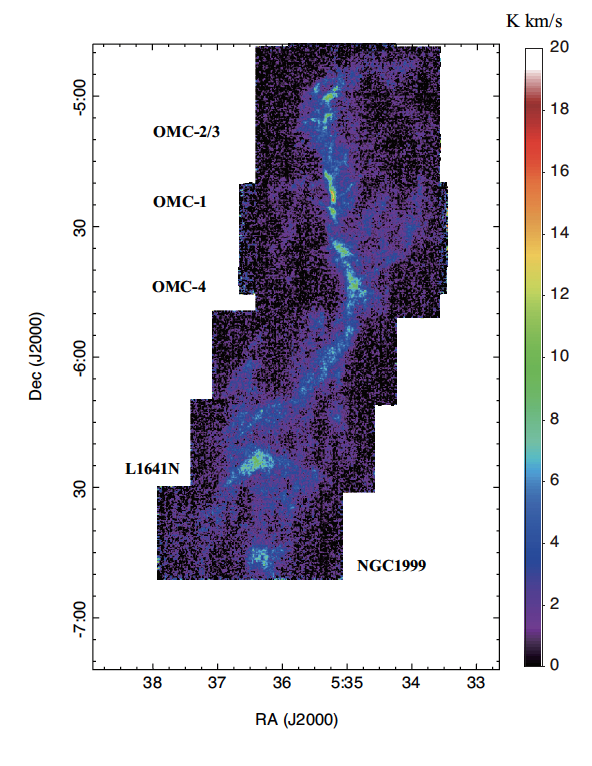}
 \includegraphics[width=8cm, bb=-100 0 442 610]{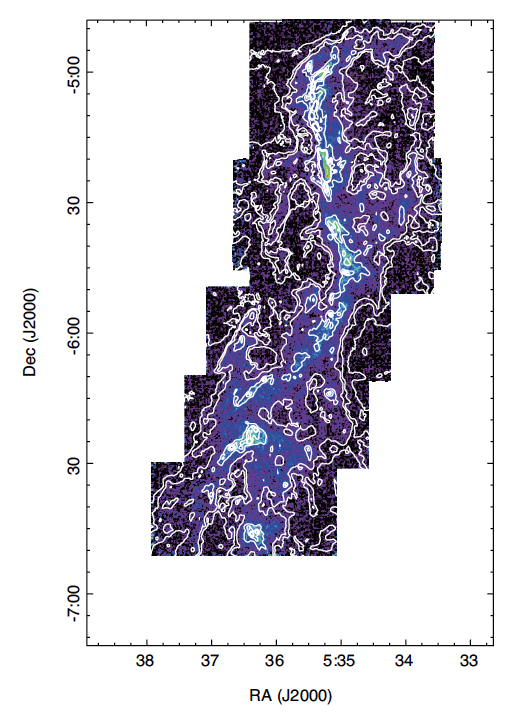}
 \end{center}
\caption{(a) C$^{18}$O ($J=1-0$) moment-0 map of Orion A, velocity-integrated from 2 km s$^{-1}$ to 20 km s$^{-1}$. (b) Same as panel (a) but the contours of the H$_2$ column density
are overlaid on the image.  The contour levels are the same as those of figure \ref{fig:orion_12co}.
The effective angular resolution of the C$^{18}$O map is 22$"$.1. 
\label{fig:orion_c18o}}
\end{figure}

\begin{figure}
 \begin{center}
 \includegraphics[width=4cm, bb=-100 0 442 610]{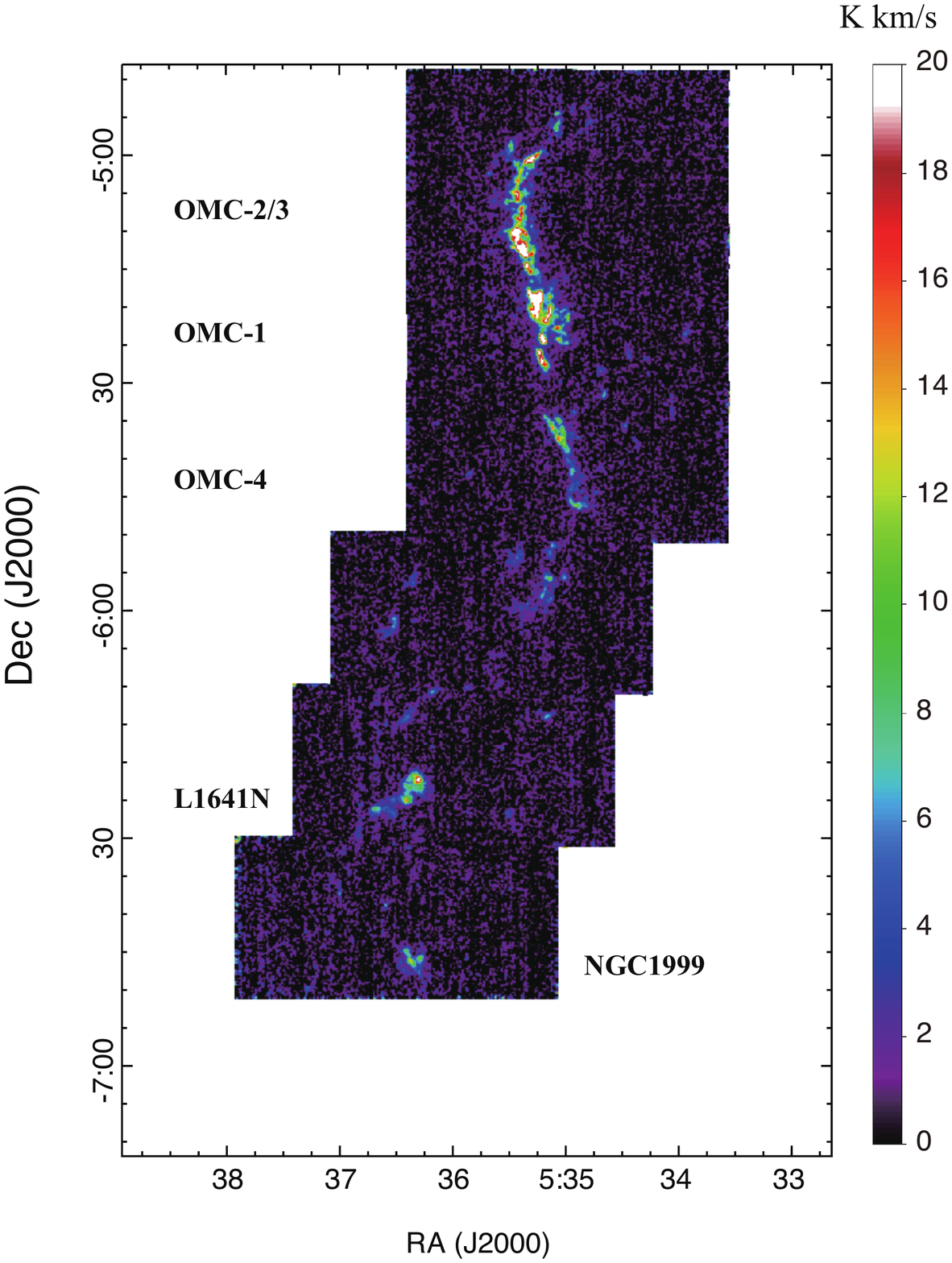}
 \includegraphics[width=8cm, bb=-100 0 442 610]{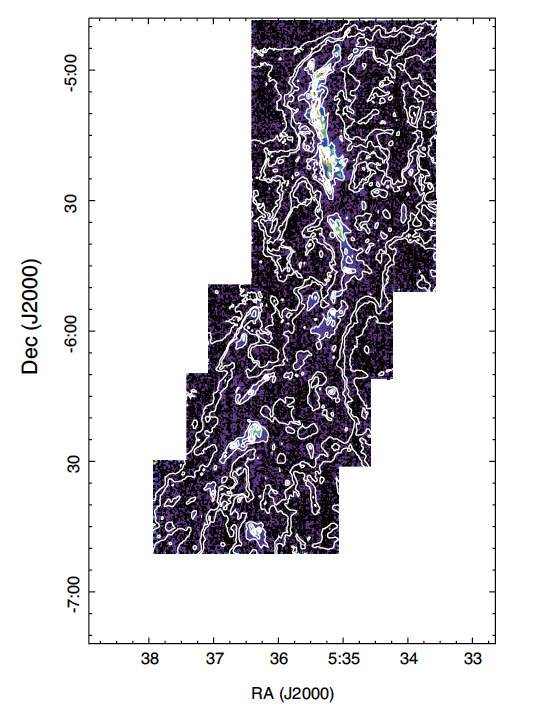}
 \end{center}
\caption{(a) N$_2$H$^+$ ($J=1-0$) velocity integrated map of Orion A. The integration range is from 0 km s$^{-1}$ to 20 km s$^{-1}$. (b) Same as panel (a) but the contours of the H$_2$ column density
are overlaid on the image.  The contour levels are the same as those of figure \ref{fig:orion_12co}.
The effective angular resolution of the N$_2$H$^+$ map is 24$"$.1.
\label{fig:orion_n2hp}}
\end{figure}

\begin{figure}
 \begin{center}
 \includegraphics[width=4cm, bb=-100 0 442 610]{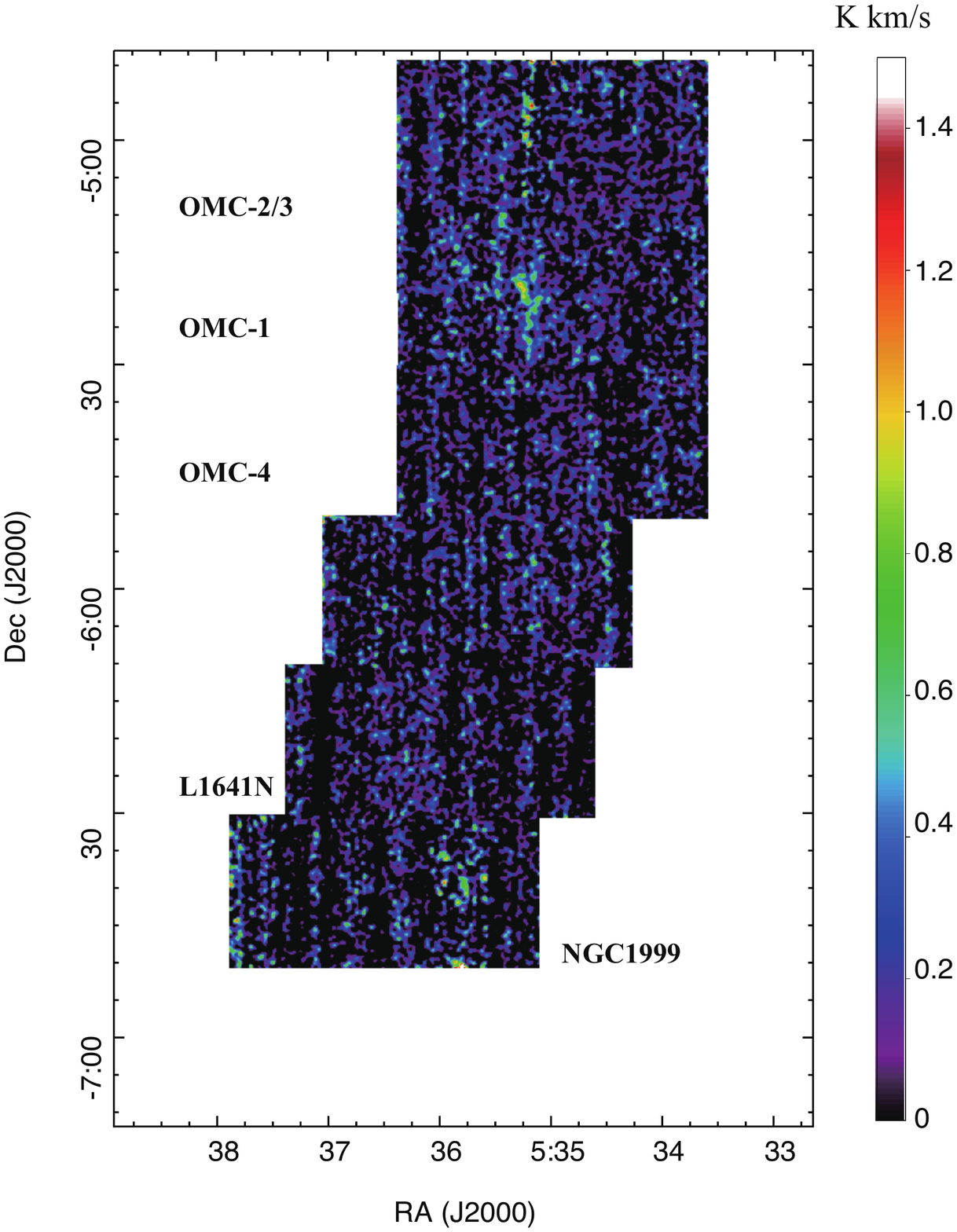}
 \includegraphics[width=10cm, bb=-100 0 442 610]{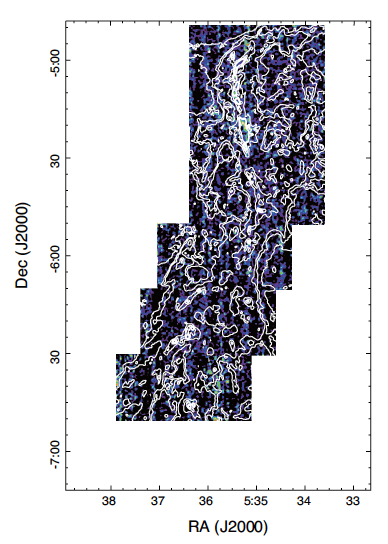}
 \end{center}
\caption{CCS integrated intensity map of Orion A.
We smoothed the CCS image with an effective angular resolution of 32$\arcsec$.
The contour levels are the same as those of figure \ref{fig:orion_12co}.
The effective angular resolution of the CCS map is 24$"$.0. 
\label{fig:orion_ccs}}
\end{figure}

Figures \ref{fig:orion_12co}, \ref{fig:orion_13co}, \ref{fig:orion_c18o}, \ref{fig:orion_n2hp}, and
\ref{fig:orion_ccs} show the integrated intensity maps of Orion A for
$^{12}$CO ($J=1-0$), $^{13}$CO ($J=1-0$), C$^{18}$O ($J=1-0$), 
N$_2$H$^+$ ($J=1-0$), and CCS ($J=8_7-7_6$), respectively.
For comparison, we overlaid   the contours of the {\it Herschel} H$_2$ column density map on the integrated intensity image 
in the right panel of each figure.
Each map except N$_2$H$^+$ is integrated from 2 km s$^{-1}$ to 20 km s$^{-1}$.
For N$_2$H$^+$, we integrated the emission from 0 km s$^{-1}$ to 22 km s$^{-1}$ so that all seven hyperfine 
components are summed up.
%We indicated several famous regions in Figure \ref{fig:orion12co}.
Our maps cover a region from OMC-3 to NGC1999, spanning about 2 degrees in declination.
%The details of the Orion A $^{12}$CO, $^{13}$CO, and C$^{18}$O maps with finer angular resolutions are presented in \citet{kong18} 
%with the CARMA+Nobeyama  combined data. 
The results of protostellar outflow survey and cloud structure analysis are given in separate papers 
(Tanabe et  al. 2019; Ishii et al. 2019; Takemura et al. 2019).

Figures \ref{fig:orion_12co}, \ref{fig:orion_13co}, \ref{fig:orion_c18o} show that the $^{12}$CO, $^{13}$CO, and C$^{18}$O emission 
trace the areas with the column density higher than $\sim 0.5 \times 10^{22}$ cm$^{-2}$, $\sim 0.75 \times 10^{22}$ cm$^{-2}$, 
and $\sim 2.5 \times 10^{22}$ cm$^{-2}$, respectively.
N$_2$H$^+$ emission comes from the area with the column density larger than $\sim 5 \times 10^{22}$ cm$^{-2}$
(see figure \ref{fig:orion_n2hp}).

In figure \ref{fig:orion_ccs}, we show the integrated intensity map of CCS, where the image was
smoothed with an effective angular resolution of 32$\arcsec$ to improve the signal to noise ratios.
The CCS emission is significantly detected only in the OMC-1 region where the strong C$^{18}$O and N$_2$H$^+$ emission is detected.
Our CCS map is the first unbiased CCS map of Orion  A with the $J_N=8_7-7_6$ line.  Previous wide-field maps were taken only toward 
the main filament with other transition lines ($J_N=4_3-3_2$ at 45 GHz and $J_N=7_6-6_5$ at 81.5 GHz) by \citet{tatematsu08,tatematsu14}.
CCS is known to trace the dense gas with densities of $10^4$ cm$^{-3}$, but the abundance of CCS
decreases very rapidly due to the destruction.  
In Orion A, the CCS emission is very weak.  Only toward the OMC-1 region, we detect the emission 
with a signal-to-noise ratio of 5 $\sigma$. 
Weaker emission is sometimes seen along the ridge. 
This weak CCS emission implies that Orion A is relatively evolved molecular cloud.
We note that CCS is detected in the OMC-2 FIR 4 region for much higher sensitivity observations \citep{nakamura19}.
The OMC-1 region may be relatively chemically-young compared to other parts in Orion A.  Recently, \citet{hacar17} proposed that the OMC-1
region is gravitationally contracting along the main filament.  Such a global infall motion can be recognized in our $^{13}$CO data \citep{ishii19}.
If OMC-1 is indeed infalling toward the center, the gas is continuously fed  along the main filament
in OMC-1. 
Thus,  the significant CCS emission in OMC-1 may come from the material newly fed from outside 
by the gravititional contraction.

Figure  \ref{fig:orion_n2hp} shows the N$_2$H$^+$ map of the Orion A molecular cloud.  The map is consistent with the image taken by \citet{tatematsu08},
who mapped the Orion A filament in a position-switch mode with full-beam sampling, using the receiver BEARS.  Our mapping area is much wider than theirs.
Also, since we mapped the Orion A in the  OTF mode, the angular resolution is somewhat better than that of \citet{tatematsu08}.
The N$_2$H$^+$ emission  is stronger in the northern part (OMC-1, OMC-2, and OMC-3) and traces the main filamentary structure 
running in the north-south direction.
Our N$_2$H$^+$ map shows that  many faint N$_2$H$^+$ cores are distributed outside the main filament as well.

\subsection{Aquila Rift}

\begin{figure}
 \begin{center}
 \includegraphics[angle=90, width=6cm,bb=0 -400 601 532]{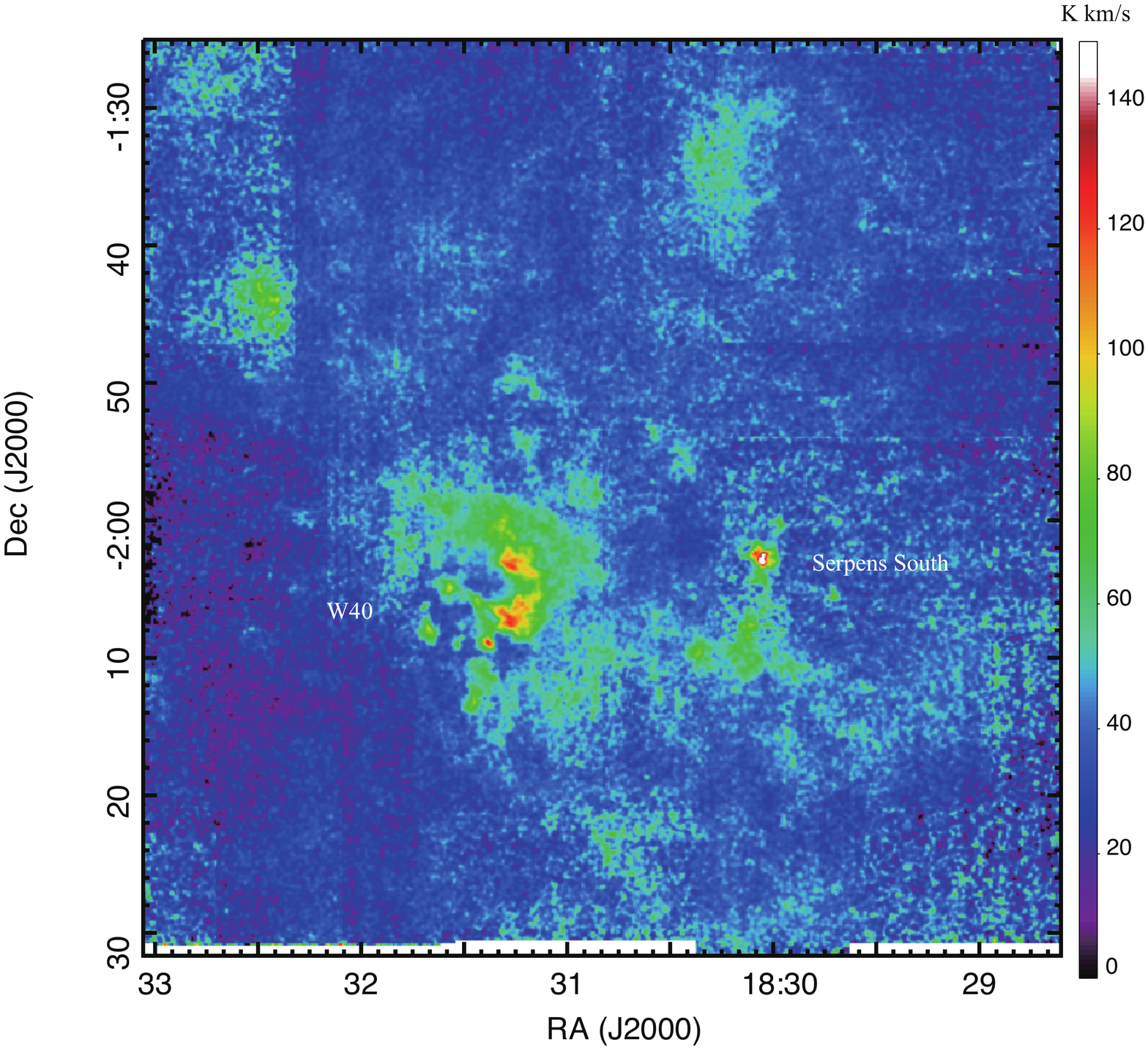}
 \includegraphics[angle=90,width=6cm,bb=0 -400 601 532]{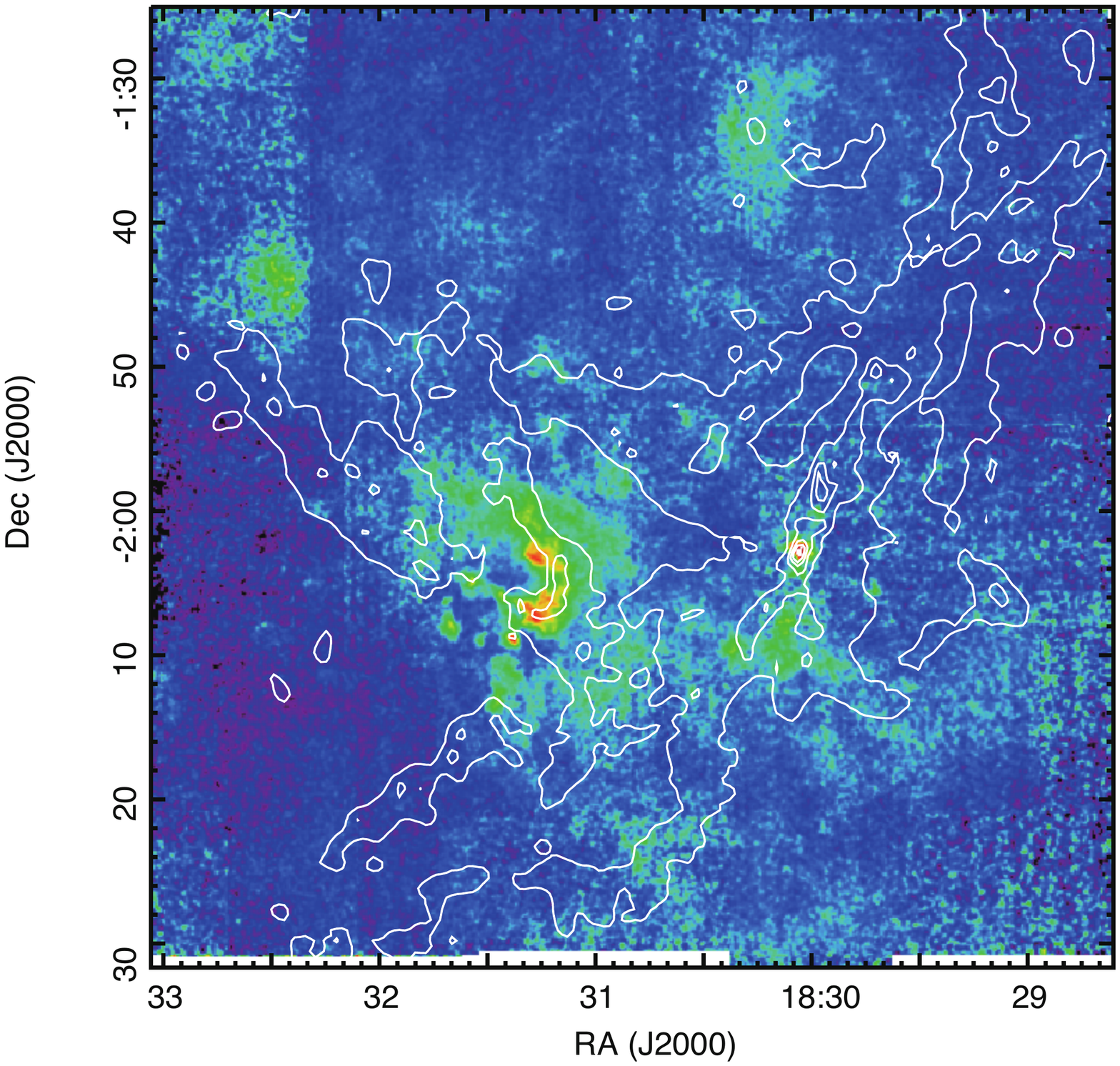}
 \end{center}
\caption{$^{12}$CO ($J=1-0$) integrated intensity map of Aquila Rift.
In the panel (b), the contour levels are drawn at $1.0\times 10^{22}$ cm$^{-2}$,
$2.5\times 10^{22}$ cm$^{-2}$, $5.0\times 10^{22}$ cm$^{-2}$, $7.5\times 10^{22}$ cm$^{-2}$,
$1.0\times 10^{23}$ cm$^{-2}$, $1.5\times 10^{23}$ cm$^{-2}$.
The effective angular resolution of the $^{12}$CO map is 21$"$.7. 
\label{fig:aquila_12co}}
\end{figure}

\begin{figure}
 \begin{center}
 \includegraphics[angle=90, width=6cm,bb=0 -400 601 532]{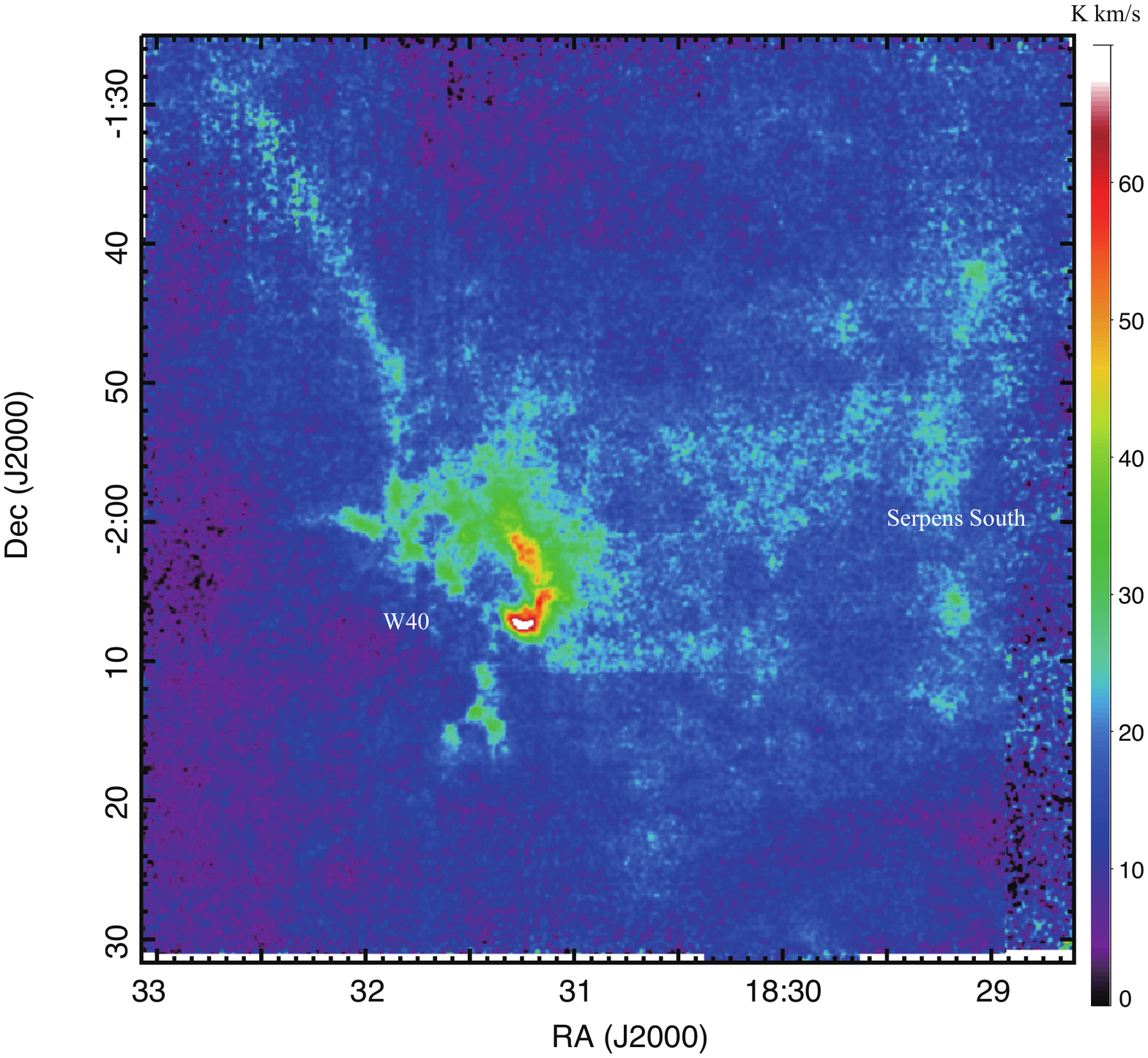}
 \includegraphics[width=4cm,bb=0 100 601 532]{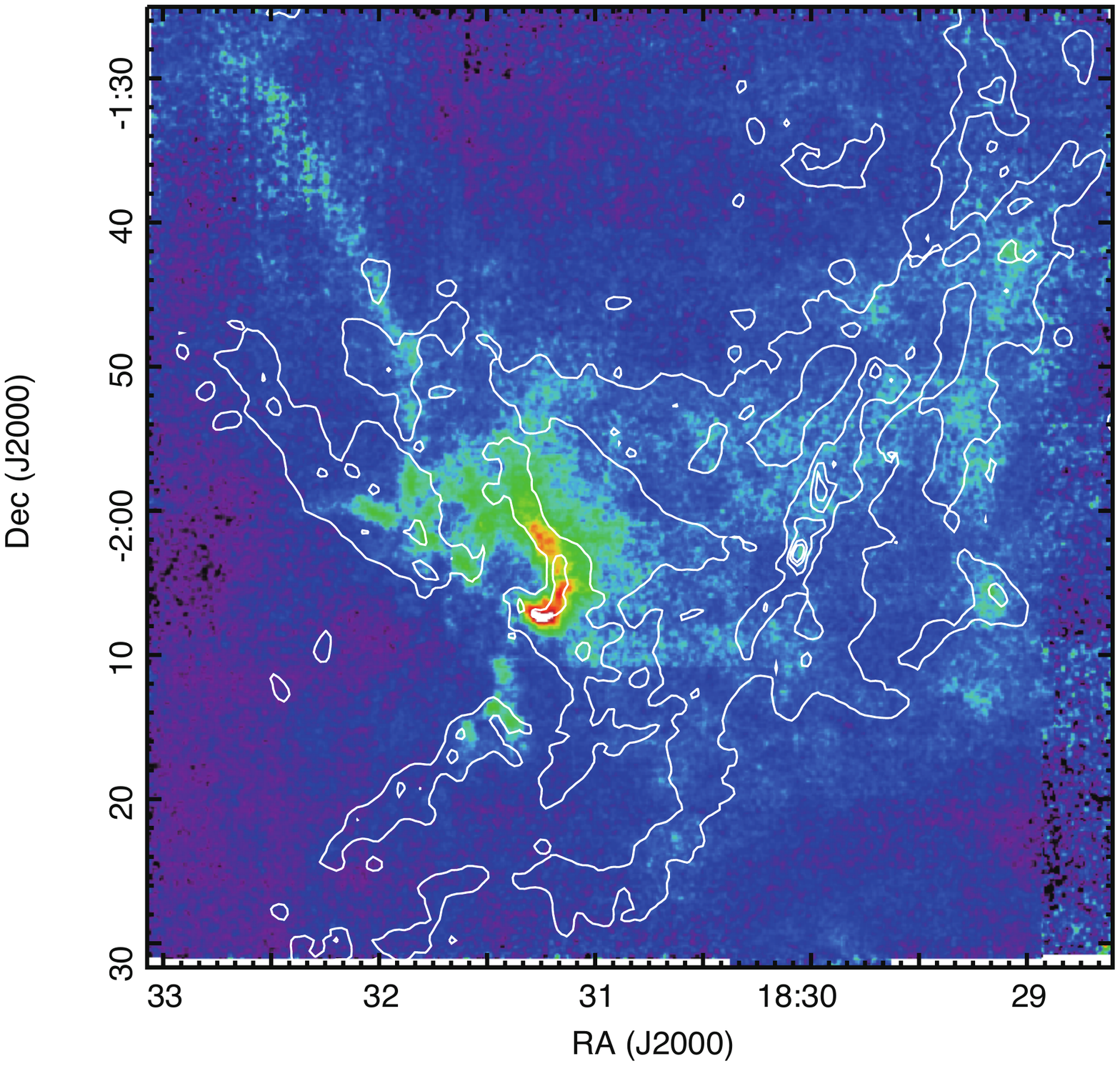}
 \end{center}
\caption{$^{13}$CO ($J=1-0$) integrated intensity map of Aquila Rift.
The contour levels are the same as those of figure \ref{fig:aquila_12co}.
The effective angular resolution of the $^{13}$CO map is 22$"$.1. 
\label{fig:aquila_13co}}
\end{figure}

\begin{figure}
 \begin{center}
 \includegraphics[width=6cm,bb=-300 0 601 532]{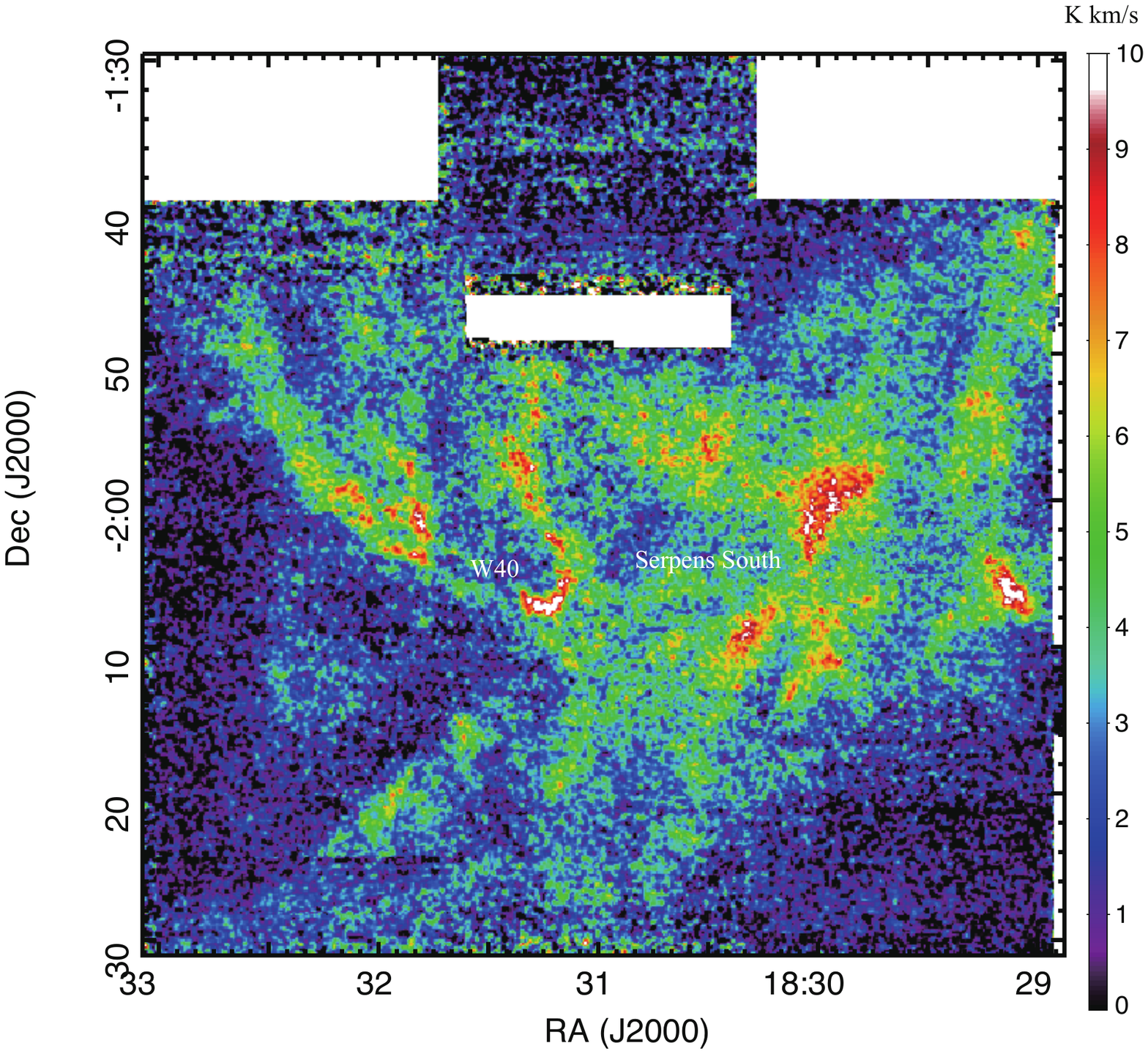}
 \includegraphics[width=6cm,bb=-300 0 601 532]{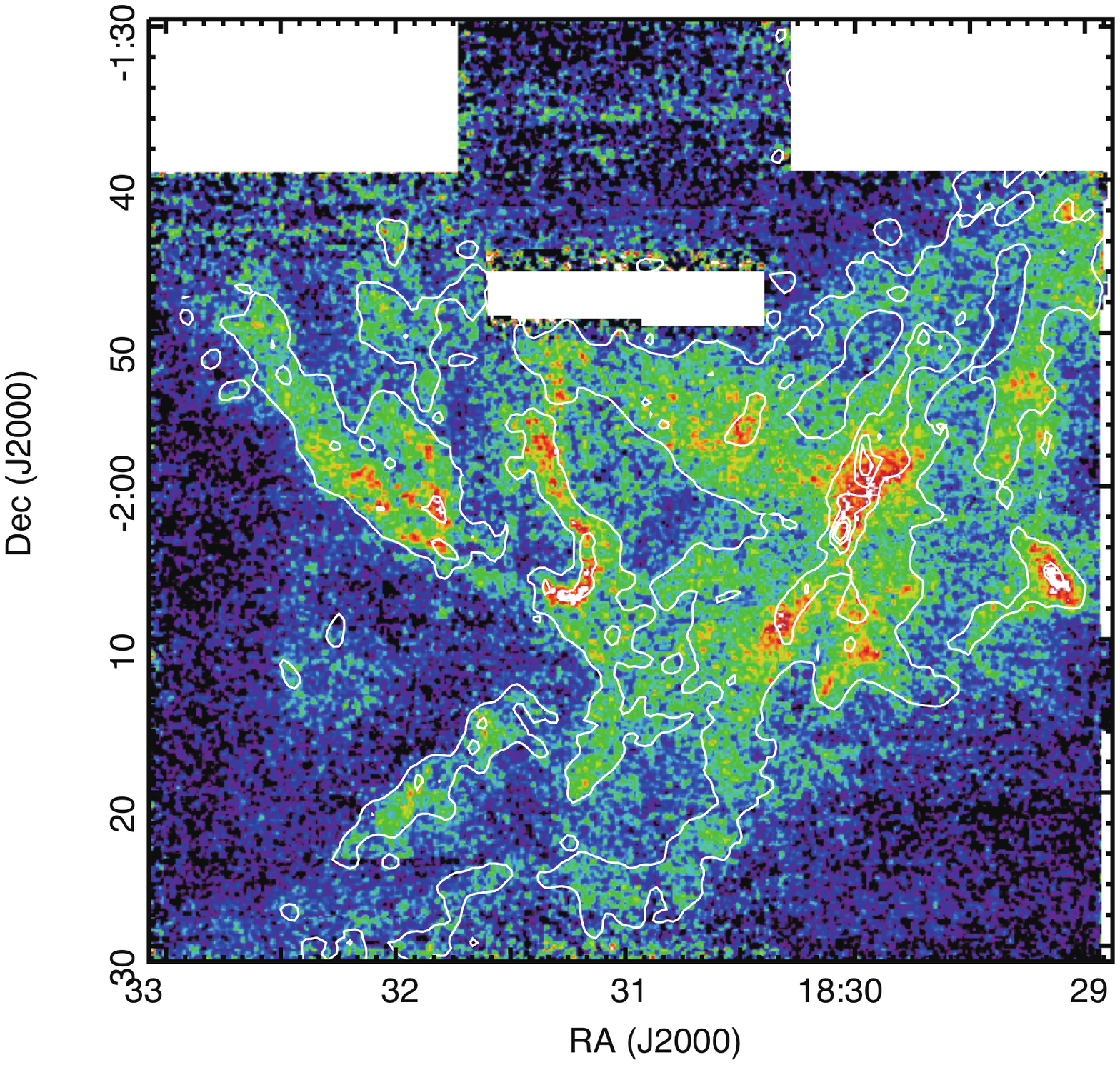}
 \end{center}
\caption{C$^{18}$O ($J=1-0$) integrated intensity map of Aquila Rift.
The contour levels are the same as those of figure \ref{fig:aquila_12co}.
The effective angular resolution of the C$^{18}$O map is 22$"$.1. 
\label{fig:aquila_c18o}}
\end{figure}

\begin{figure}
 \begin{center}
  \includegraphics[width=6cm,bb=-200 0 601 532]{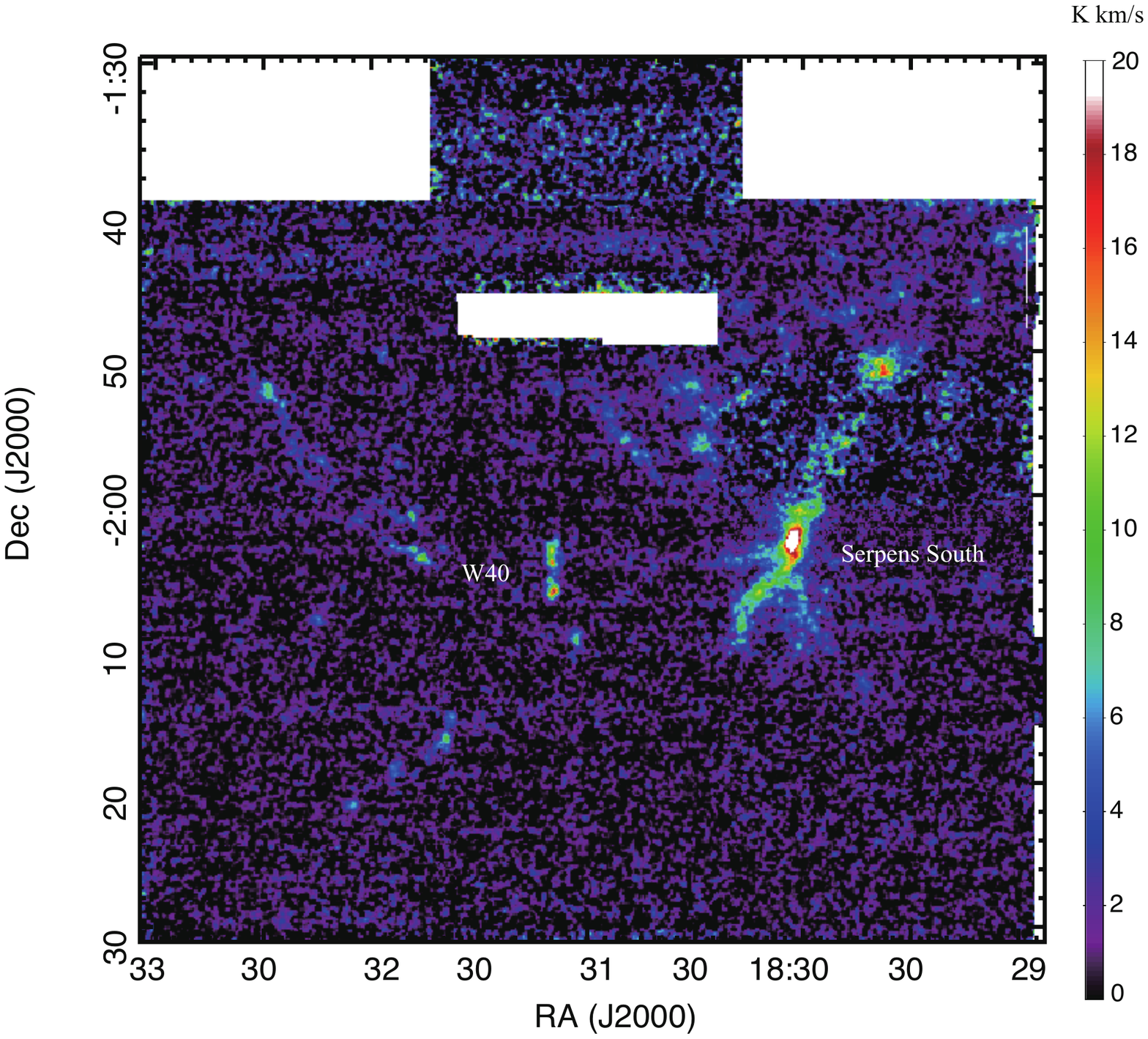}
 \includegraphics[width=6cm,bb=-200 0 601 532]{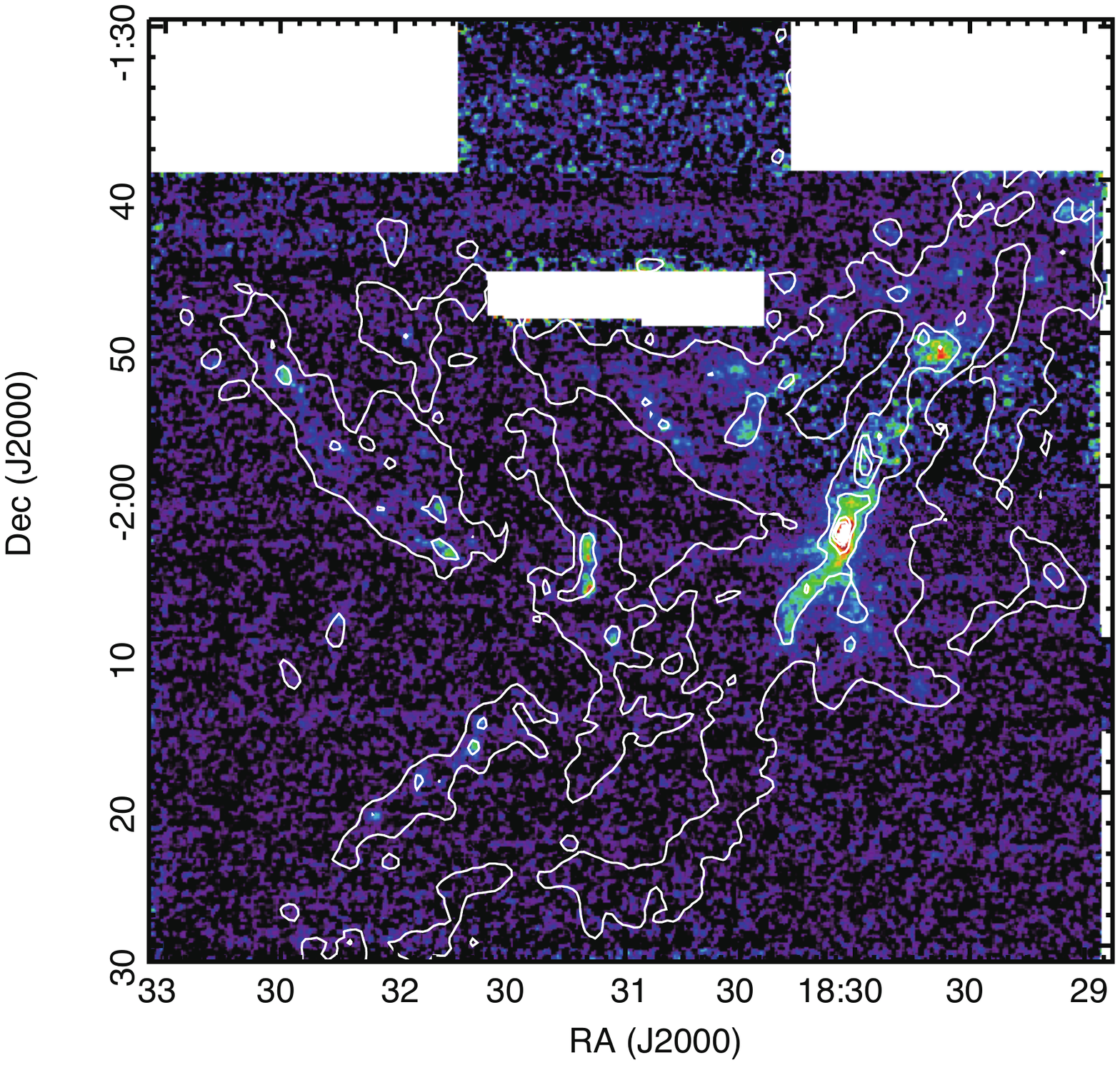}
 \end{center}
\caption{N$_2$H$^+$ ($J=1-0$) integrated intensity map of Aquila Rift.
The contour levels are the same as those of figure \ref{fig:aquila_12co}.
The effective angular resolution of the N$_2$H$^+$ map is 24$"$.1.
\label{fig:aquila_n2hp}}
\end{figure}

\begin{figure}
 \begin{center}
 \includegraphics[width=6cm,bb=-200 0 601 532]{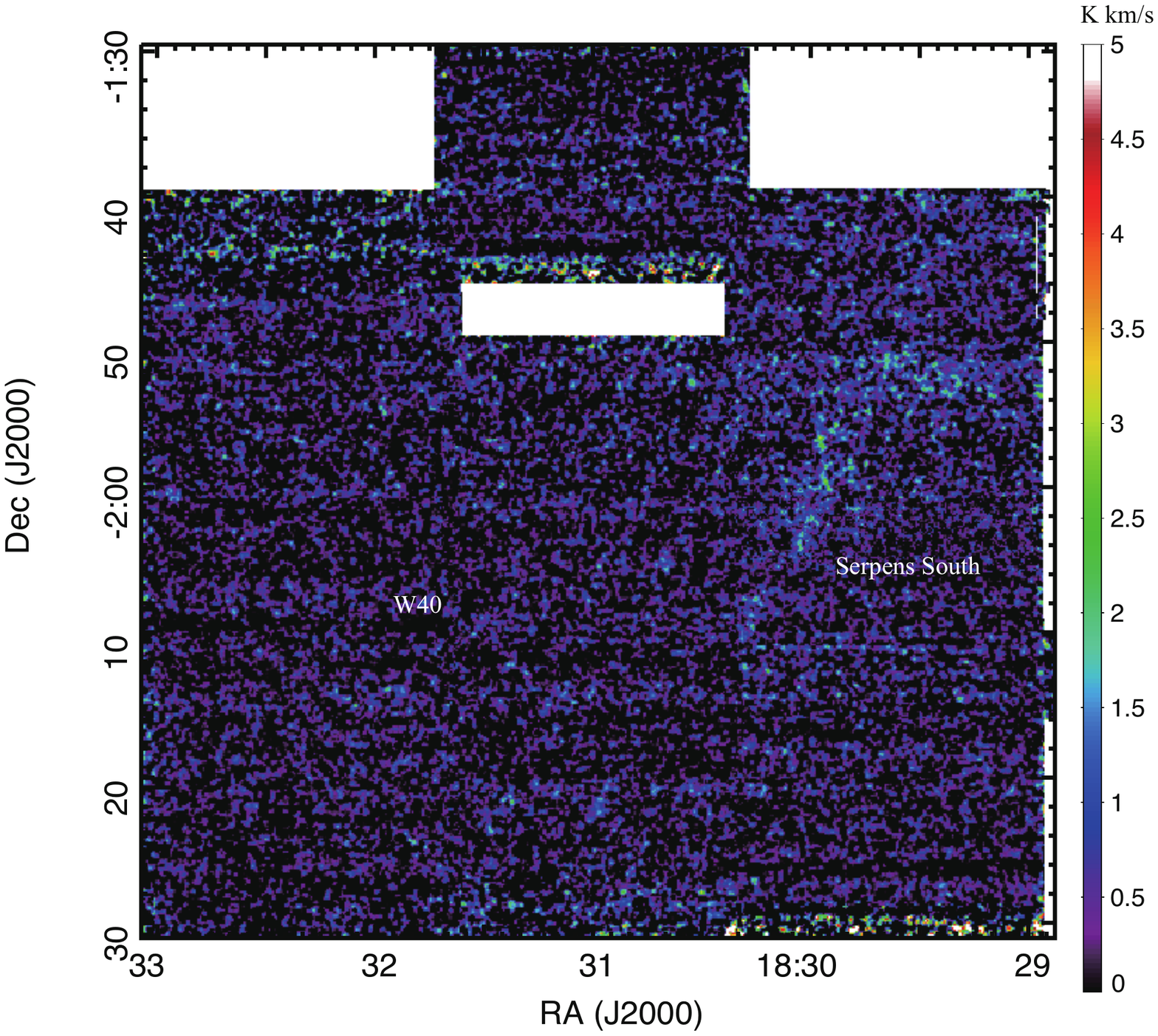}
 \includegraphics[width=6cm,bb=-200 0 601 532]{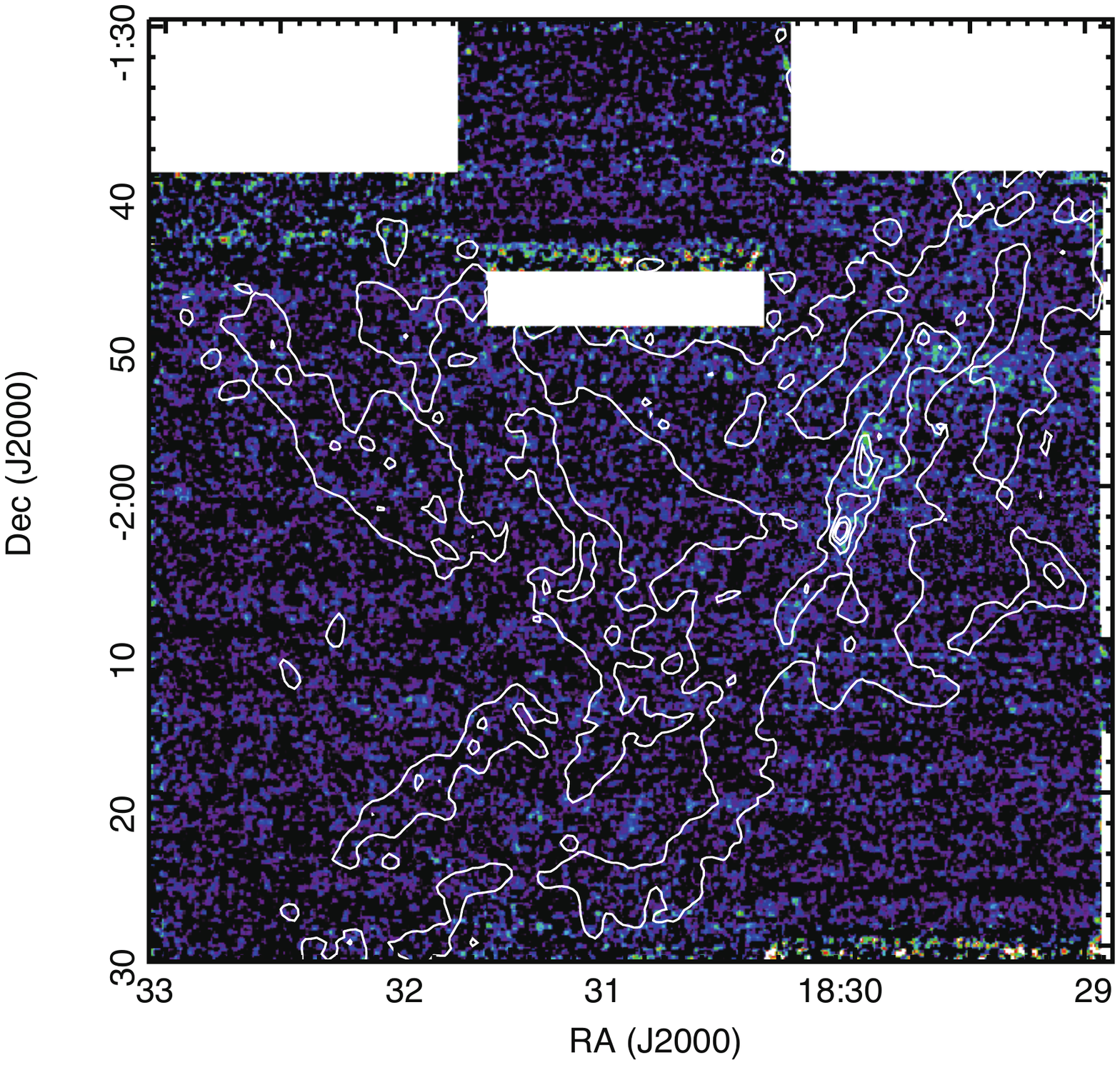}
 \end{center}
\caption{CCS ($J=8_7-7_6$) integrated intensity map of Aquila Rift.
The contour levels are the same as those of figure \ref{fig:aquila_12co}.
The effective angular resolution of the CCS map is 24$"$.0. 
\label{fig:aquila_ccs}}
\end{figure}

Figure \ref{fig:aquila_12co}, \ref{fig:aquila_13co}, \ref{fig:aquila_c18o}, \ref{fig:aquila_n2hp}, and
\ref{fig:aquila_ccs} show the integrated intensity maps of Aquila Rift for
$^{12}$CO ($J=1-0$), $^{13}$CO ($J=1-0$), C$^{18}$O ($J=1-0$), 
N$_2$H$^+$ ($J=1-0$), and CCS ($J=7_6-6_5$), respectively.
Each map except N$_2$H$^+$ is integrated from $-$10 km s$^{-1}$ to 45 km s$^{-1}$.
For N$_2$H$^+$, we integrated all seven hyperfine components to produce the intensity map.

The $^{12}$CO emission is strongest toward the W40 region and the Serpens South cluster.
The $^{13}$CO emission traces an arc-like structure in the W40 region.
However, the filamentary structures particularly toward the Serpens South region  are difficult to be recognized 
in the $^{12}$CO and $^{13}$CO images.  From the C$^{18}$O image, we can vaguely find the filamentary structures
in Serpens South. 
The filamentary structures detected by the {\it Herschel} map are prominent toward Serpens South in N$_2$H$^+$

\begin{figure}
 \begin{center}
 \includegraphics[width=8cm,bb=0 0 601 700]{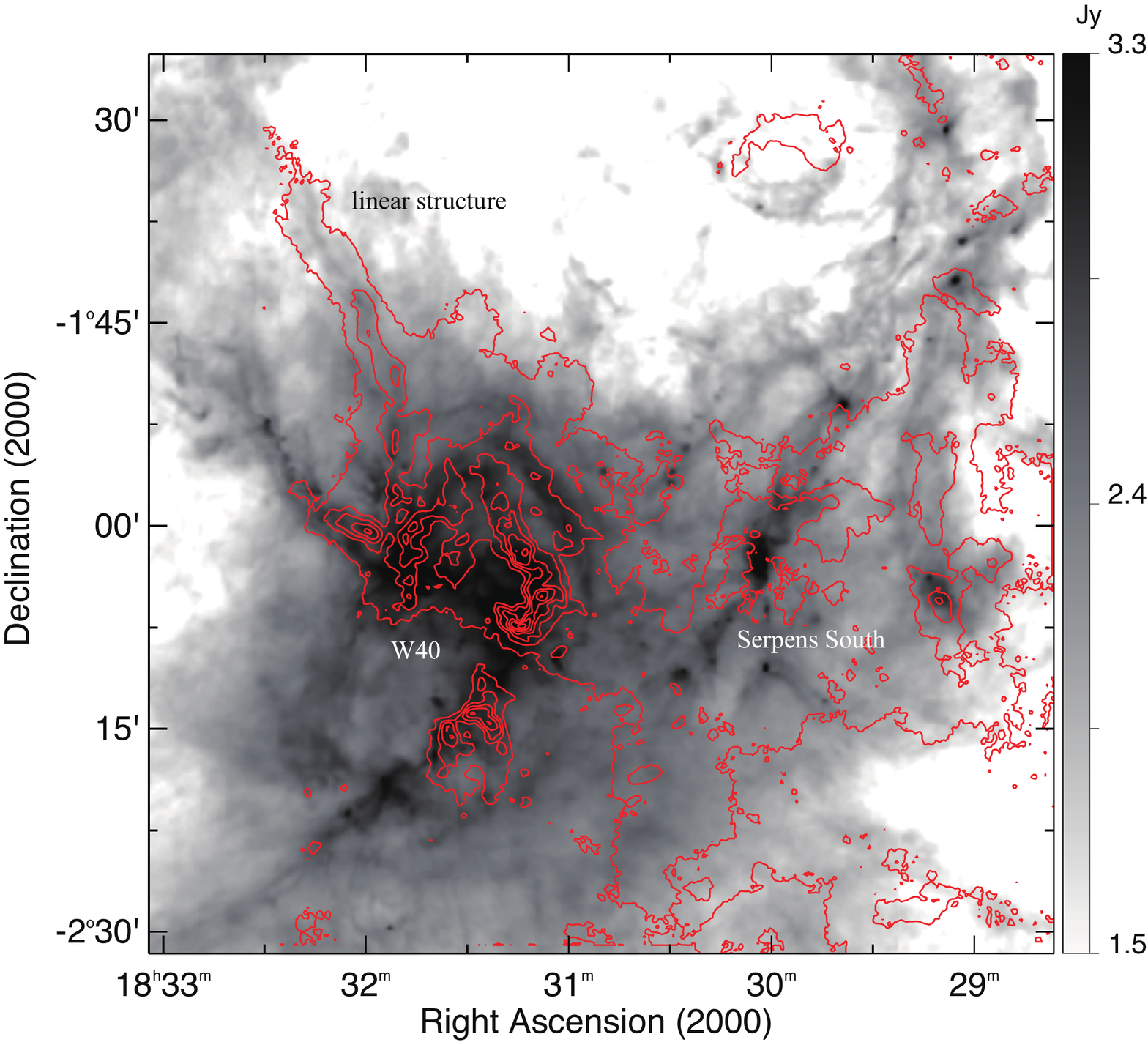}
 \end{center}
\caption{The $^{13}$CO intensity map overlaid on the Herschel 250 $\mu$m image. The integration range of the $^{13}$CO
is from 5.0 km s$^{-1}$ to 6.6 km s$^{-1}$. The coutours start at 4.0 K km s$^{-1}$ with an interval of 4.0 K km s$^{-1}$.
\label{fig:13co_filament}}
\end{figure}

A prominent linear structure is seen in the north east part of the observed area from W40 in $^{13}$CO map. 
This $^{13}$CO structure is less prominent in the Herschel column density map.
This structure may be created by the interaction of the molecular cloud and  the W40 HII region. 
Similar linear structures are seen in C$^{18}$O at different parts of the mapped area. In figure \ref{fig:13co_filament},
we compare the 250$\mu$m Herschel image (gray scale) with the $^{13}$CO image velocity-integrated from 5.0 km s$^{-1}$ to 6.6 km s$^{-1}$
(contours). Weak 250 $\mu$m emission appears to trace the $^{13}$CO linear structure.
See \citet{shimoikura19} for the details of the cloud structure and properties of the Aquila Rift.
The $^{12}$CO integrated intensity map indicates the presence of protostellar outflows particularly in the Serpens South 
protocluster region [see also \citet{nakamura11,shimoikura15}].
In Section \ref{sec:outflow}, we present the results of outflow survey toward the Serpens South region.
The CCS emission is weak but significant emission comes along the Serpens South filament.
This indicates that the Serpens South filaments are relatively chemically-young.

\subsubsection{M17}

\begin{figure}
 \begin{center}
 \includegraphics[angle=90,width=8cm,bb=0 0 601 532]{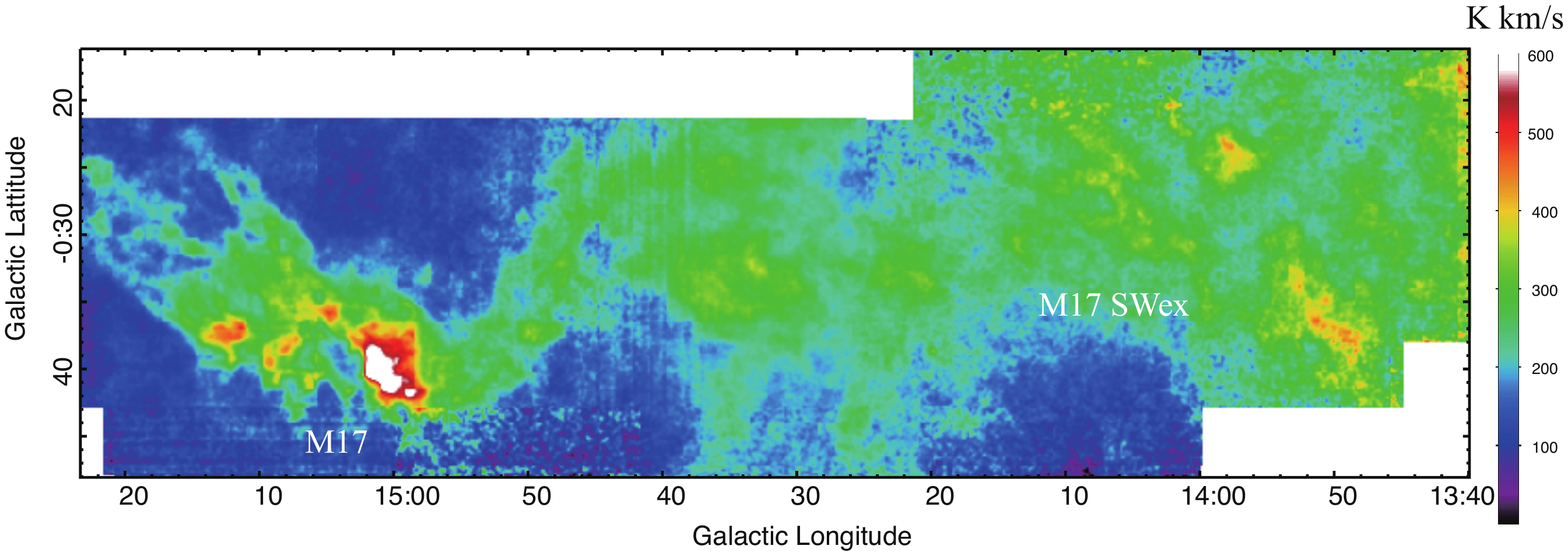}
 \end{center}
\caption{$^{12}$CO ($J=1-0$) integrated intensity map of M17.
The effective angular resolution of the $^{12}$CO map is 21$"$.7. 
\label{fig:m1712co}}
\end{figure}

\begin{figure}
 \begin{center}
 \includegraphics[angle=90,width=8cm,bb=0 0 601 532]{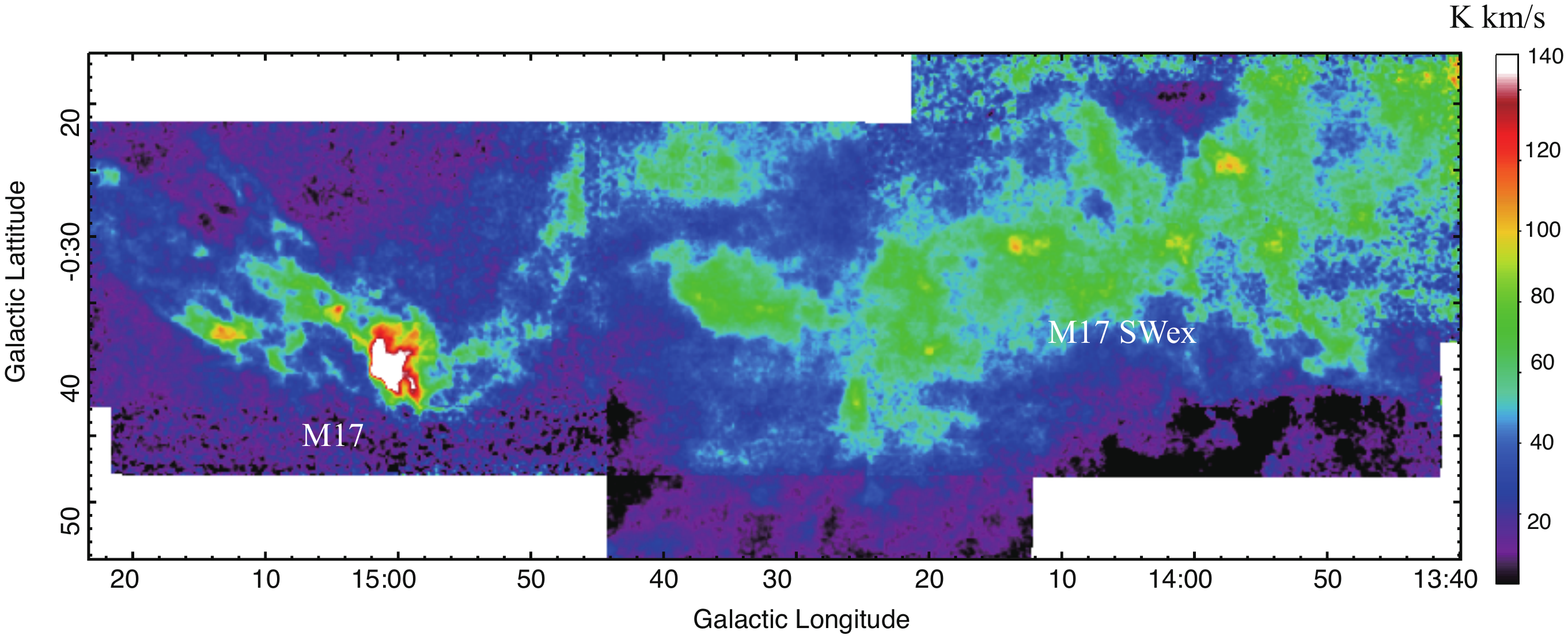}
 \end{center}
\caption{$^{13}$CO ($J=1-0$) integrated intensity map of M17.
The effective angular resolution of the $^{13}$CO map is 22$"$.1. 
\label{fig:m1713co}}
\end{figure}

\begin{figure}
 \begin{center}
 \includegraphics[width=8cm,bb=0 0 601 532]{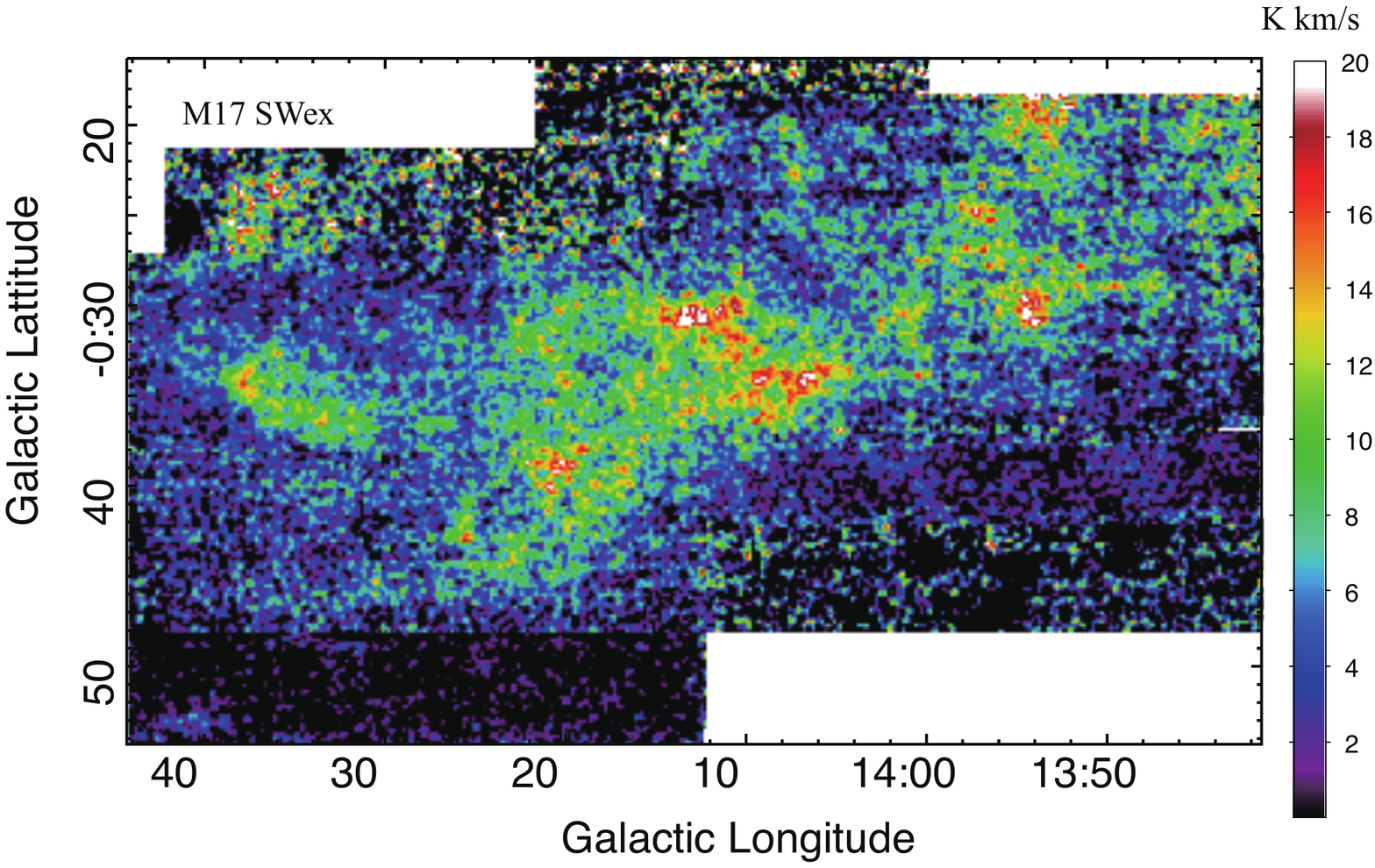}
 \end{center}
\caption{C$^{18}$O ($J=1-0$) integrated intensity map of M17.
The effective angular resolution of the C$^{18}$O map is 22$"$.1. 
\label{fig:m17c18o}}
\end{figure}

\begin{figure}
 \begin{center}
 \includegraphics[width=8cm,bb=0 0 601 532]{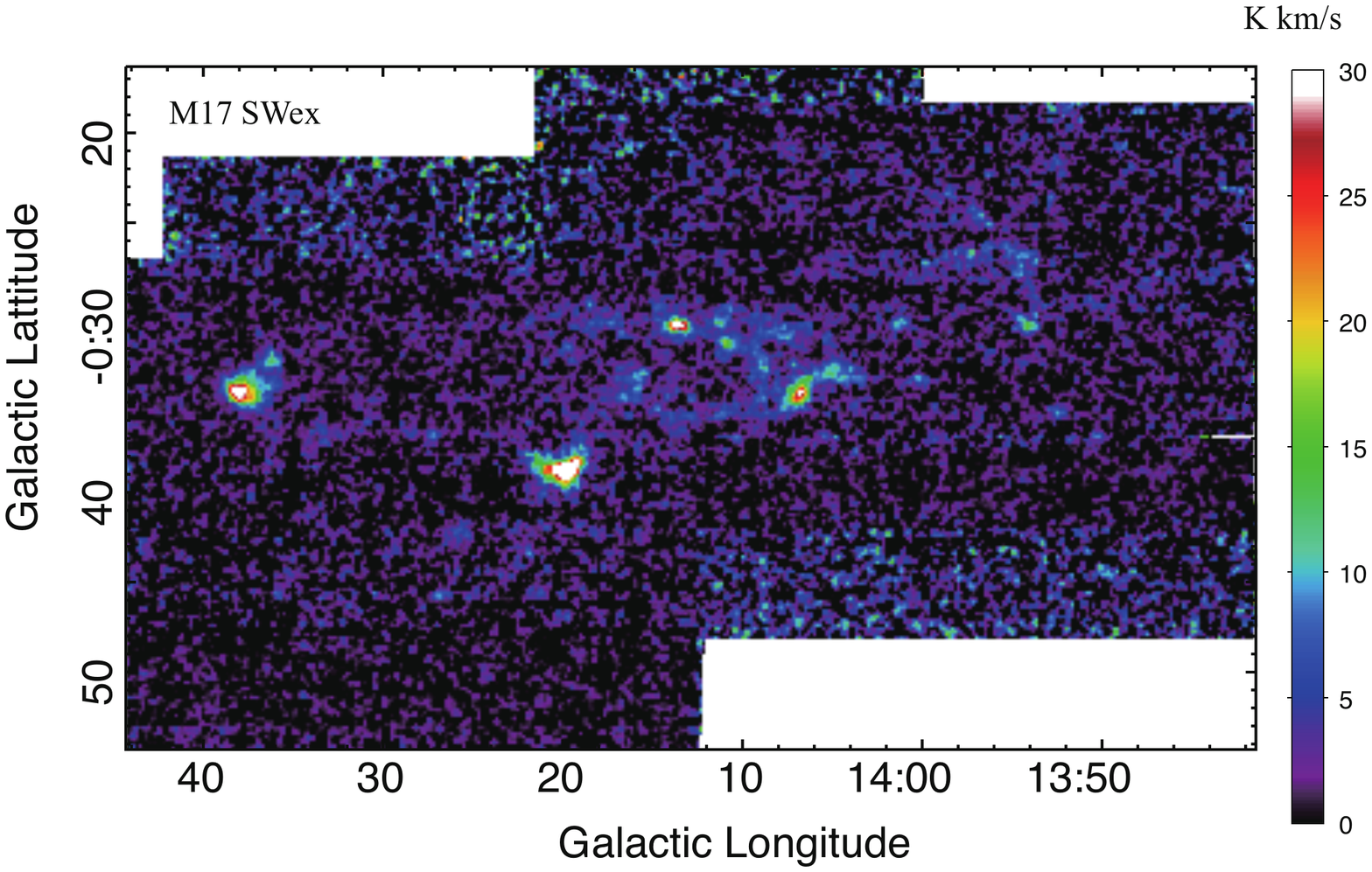}
  \end{center}
\caption{N$_2$H$^+$ ($J=1-0$) integrated intensity map of M17.
The effective angular resolution of the N$_2$H$^+$ map is 24$"$.1.
\label{fig:m17n2hp}}
\end{figure}

Figure \ref{fig:m1712co}, \ref{fig:m1713co}, \ref{fig:m17c18o}, and \ref{fig:m17n2hp} show the integrated intensity maps of M17 for
$^{12}$CO ($J=1-0$), $^{13}$CO ($J=1-0$), C$^{18}$O ($J=1-0$), and 
N$_2$H$^+$ ($J=1-0$), respectively.
Each map except N$_2$H$^+$ is integrated from $-$10 to 60 km s$^{-1}$.
We could not detect the significant CCS emission, and thus we do not show the CCS integrated intensity map.
For N$_2$H$^+$, we integrated all seven hyperfine components to make an integrated intensity map.

The $^{12}$CO and $^{13}$CO emission is strongest toward the M17 HII region.
The emission lines are spatially extended toward the M17 SWex region whose dense parts are detected in C$^{18}$O.
The N$_2$H$^+$ emission is spatially localized and seen as blobs.
The detailed cloud structure and kinematics will be discussed in separate papers 
\citep[e.g.,][]{shimoikura19}.

\section{Spatial Variation of the fractional abundances of $^{13}$CO and C$^{18}$O}
\label{sec:abundance}

Here, using $^{12}$CO, $^{13}$CO, and C$^{18}$O, we derive the physical quantities such as 
the excitation temperature, fractional abundances and optical depth toward Orion A.
The detailed analysis is presented in Ishii et al. (2018).
See separate papers for the results of the other regions.

\subsection{Derivation of Excitation Temperature, Optical Depth, and column density of CO}

Here, we derive excitation temperature, optical depth, and $^{13}$CO and C$^{18}$O column densities toward Orion A.
First, we describe how we derive the physical quantities from our data, and then briefly present the spatial
distributions of the physical quantities for Orion A. 

Assuming that the $^{12}$CO ($J=1-0$) line is optically thick, we derive the excitation temperature $T_{\rm ex}$
using the following equation (see e.g., \citet{pineda08}, \citet{shimajiri11}, and \citet{ishii19}),
\begin{equation}
T_{\rm ex} = { 5.53  \over \ln \{ 1+ 5.53  / (T_{\rm peak} + 0.819)\}  }  \ {\rm K}
\end{equation}
where $T_{\rm peak} $ is the maximum intensity along the line-of-sight. 
We also assumed that the excitation temperatures of $^{13}$CO and C$^{18}$O are the same as $T_{\rm ex}$ calculated with the above equation.
We note that the excitation temperature of $^{12}$CO, which is likely to be close to LTE,
can be used as a good measure of the gas temperature.

The optical depths and column densities at each channel are derived with the following formula,
\begin{equation}
\tau_{\rm ^{13}CO} = - \ln \left[ 1 -  {\ T_{\rm ^{13}CO}    \over 5.29 (J  (T_{\rm ex}) - 0.164)}\right]
\end{equation}
\begin{equation}
N_{\rm ^{13}CO}  = 2.42 \times 10^{14} \left[ \tau_{\rm ^{13}CO}  T_{\rm ex} \Delta V \over 1 - \exp (-5.29 / T_{\rm ex} ) \right]  \ {\rm cm^{-2}}
\end{equation}
\begin{equation}
\tau_{\rm C^{18}O} = - \ln \left[ 1 -  {\ T_{\rm C^{18}O}   \over 5.27 (J (T_{\rm ex}) - 0.164)}\right]
\end{equation}
and
\begin{equation}
N_{\rm C^{18}O}  =   2.42 \times 10^{14} \left[ \tau_{\rm C^{18}O}  T_{\rm ex} \Delta V \over 1 - \exp (-5.27 / T_{\rm ex} ) \right]  \ {\rm cm^{-2}}
\end{equation}
where we assumed that the rotational transition is in the LTE.
Here, $\Delta V $ is the velocity resolution. The beam filling factors of  $^{13}$CO and C$^{18}$O
are set to unity. 
The function $J_X (T)$ is given by $[\exp (5.29/T) -1]^{-1}$ and $[\exp (5.27/T) -1]^{-1}$
for $^{13}$CO and C$^{18}$O, respectively.
To derive the opacity-corrected total column densities \citep{pineda08}, 
we sum up the column density at each channel  along the line of sight
and multiplied the column densities by the correction factor of $\tau_i/[1-\exp ({-\tau_i})]$, where $i$ is 
either for $^{13}$CO or C$^{18}$O.
The opacity-corrected total H$_2$ column density is then derived from the $^{13}$CO integrated intensity 
assuming the constant abundance of $2\times 10^{-6}$ \citep{dickman78}.
We also derived the $^{13}$CO and C$^{18}$O fractional abundances relative to H$_2$  by dividing the column densities by
the {\it Herschel} H$_2$ column density. 
See also Ishii et al. (2019) for these derivations for Orion A.

We summarize the molecular gas mass derived from $^{13}$CO in table \ref{tab:mass}. The total masses of the observed areas are
estimated to be $2.67 \times 10^4 M_\odot$, $3.86 \times 10^4 M_\odot$, and $8.1 \times 10^5 M_\odot$ for Orion A, Aquila Rift, and M17, respectively.

\subsubsection{Orion A}

\begin{figure}
 \begin{center}
 \includegraphics[width=6cm, bb=-100 0 442 800]{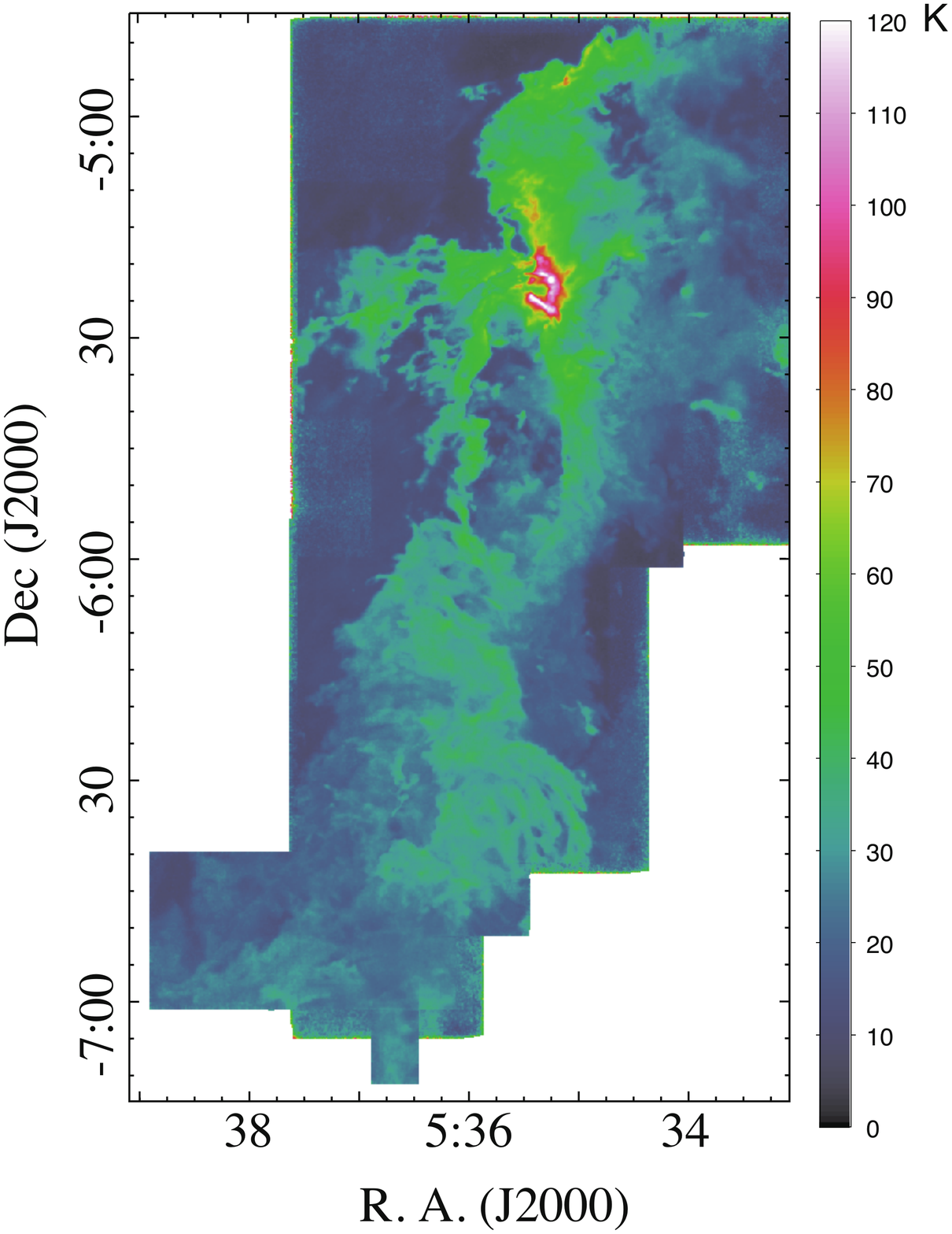}
 \end{center}
\caption{The excitation temperature map of Orion A.
\label{fig:orionatex}}
\end{figure}

In figure \ref{fig:orionatex}, we present the excitation temperature map of Orion A determined by the $^{12}$CO peak intensity.
The excitation temperature ranges from $\sim 8$ K to $\sim$ 100 K, and takes its maximum near Orion KL.  The main filament
has somewhat higher temperatures at $\sim$ 50 K in the northern region.  
The southern part including L1641N and NGC 1999 has somewhat lower temperature at 10$-$20 K.
This temperature distribution is consistent with that of \citet{kong18} with a finer angular resolution.

\begin{figure}
 \begin{center}
\includegraphics[width=6cm, bb=-100 0 442 800]{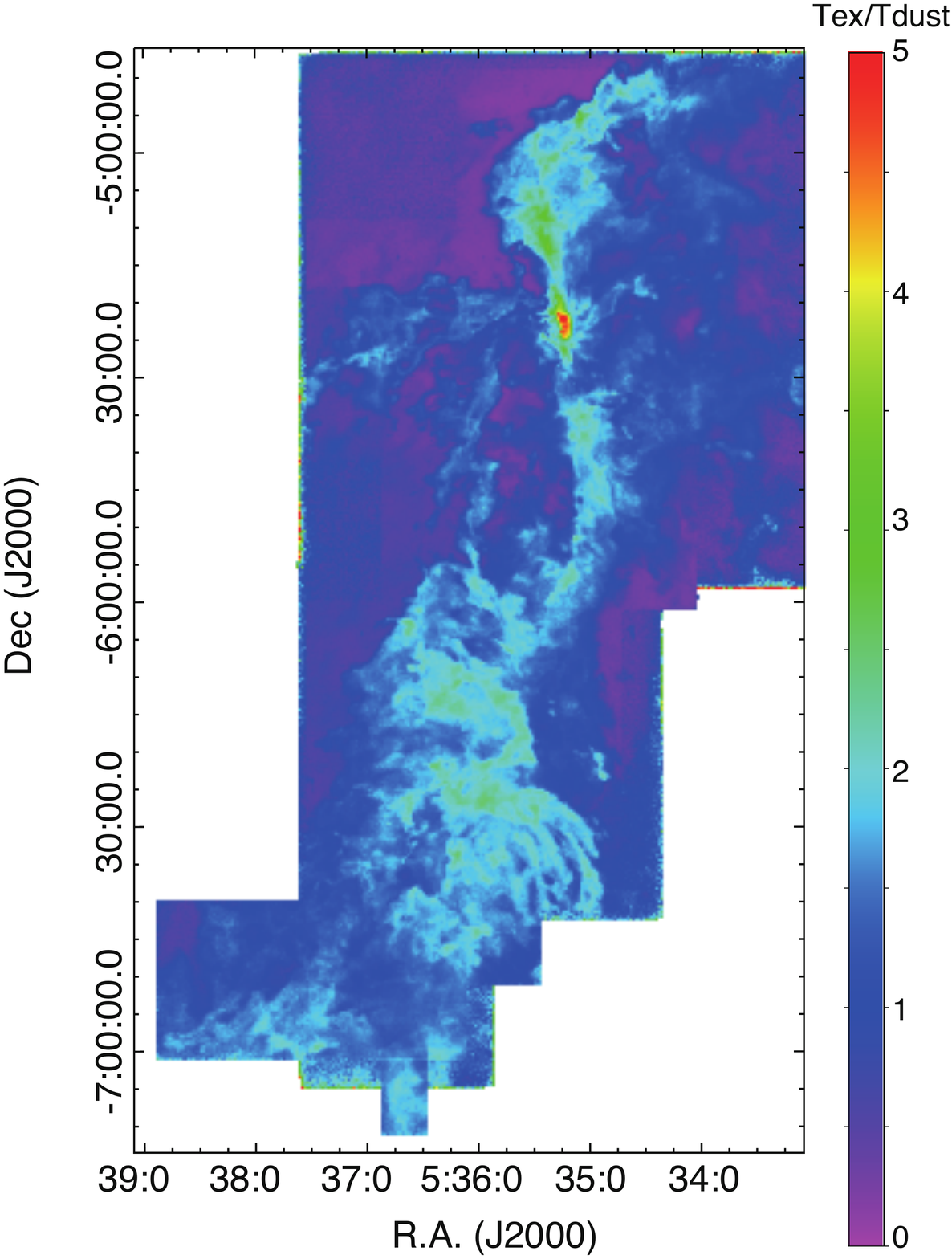}
 \end{center}
\caption{The ratio of the excitation temperature derived from $^{12}$CO to the dust temperature derived from the Herschel data.
The ratio is around 1 -- 2 along the main ridge of Orion A, except in OMC-1.
\label{fig:orionTdust}}
\end{figure}

The excitation temperature derived from $^{12}$CO tends to be somewhat higher than the dust temperature
derived from the Spectral Energy Distribution (SED) of the Herschel data.  
In figure \ref{fig:orionTdust}, we show the ratio of the excitation temperature derived from $^{12}$CO to the dust temperature. 
Here, we smoothed the excitation temperature map with an effective angular resolution of
36\arcsec to match the effective resolution of the dust temperature map. The ratio stays at around 2 along the main filament, except in OMC-1 where the ratio is about 5.

\begin{figure}
 \begin{center}
 \includegraphics[width=6cm, bb=-100 0 442 800]{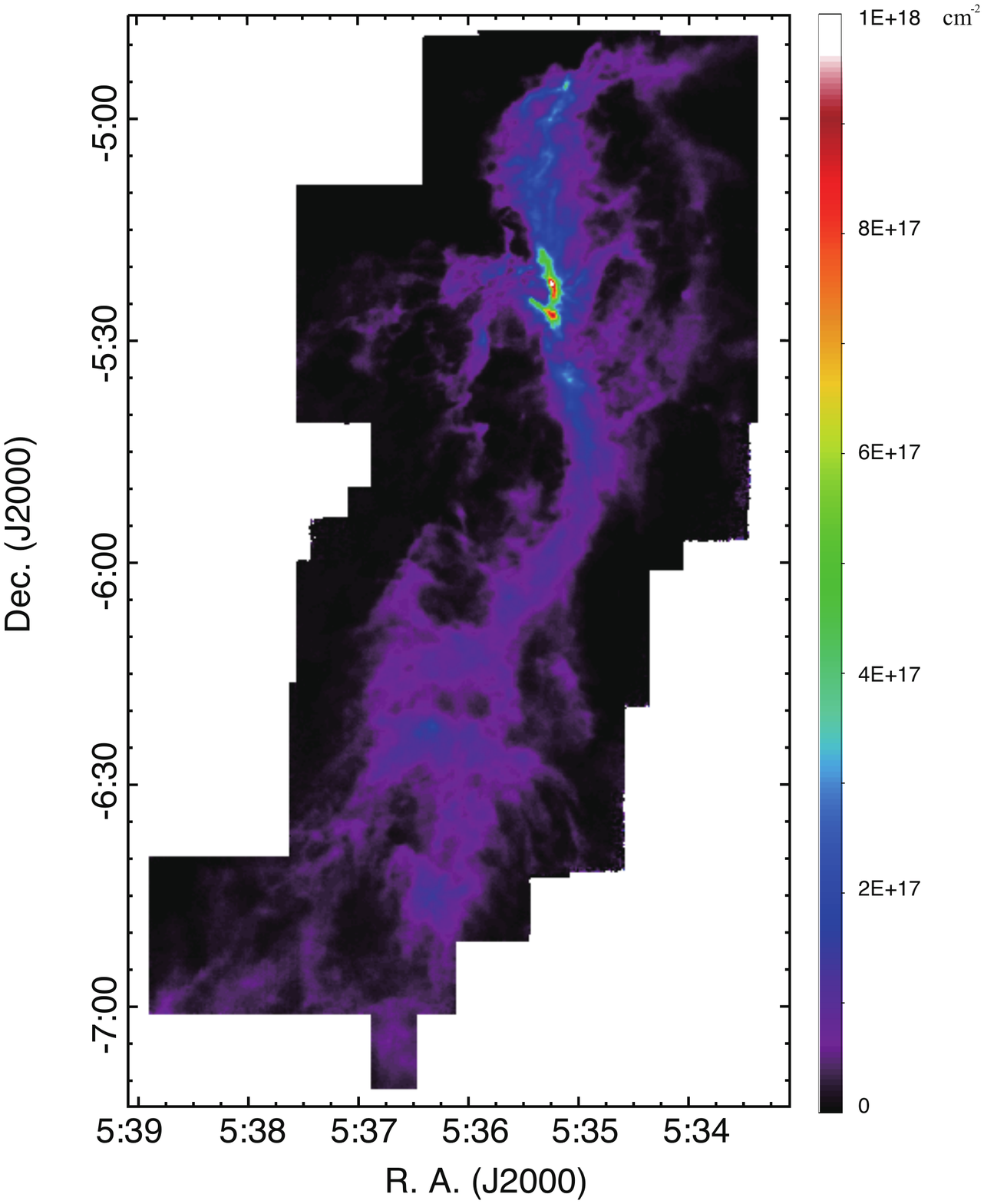}
 \end{center}
\caption{The opacity-corrected $^{13}$CO column density map of Orion A.
\label{fig:orion13cocolumndensity}}
\end{figure}

\begin{figure}
 \begin{center}
 \includegraphics[width=6cm, bb=-100 0 442 800]{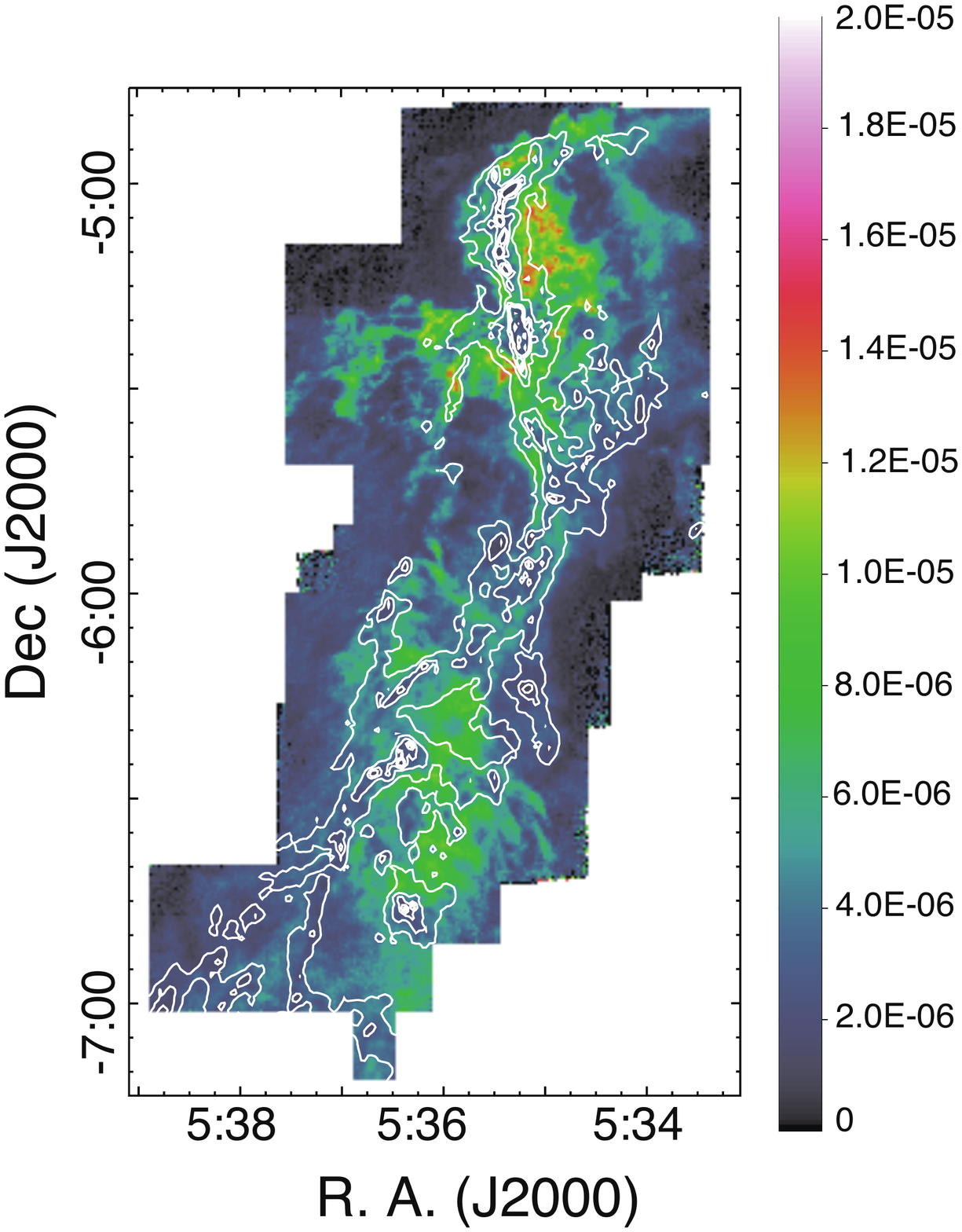}
 \includegraphics[width=5cm, bb=-100 0 442 800]{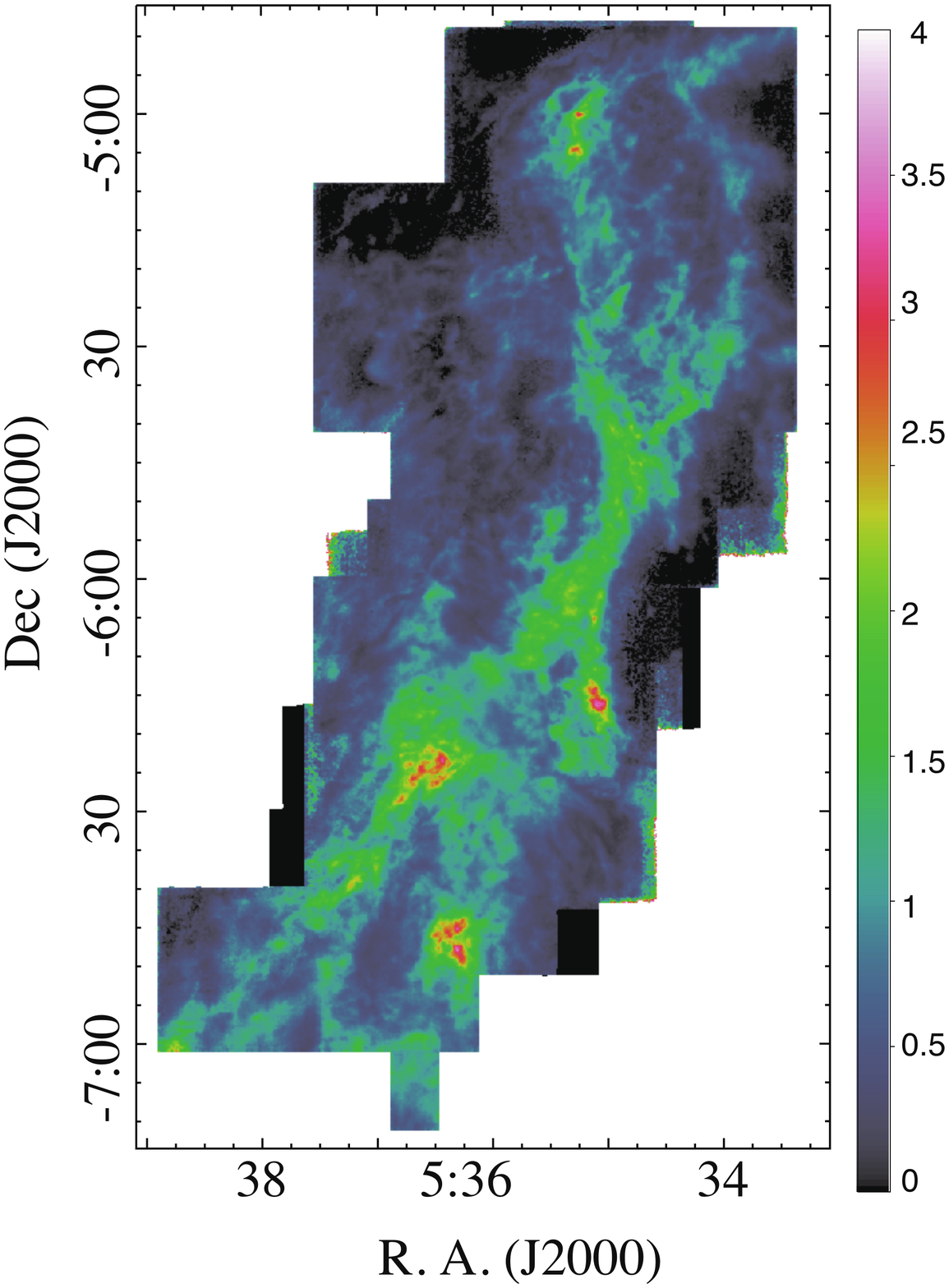}
 \end{center}
\caption{(a) The  $^{13}$CO fractional abundance and (b) its optical depth maps of Orion A.
\label{fig:oriona13coabundance}}
\end{figure}

\begin{figure}
 \begin{center}
\includegraphics[width=6cm, bb=-200 0 442 800]{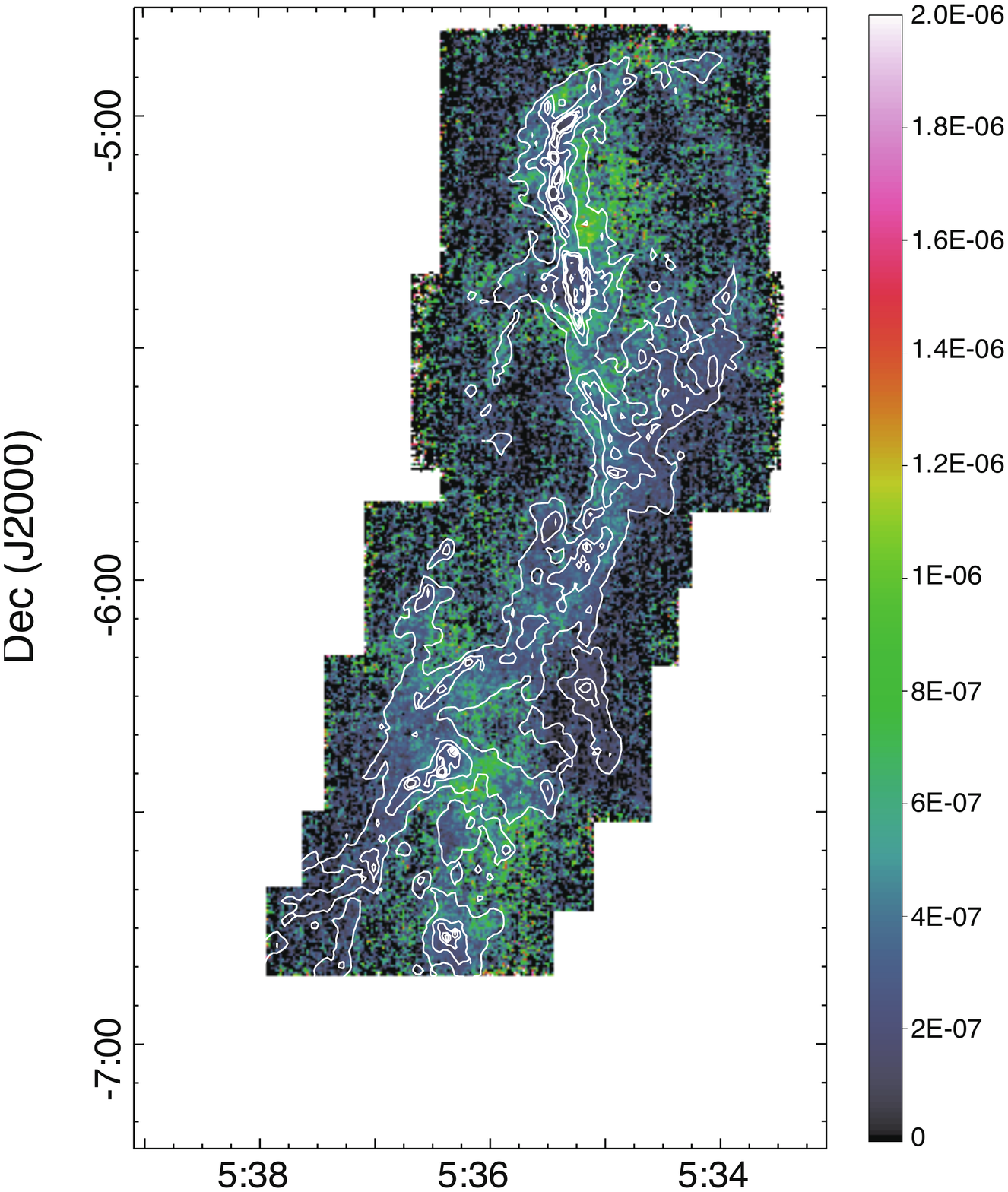}
\includegraphics[width=6cm, bb=-200 0 442 800]{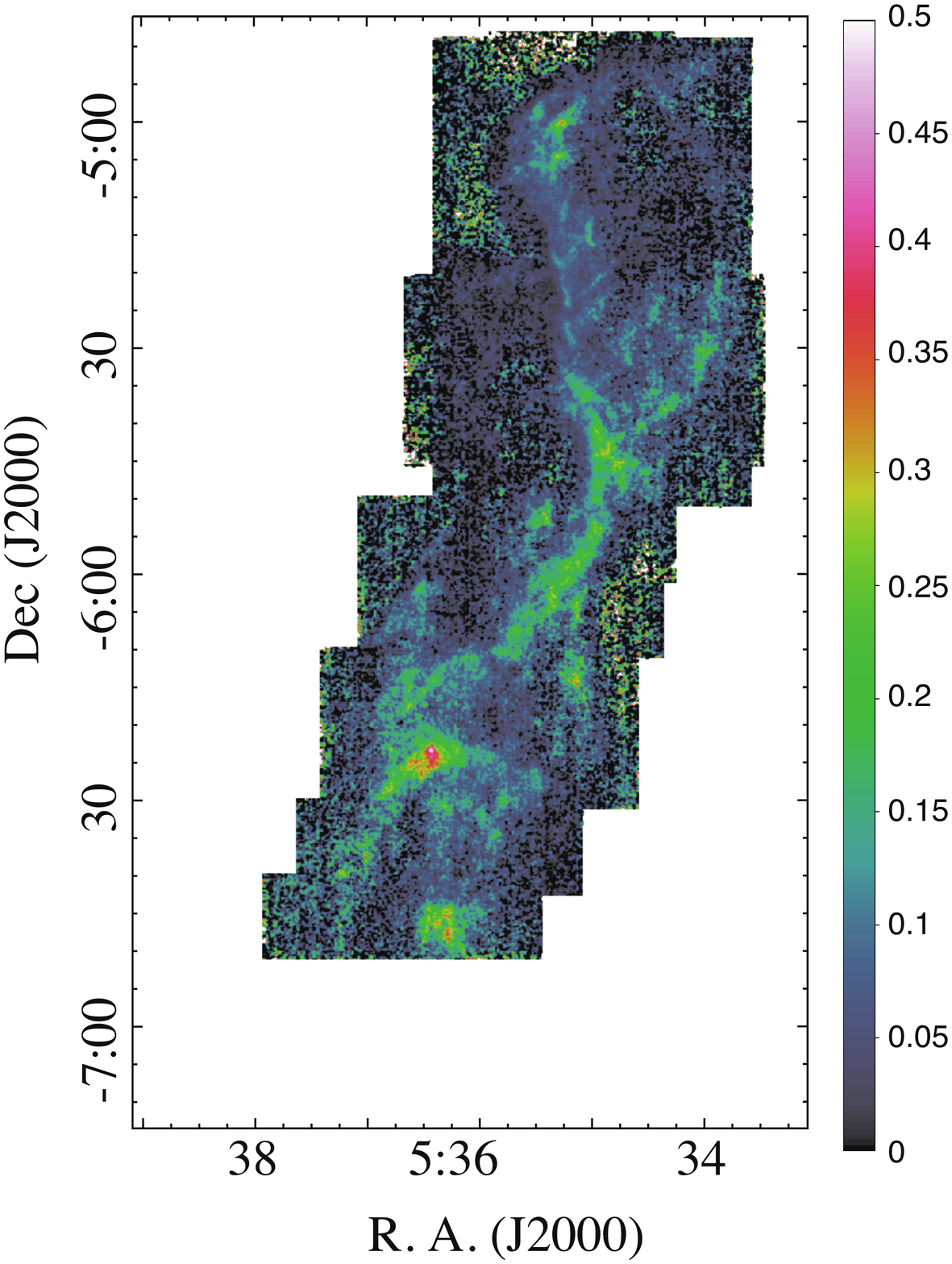}
 \end{center}
\caption{(a) The C$^{18}$O fractional abundance and (b) its optical depth maps of Orion A.
\label{fig:orionac18oabundance}}
\end{figure}

Figures \ref{fig:orion13cocolumndensity}, \ref{fig:oriona13coabundance}, 
and \ref{fig:orionac18oabundance} show the 
    spatial distributions of the opacity-corrected $^{13}$CO column density, the fractional abundances
relative to H$_2$ and the optical depth for $^{13}$CO ($J=1-0$) and C$^{18}$O ($J=1-0$), respectively.  
% \textcolor{red}{As we mentioned above, we assumed a constant $^{13}$CO abundance of $2\times 10^{-6}$
% to derive the H$_2$ column densities.}
The optical depth of $^{13}$CO ranges from a few to unity. It is about unity along the filament.
Several compact regions such as L1641N has larger optical depth at $\sim$ 3.
On the other hand, the optical depth of C$^{18}$O is less than unity for almost all areas.
Thus, the C$^{18}$O emission is reasonably optically-thin for the entire cloud.

The fractional abundance of $^{13}$CO ranges from a few $\times 10^{-7}$ to a few $\times 10^{-6}$.
It tends to be small at being $5\times 10^{-7}$ toward some dense areas such as the main filament of 
OMC-1/2/3 and L1641N.  
Outside the dense areas, the abundance goes up to a few $\times 10^{-6}$.
For C$^{18}$O, the fractional abundance ranges from $1 \times 10^{-8}$ to $1.5\times 10^{-7}$.
Similar to the $^{13}$CO abundance, it is small at some dense areas such as OMC-1/2/3 and L1641N. 

\begin{figure}
 \begin{center}
 \includegraphics[width=6cm, bb=-200 0 442 800]{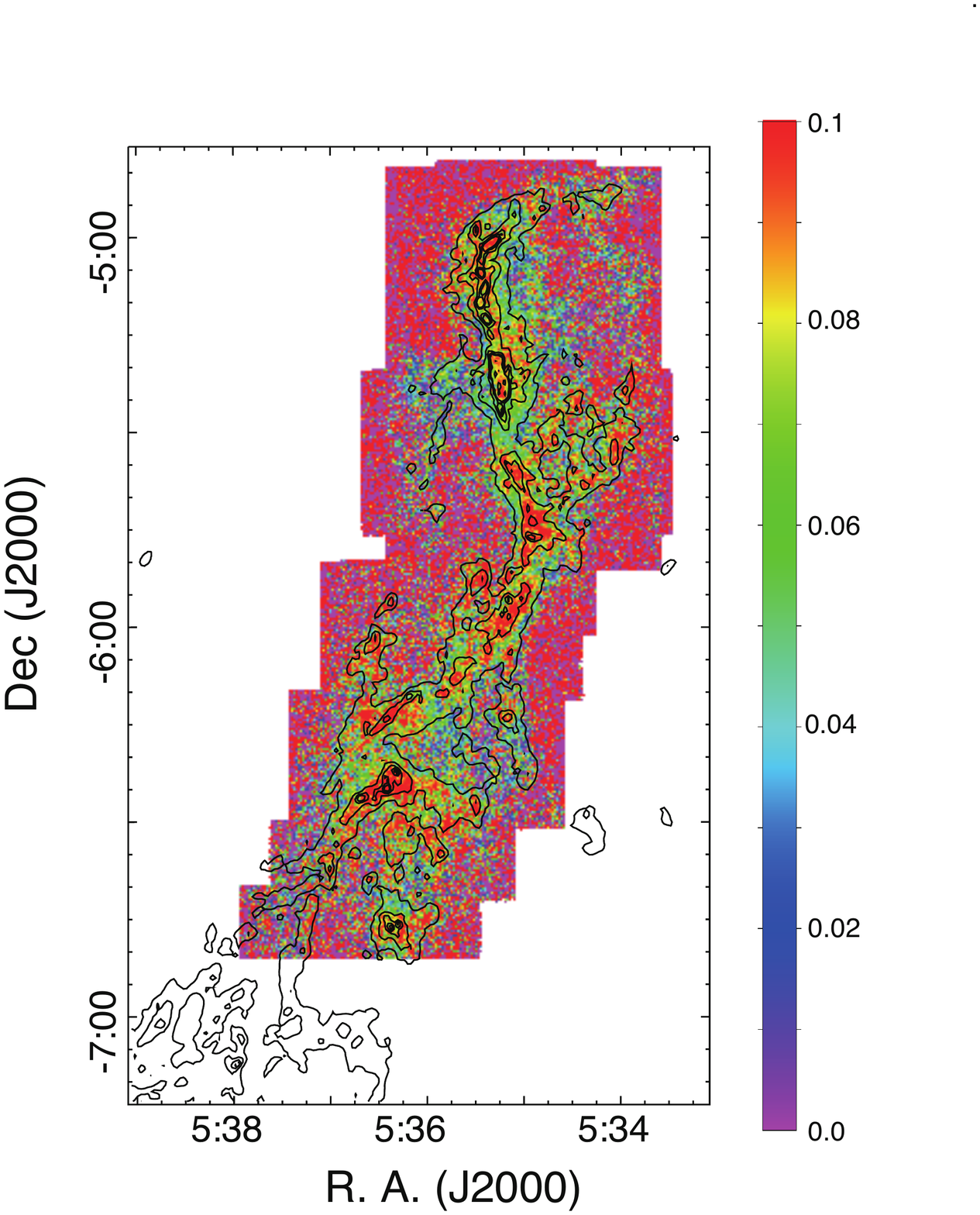}
 \end{center}
\caption{The $^{13}$CO-to-C$^{18}$O fractional abundance ratio of Orion A.
\label{fig:oriona13co-c18o}}
\end{figure}

Our map indicates that the fractional abundance varies from region to region within a factor of $\sim$ 10
(see figure \ref{fig:oriona13co-c18o}).
Toward the dense regions, the ratio seems to approach to $\sim$ 5, close to the standard interstellar value.
This variation may be related partly to the selective dissociation due to the FUV radiation discussed 
by \citet{shimajiri11} and \citet{ishii19}.  See Ishii et al. (2018) for more details of the analysis.

\begin{figure}
 \begin{center}
 \includegraphics[width=6cm, bb=-200 0 442 800]{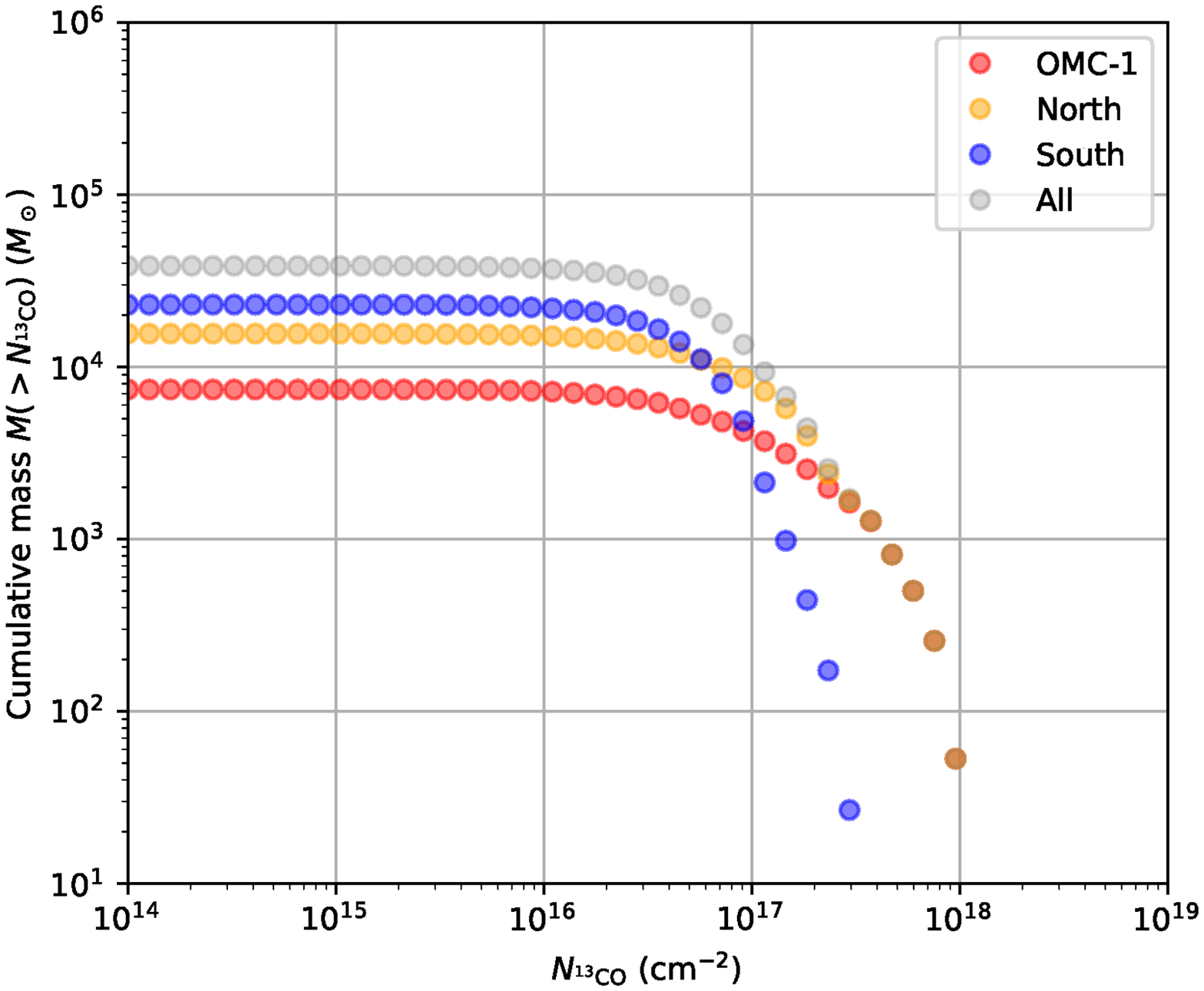}
 \includegraphics[width=6cm, bb=-200 0 442 800]{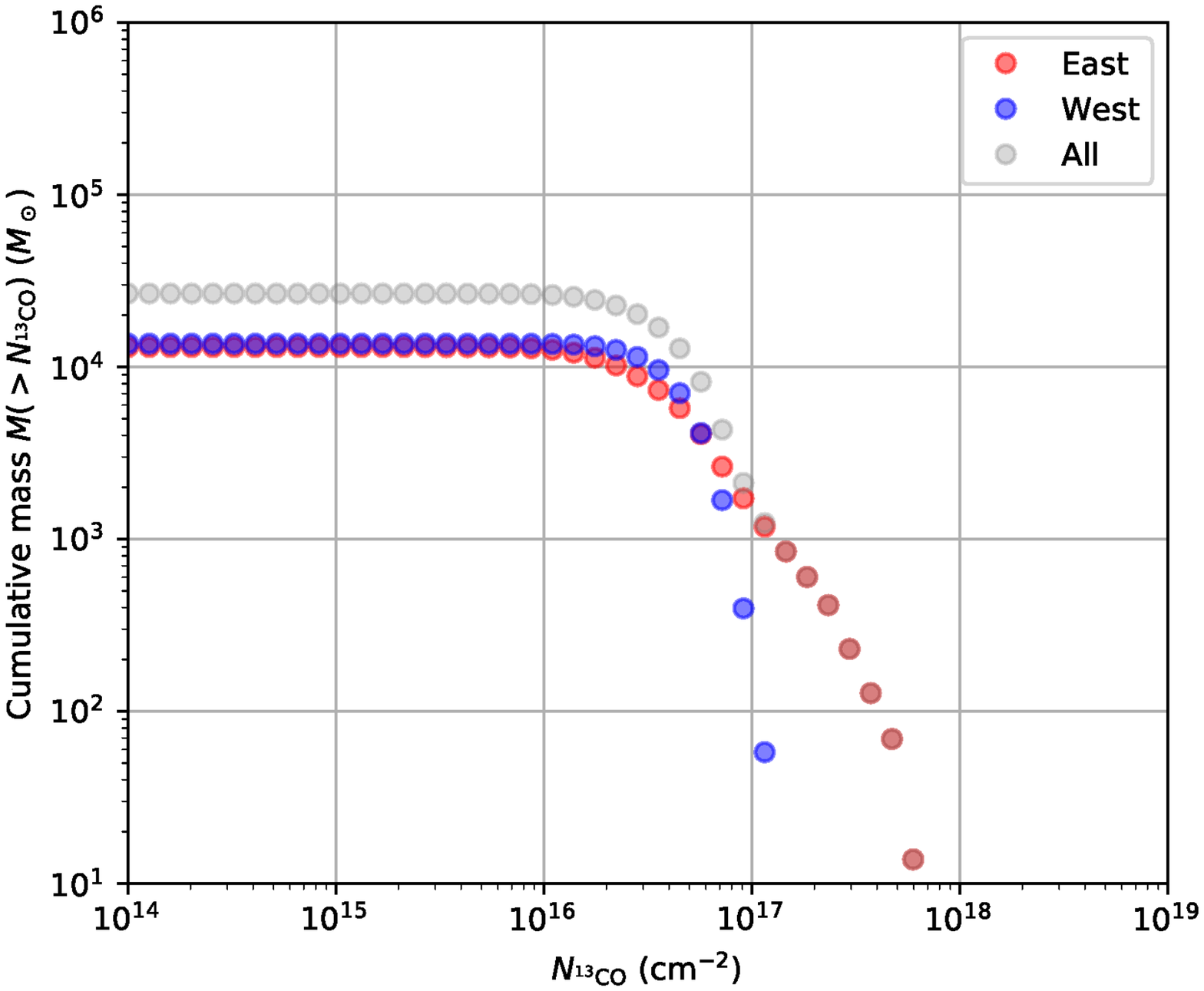}
  \includegraphics[width=6cm, bb=-200 0 442 800]{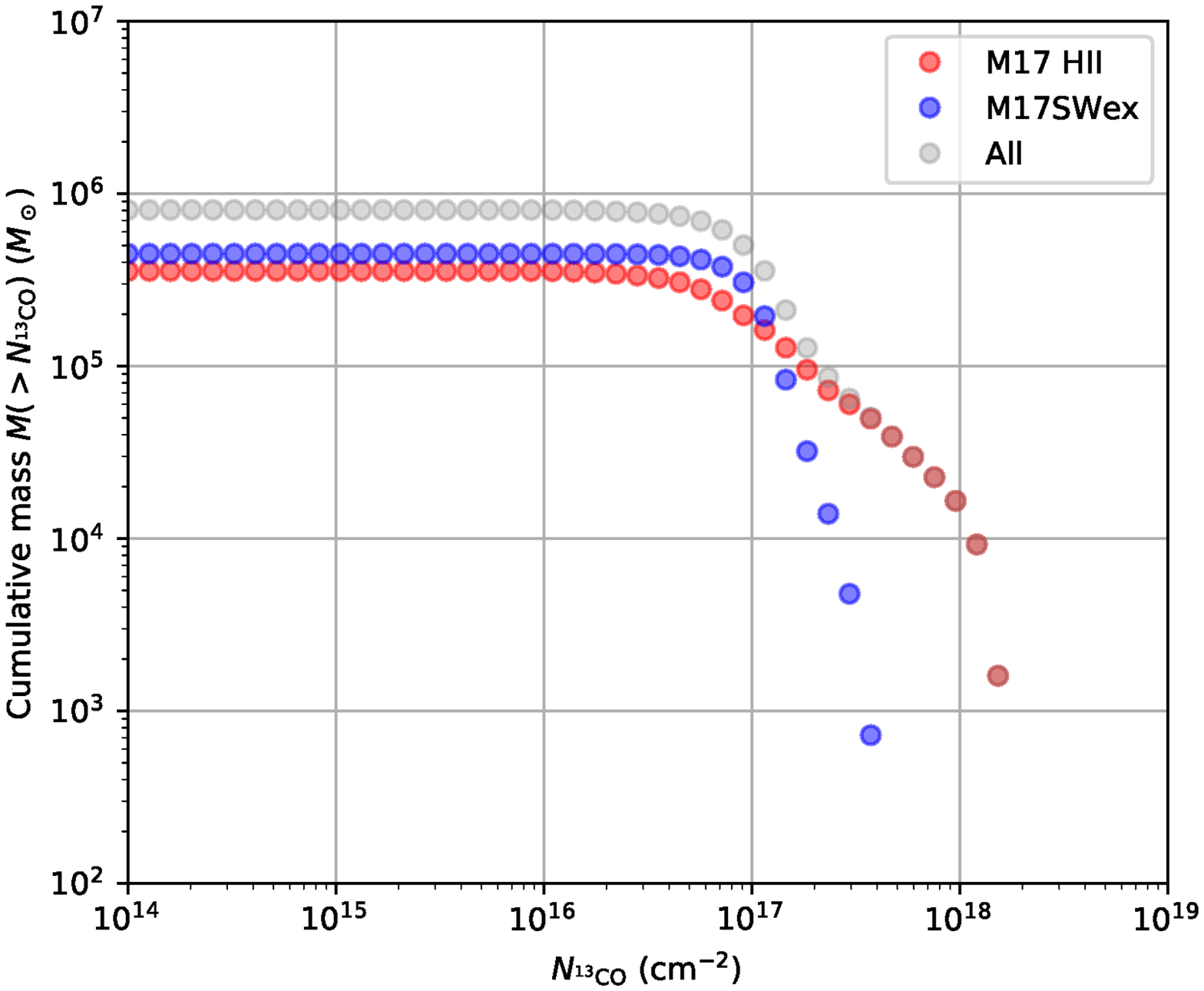}
 \end{center}
\caption{The cumulative column density distributions in (a) Orion A, (b) Aquila Rift, and (c) M17. 
In Orion A, the area is divided into two (North and South) at Dec. = --05:32:56.0. 
OMC-1 is a part of North and within --05:30:26.0 $\le$ Dec $\le$ --05:16:03.5.
In Aquila Rift, the area is divided into two (West and East) at R.A. = 18:30:39.6.  
In M17, the area is divided into two (HII and SWex) at l = 14:26:22.5.
\label{fig:cumulative}}
\end{figure}

In figure \ref{fig:cumulative}a, we present the cumulative $^{13}$CO mass distributions of Orion A. 
We divide the cloud into three areas: North (Dec. $\ge $ -05:32:56.0), 
OMC-1 (--05:30:26.0 $\le  $ Dec. $\le $ -05:16:03.5), and South (Dec $\le$ --05:32:56.0), where 
North area includes OMC-1.
This indicates that the OMC-1 area contains a fraction of molecular gas with column density higher than 
$\sim$ 1 g cm$^{-2}$, the threshold for high mass star formation proposed by \citet{krumholz08}, 
beyond which further fragmentation of clouds can be avoided to form lower-mass cores, where
$\sim$ 1 g cm$^{-2}$ corresponds to the $^{13}$CO column density of $\sim 5\times 10^{17}$ cm$^{-2}$,
assuming the $^{13}$CO fractional abundance of $2\times 10^{-6}$ \citep{dickman78}.
In contrast, in the South, the molecular gas with column densities higher than the threshold is deficient.
The total $^{13}$CO mass is estimated to be $3.9 \times 10^4 M_\odot$ with $7.4 \times 10^3$ M$_\odot$, 
$1.6\times 10^4$ M$_\odot$, and $2.3\times 10^4$ M$_\odot$ for OMC-1, North, and South, respectively.

Below, we repeat the same analysis shown above to the other two regions.

\subsubsection{Aquila Rift}

\begin{figure}
 \begin{center}
 \includegraphics[angle=90,width=6cm, bb=-200 -200 442 900]{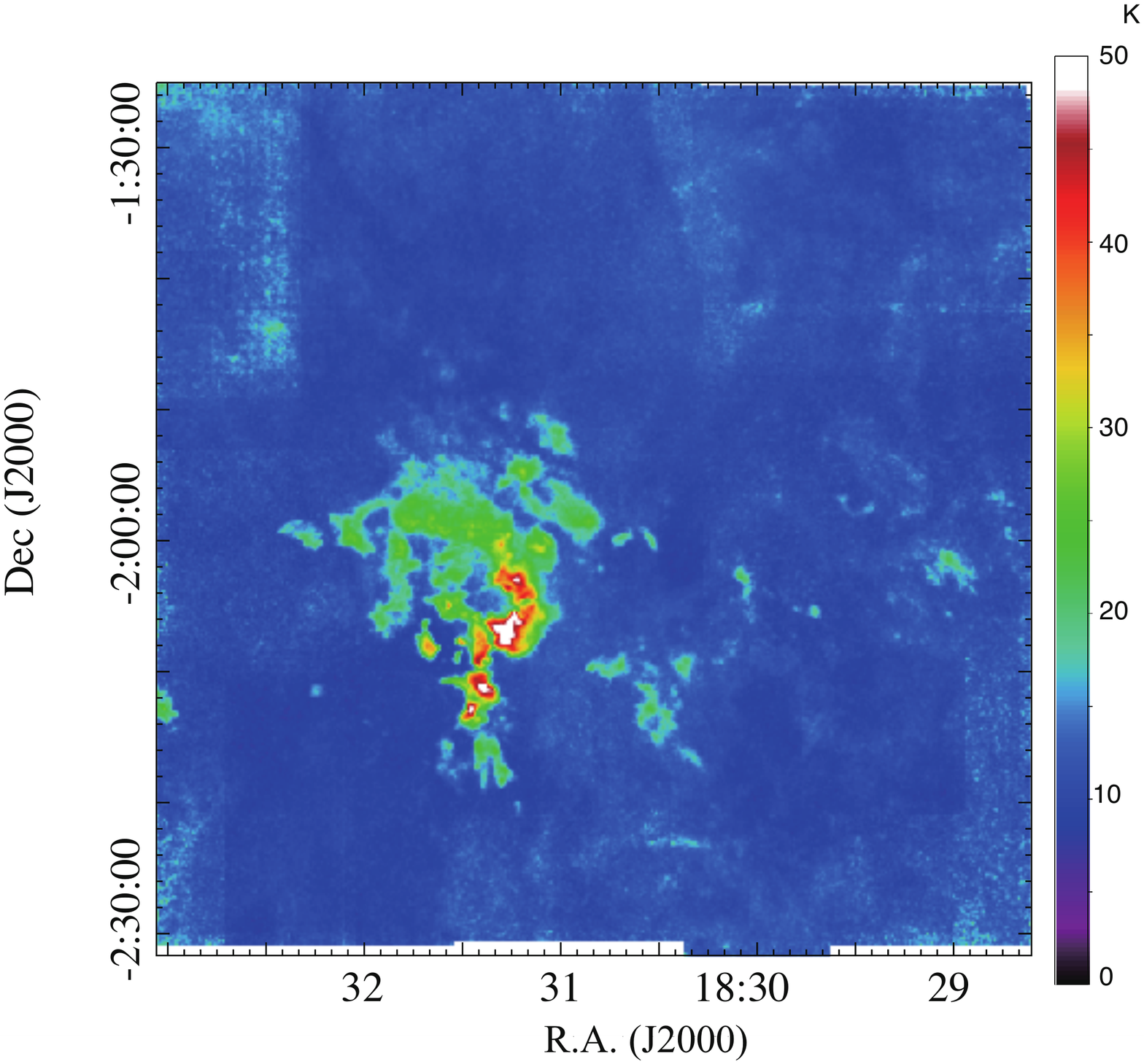}
 \end{center}
\caption{The excitation temperature map of Aquila Rift.
\label{fig:aquilatex}}
\end{figure}

\begin{figure}
 \begin{center}
\includegraphics[width=6cm, bb=-200 0 442 800]{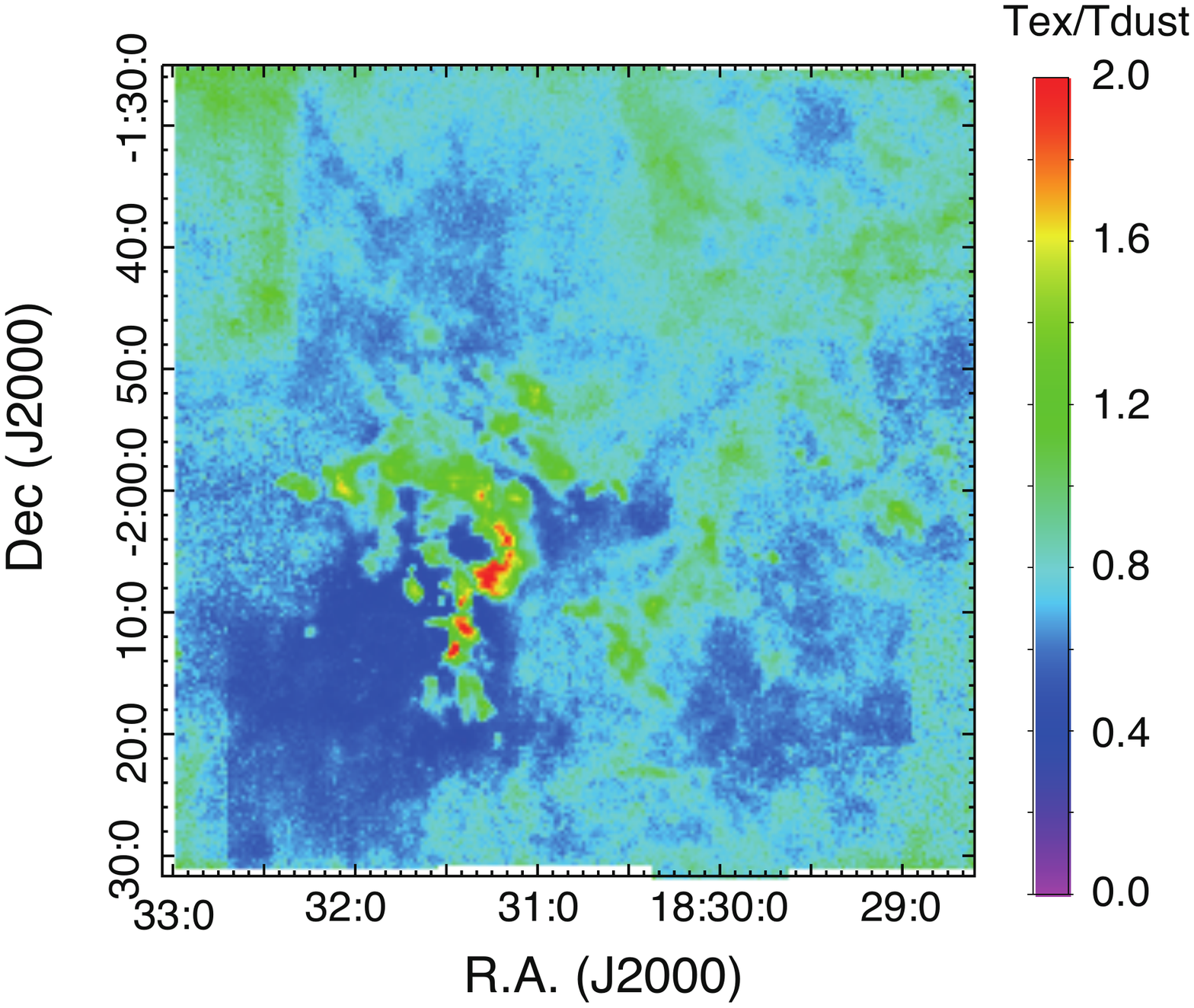}
 \end{center}
\caption{The ratio of the excitation temperature derived from $^{12}$CO to the dust temperature derived from the Herschel data.
The ratio stays at around unity in the molecular clouds.  In W40, it goes up to $\sim$ 2.
\label{fig:aquila_Tdust}}
\end{figure}

Figure \ref{fig:aquilatex} shows the spatial distribution of the CO excitation temperature. 
The excitation temperature is high toward the W40 HII region. 
Serpens South has a relatively high excitation temperature of 25 K, but in other parts around Serpens South
it is very low at $\sim$10 K, indicating that the star formation may not be active.
In figure \ref{fig:aquila_Tdust}, we show the ratio of the excitation temperature derived from $^{12}$CO to the dust temperature. 
Here, we smoothed the excitation temperature map with an effective angular resolution of
36\arcsec to match the effective resolution of the dust temperature map. The excitation temperature is nearly the same as the dust temperature
in Aquila Rift except in W40 HII region where the excitation temperature goes up to about 2.

\begin{figure}
 \begin{center}
\includegraphics[angle=90,width=6cm, bb=-200 0 442 800]{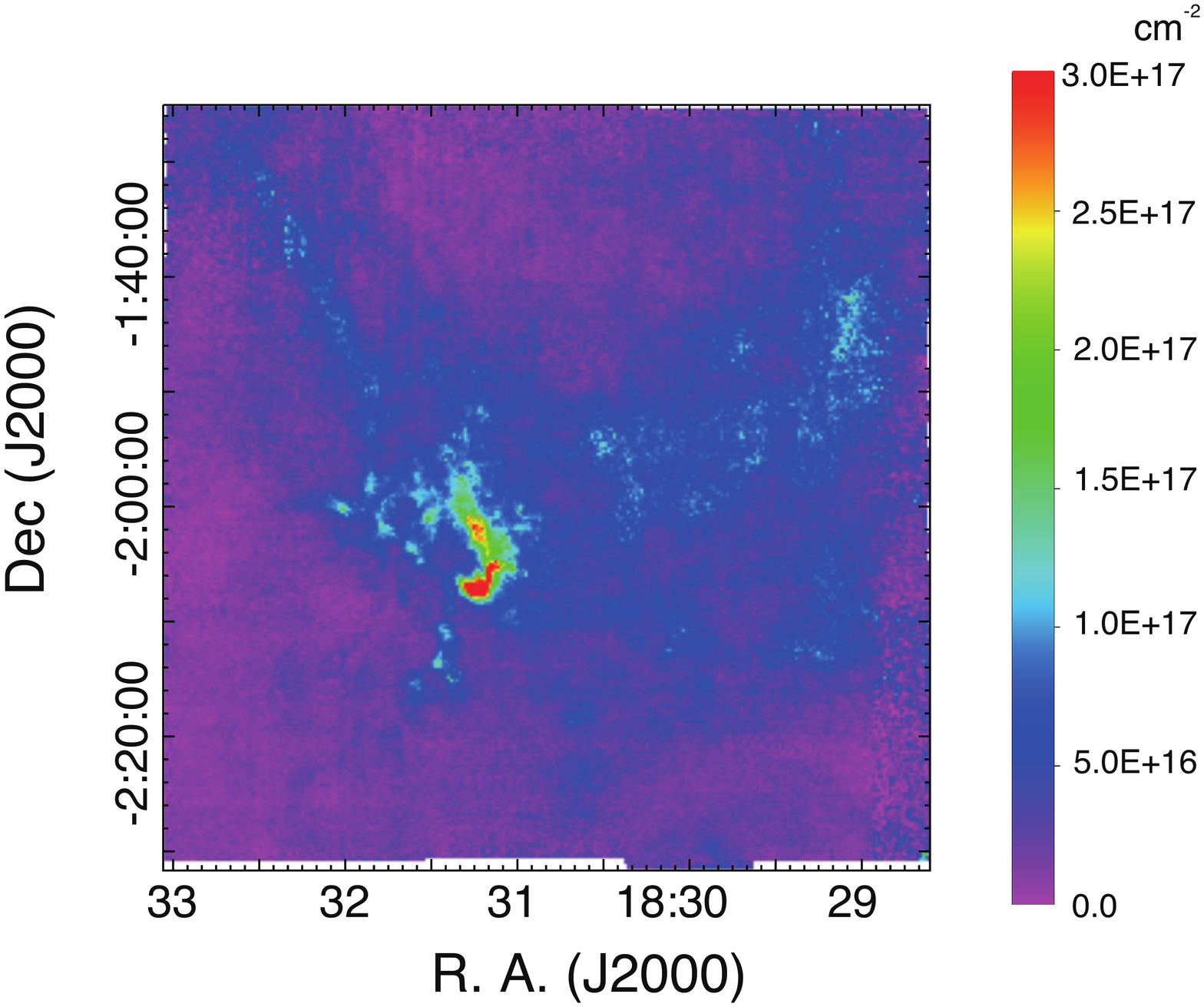}
 \end{center}
\caption{The opacity-corrected $^{13}$CO column density map of Aquila Rift.
\label{fig:aquila13cocolumndensity}}
\end{figure}

\begin{figure}
 \begin{center}
 \includegraphics[width=6cm, bb=-200 0 442 800]{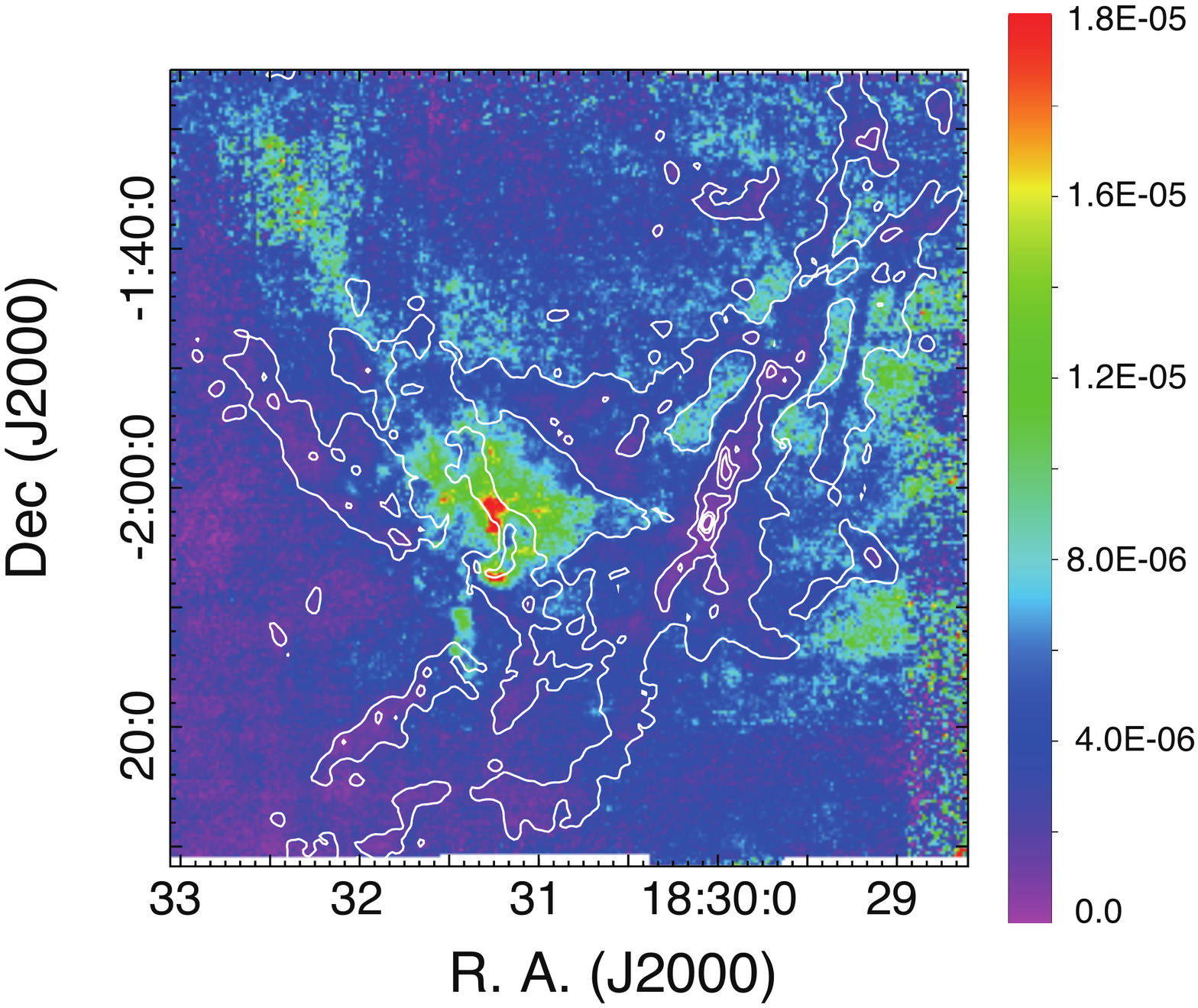}
  \includegraphics[width=6cm, bb=-200 0 442 800]{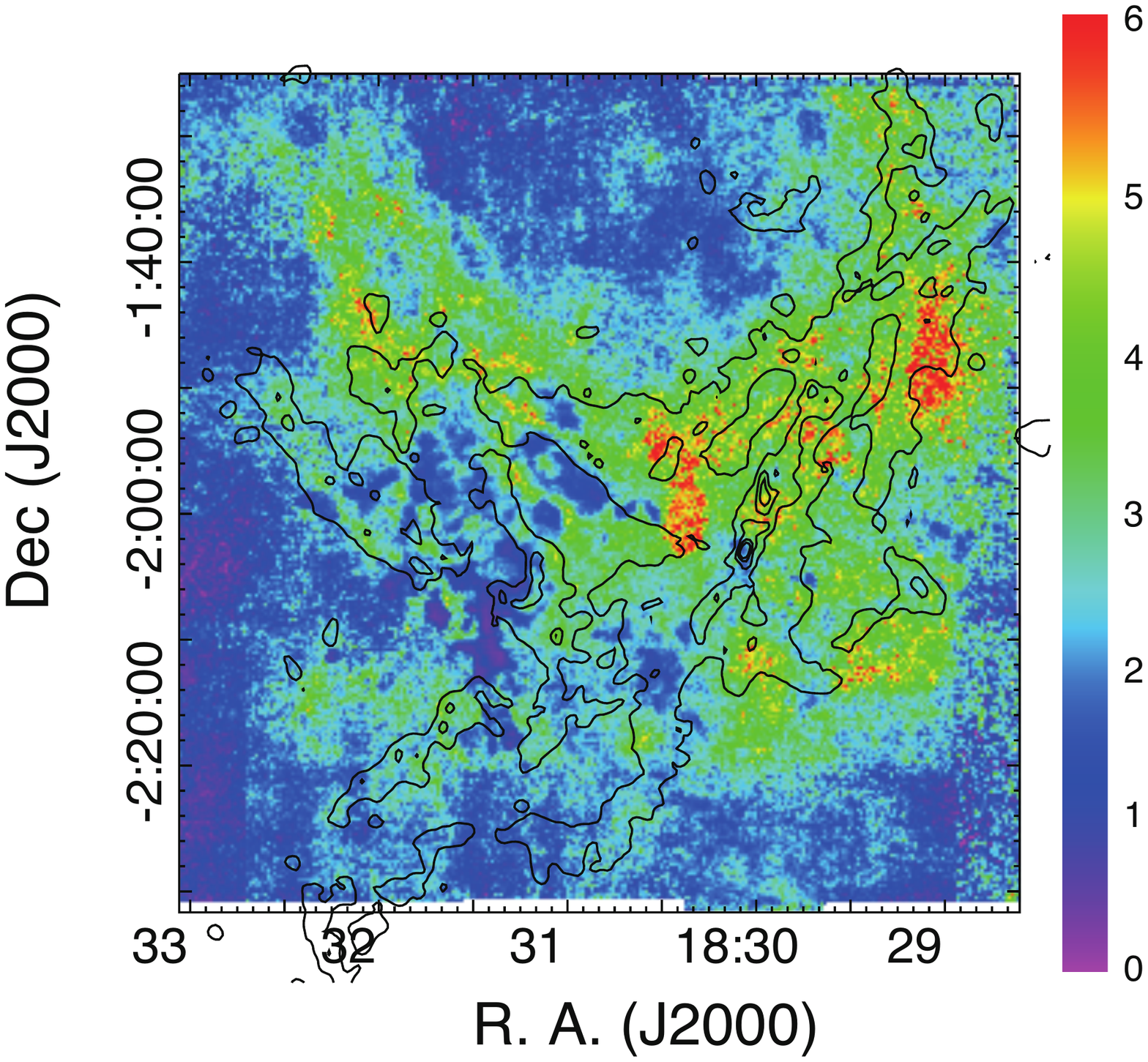}
 \end{center}
\caption{(a) The  $^{13}$CO fractional abundance and (b) its optical depth maps of Aquila Rift.
\label{fig:aquila13coabundance}}
\end{figure}

\begin{figure}
 \begin{center}
\includegraphics[width=6cm, bb=-200 0 442 800]{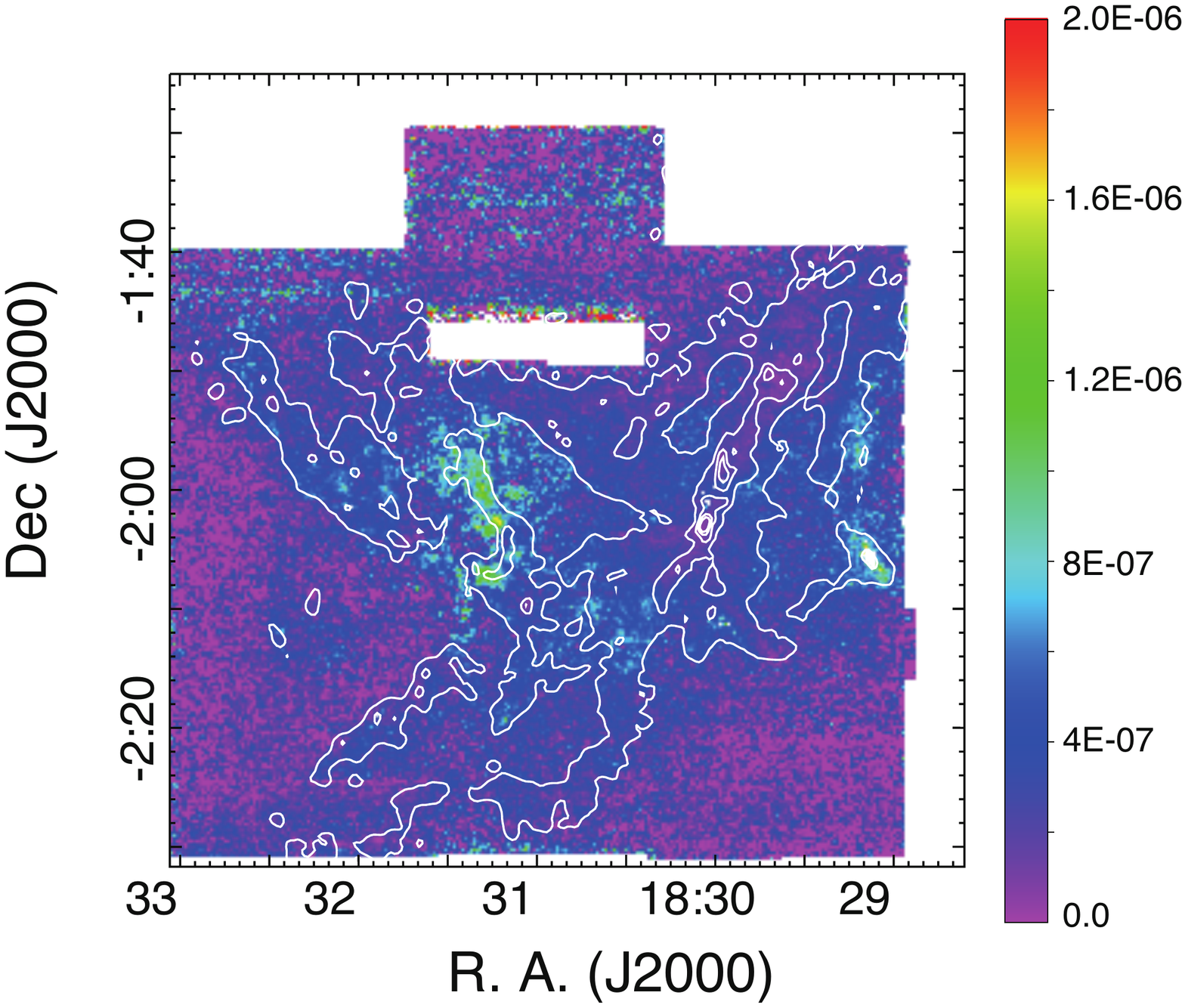}
\includegraphics[width=6cm, bb=-200 0 442 800]{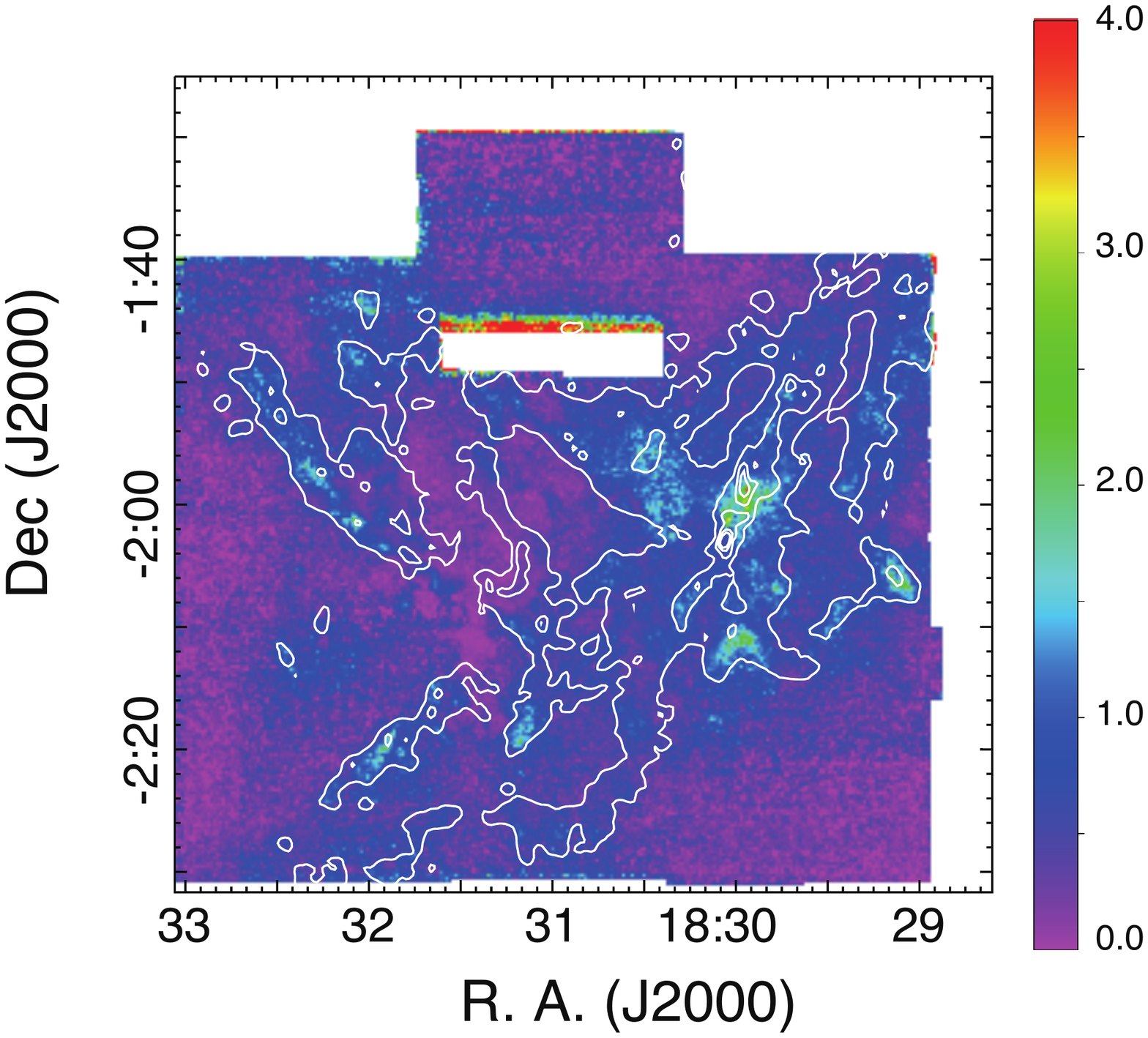}
 \end{center}
\caption{(a) The C$^{18}$O fractional abundance and (b) its optical depth maps of Aquila Rift.
\label{fig:aquilac18oabundance}}
\end{figure}

\begin{figure}
 \begin{center}
 \includegraphics[width=6cm, bb=-200 0 442 800]{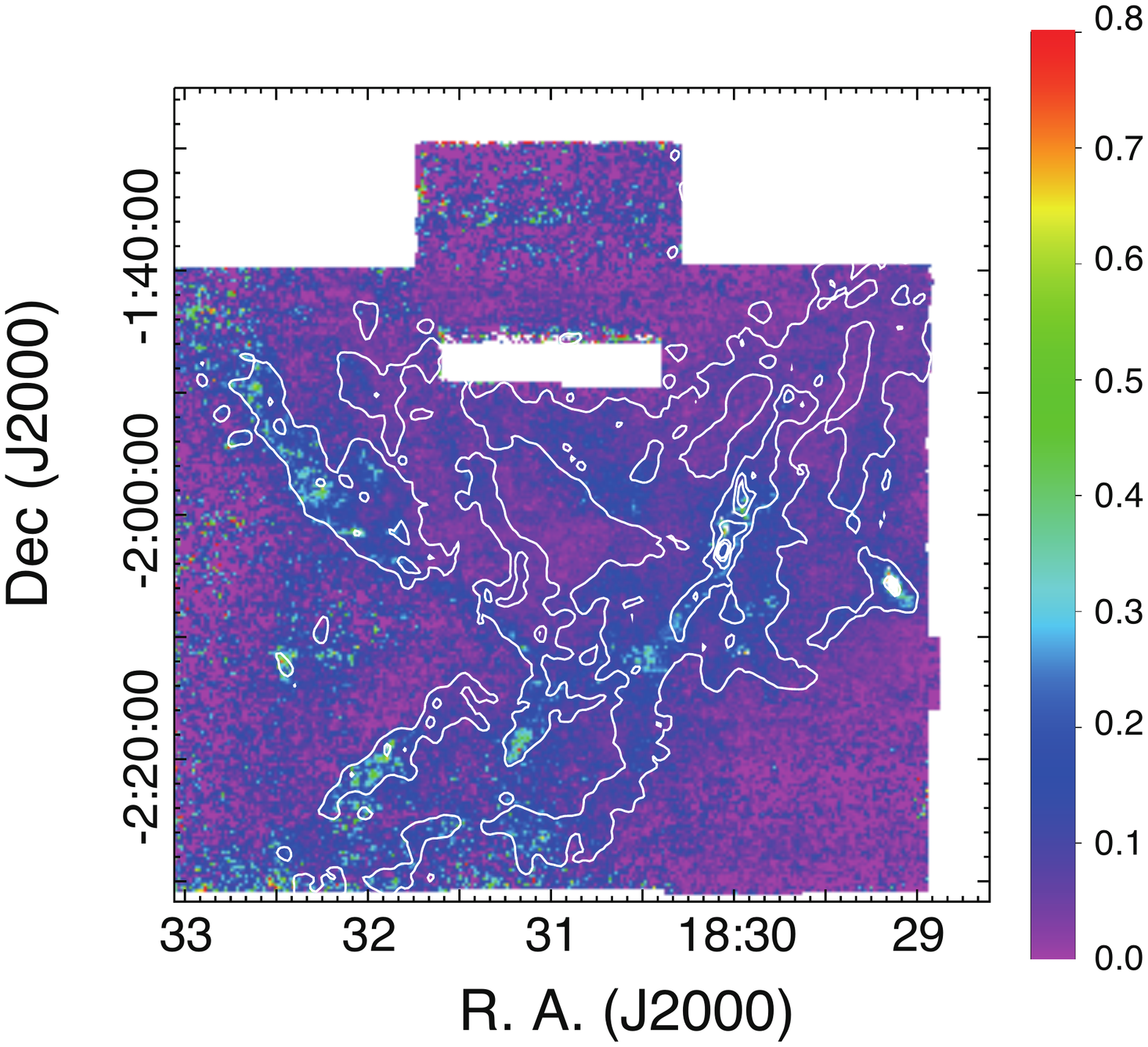}
 \end{center}
\caption{The C$^{18}$O-to-$^{13}$CO fractional abundance ratio of Aquila Rift.
\label{fig:aquila13co-c18o}}
\end{figure}

Figures \ref{fig:aquila13cocolumndensity}, \ref{fig:aquila13coabundance} and \ref{fig:aquilac18oabundance} show the spatial distribution of the $^{13}$CO column density, the fractional abundances
relative to H$_2$ and the optical depth for $^{13}$CO ($J=1-0$) and C$^{18}$O ($J=1-0$), respectively.  
The optical depth of $^{13}$CO tends to be higher than Orion A, and sometimes larger than 2--3 toward the  Serpens South region. 
%The filamentary structures are not so prominent both in $^{13}$CO ($J=1-0$) and C$^{18}$O ($J=1-0$).
The fractional abundances of $^{13}$CO and C$^{18}$O seem to be low in the filamentary structures seen in the Herschel map.
This may indicate that CO molecules are depleted in the cold dense gas in Serpens South.
The ratio of $^{13}$CO and C$^{18}$O abundances stays at around 0.1 (see figure \ref{fig:aquila13co-c18o}).

In figure \ref{fig:cumulative}b, we present the cumulative $^{13}$CO mass distributions of Aquila Rift. 
We divide the cloud into two areas: East (Dec. $\ge $ 18:30:39.6) and West (Dec $\le$ 18:30:39.6).
This indicates that the Eastern area contains a fraction of molecular gas with column density higher than 
$\sim$ 1 g cm$^{-2}$, the threshold for high mass star formation proposed by \citet{krumholz08}.
In contrast, in the Western area which contains Serpens South, the molecular gas with column densities higher than the threshold 
is deficient.
The total $^{13}$CO mass is estimated to be $3.9 \times 10^4 M_\odot$ with $7.4 \times 10^3$ M$_\odot$, 
$1.6\times 10^4$ M$_\odot$, and $2.3\times 10^4$ M$_\odot$ for OMC-1, North, and South, respectively.

    \subsection{M17}

\begin{figure}
 \begin{center}
 \includegraphics[angle=90,width=6cm, bb=-200 100 442 600]{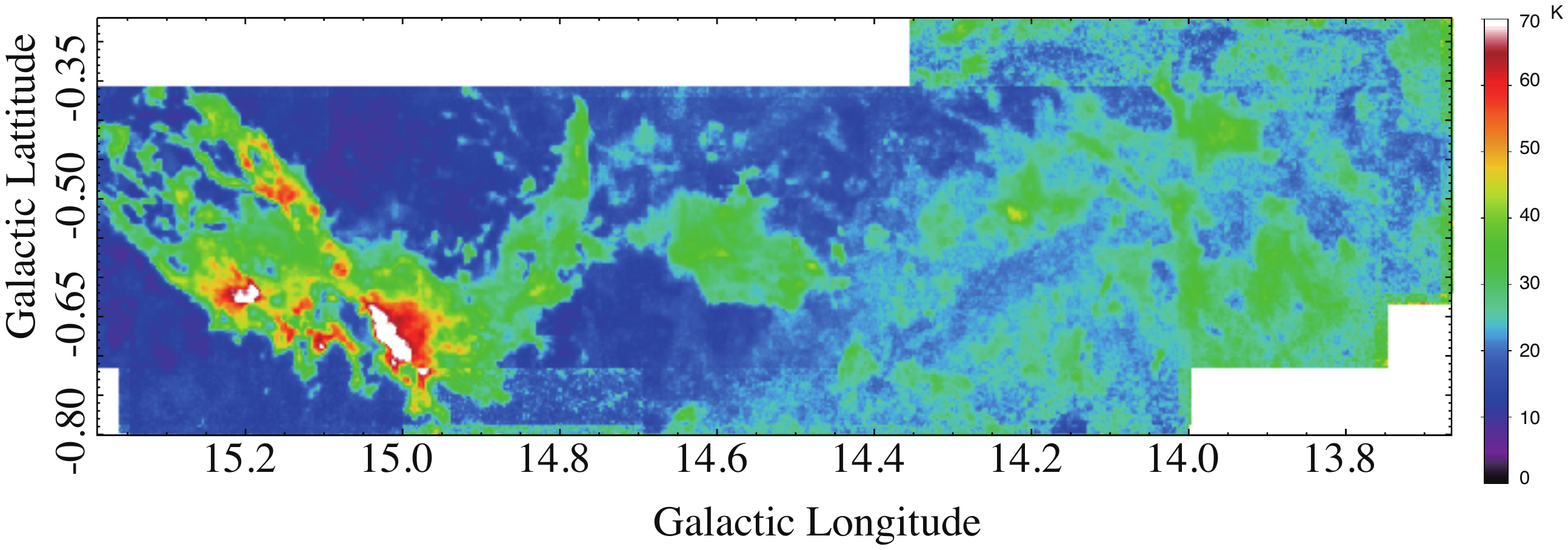}
 \end{center}
\caption{The excitation temperature map of M17.
\label{fig:M17tex}}
\end{figure}

\begin{figure}
 \begin{center}
   \includegraphics[angle=90,width=5cm, bb=-200 -200 442 600]{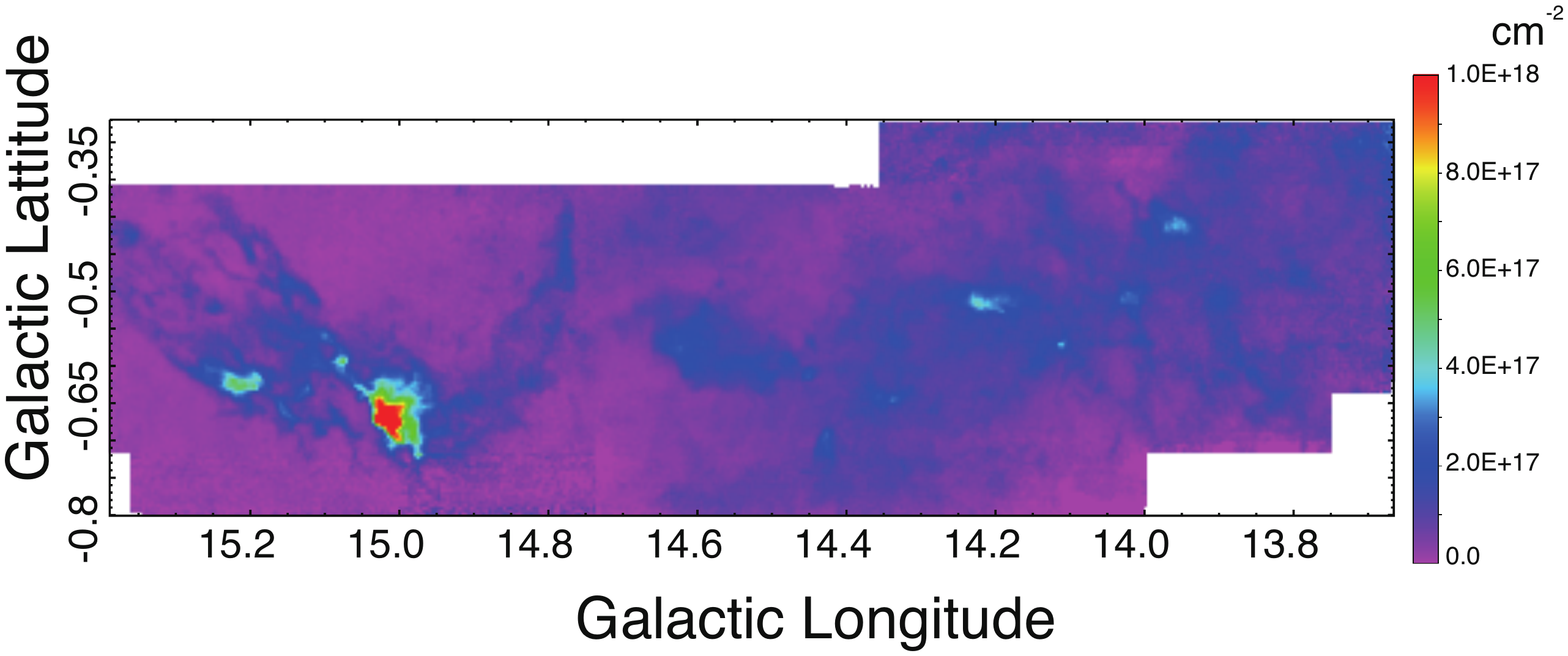}
   \includegraphics[angle=90,width=5cm, bb=-200 -200 442 600]{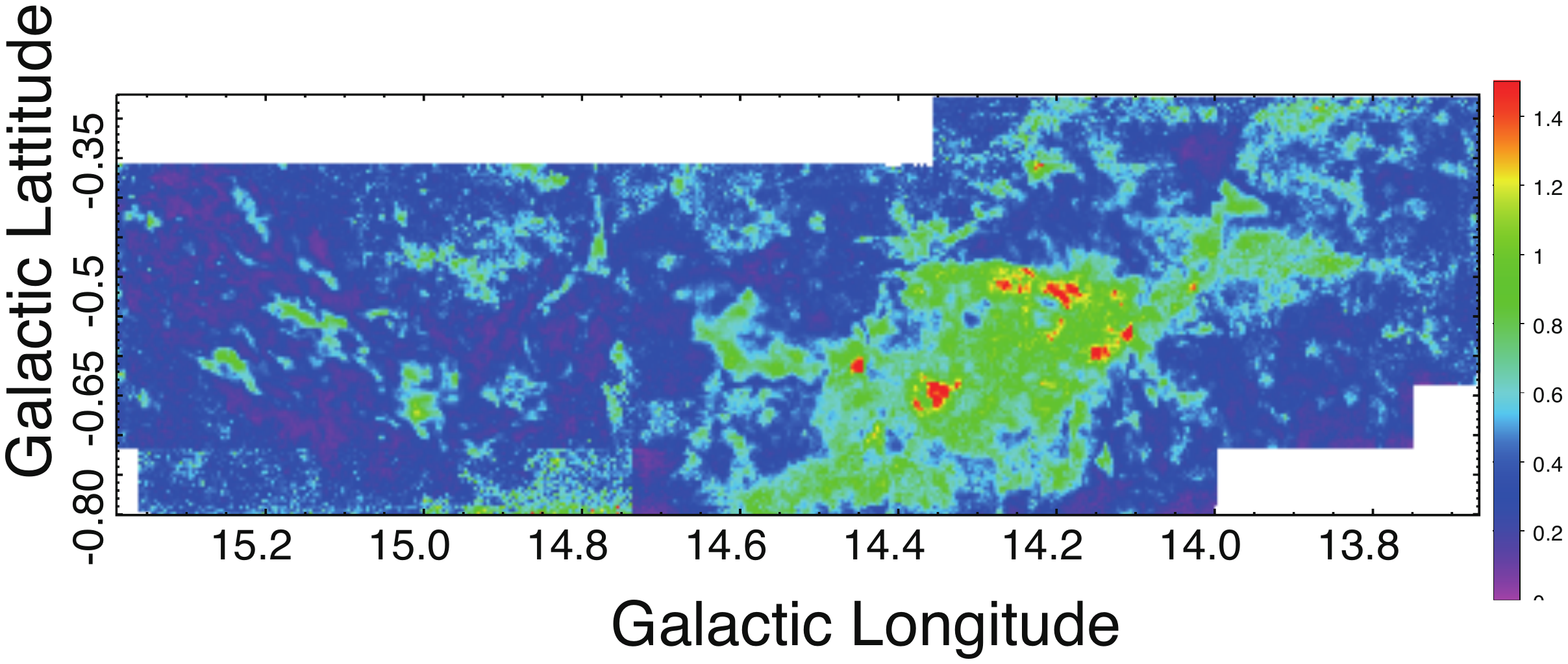}
 \end{center}
\caption{(a) The  $^{13}$CO fractional abundance and (b) its optical depth maps of M17.
\label{fig:M1713coabundance}}
\end{figure}

\begin{figure}
 \begin{center}
\includegraphics[angle=90,width=7cm, bb=-200 -200 500 600]{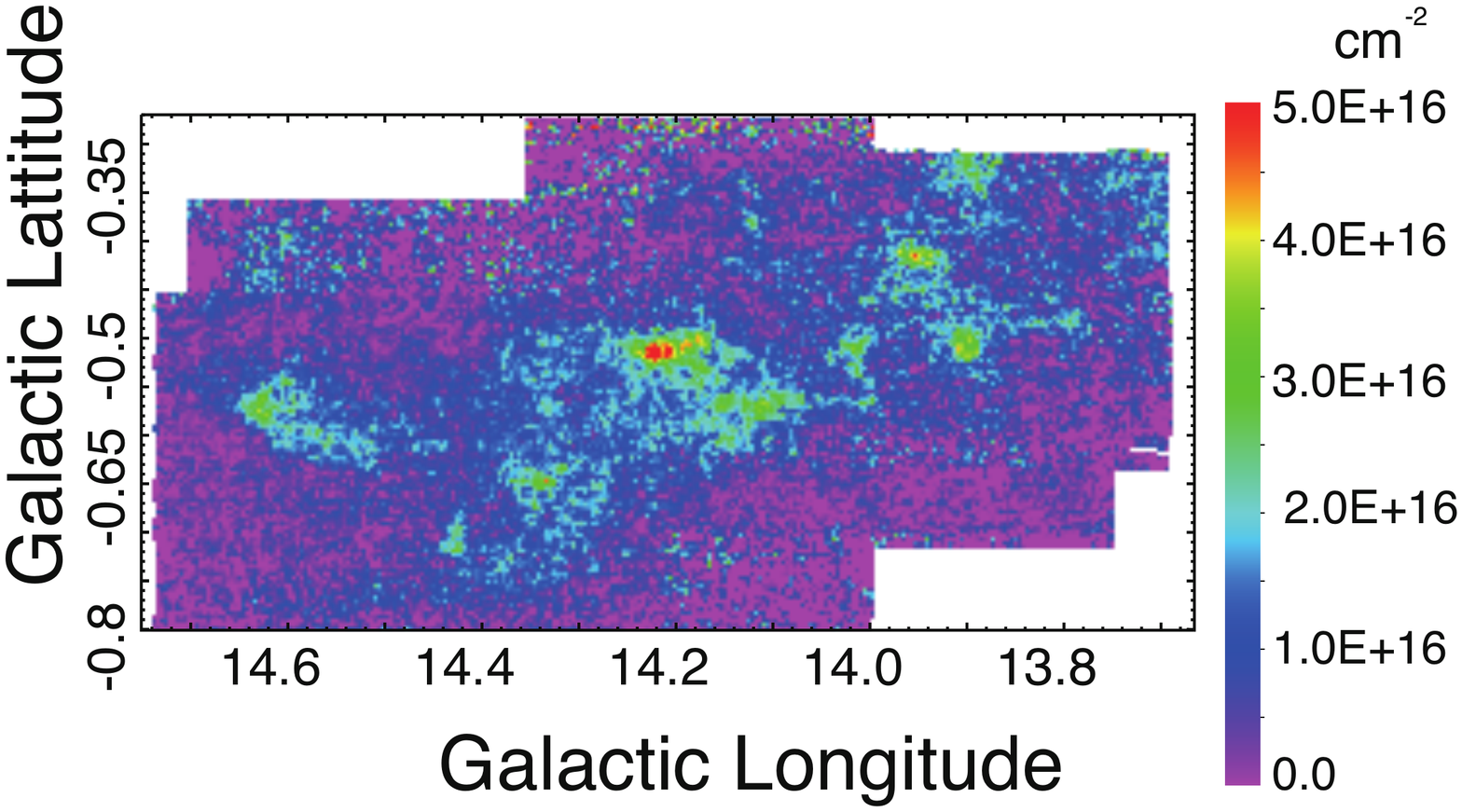}
 \includegraphics[angle=90,width=7cm, bb=-200 -200 500 600]{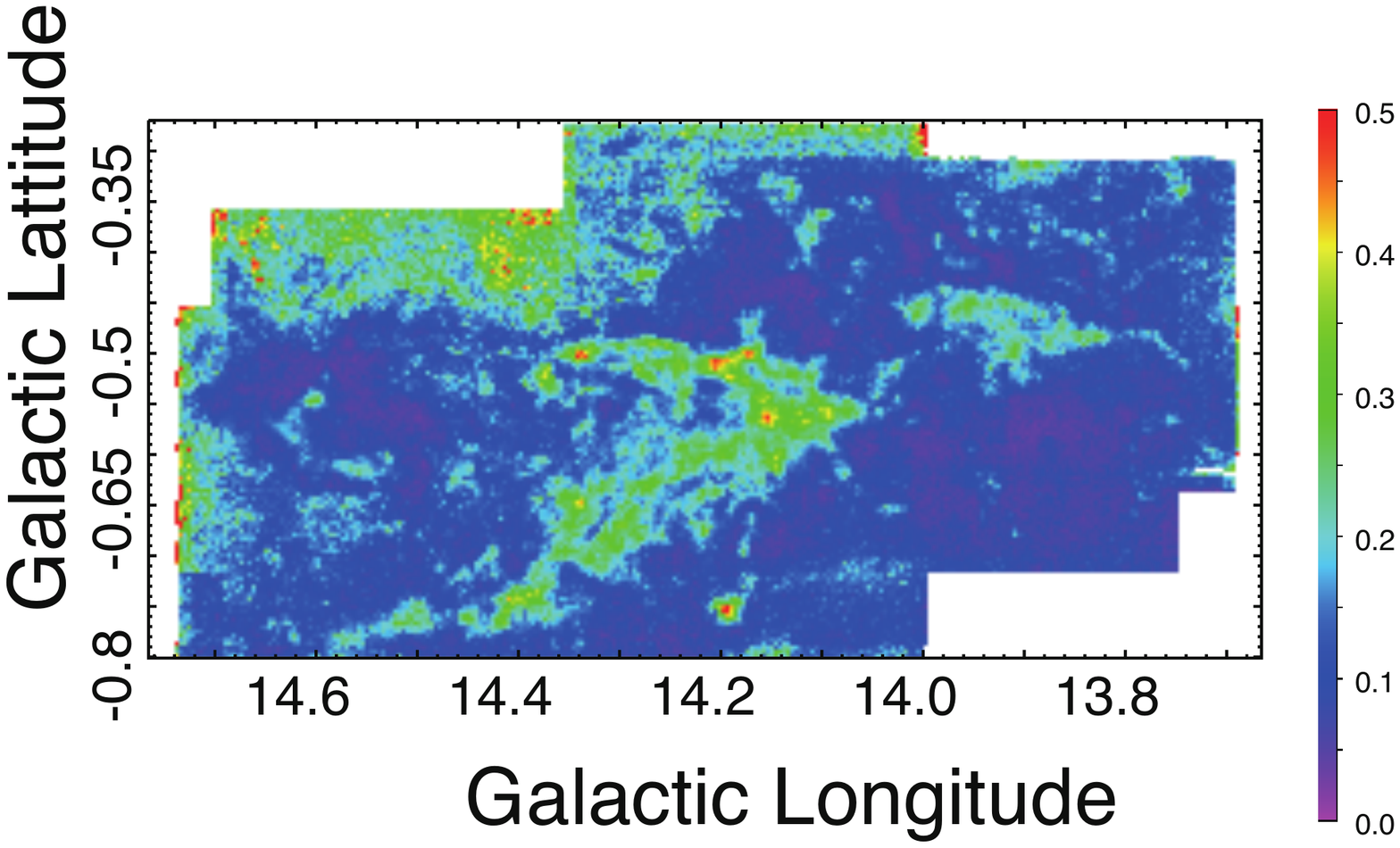}
 \end{center}
\caption{(a) The C$^{18}$O fractional abundance and (b) its optical depth maps of M17.
\label{fig:M17c18oabundance}}
\end{figure}

Figure \ref{fig:M17tex} shows the spatial distribution of the CO excitation temperature. 
The excitation temperature is high toward the M17 HII region.
In contrast, in the infrared dark cloud, the excitation temperature
stays  at 30 $-$ 40 K.  Te excitation temperatures are not so low as those of Serpens South region.
Figures \ref{fig:M1713coabundance} and \ref{fig:M17c18oabundance} show the spatial distribution of the column density
 and the optical depth for $^{13}$CO ($J=1-0$) and C$^{18}$O ($J=1-0$), respectively.  
The optical depths of $^{13}$CO and C$^{18}$O tend to be higher toward the M17 SWex.  
 See Nuygen Luong et al. (2019)  for the details of the global molecular gas distributions.

In figure \ref{fig:cumulative}c, we present the cumulative $^{13}$CO mass distributions of M17. 
We divide the region into two areas: M17 HII area ($l$ $\ge $ 14.9) and M17 SWex ($l$ $\le$ 14.9).
This indicates that the M17 HII contains a fraction of molecular gas with column density higher than 
$\sim$ 1 g cm$^{-2}$, the threshold for high mass star formation proposed by \citet{krumholz08}.
In contrast, in the M17 SWex, the molecular gas with column densities higher than the threshold 
is deficient.  This may be a reason why the high-mass star formation is not active in M17 SWex \citep{povich16}.
The total $^{13}$CO mass is estimated to be $3.6 \times 10^5 M_\odot$ and $4.5 \times 10^5$ M$_\odot$, 
for M17 HII and M17 SWex, respectively. 
M17 SWex area contains about two time more massive molecular gas than M17 HII.

\section{Molecular Outflows}
\label{sec:outflow}

Molecular outflow feedback is one of the important stellar feedback mechanisms. 
Molecular outflows can inject significant energy and momentum in molecular clouds.
Our $^{12}$CO data are useful to identify the outflows and gauge how much energy and momentum
are injected into the clouds.  Molecular outflow survey is one of the main sciences we will conduct.
Here we briefly present the result of the molecular outflow survey toward small areas in Orion A 
(L1641N) and Aquila Rift (Serpens South).  
See Tanabe et al. (2018) for the results of comprehensive outflow surveys 
toward Orion A.

\subsection{Orion A (L1641N)}

\begin{figure}
 \begin{center}
 \includegraphics[angle=90,width=7cm, bb=-200 0 500 600]{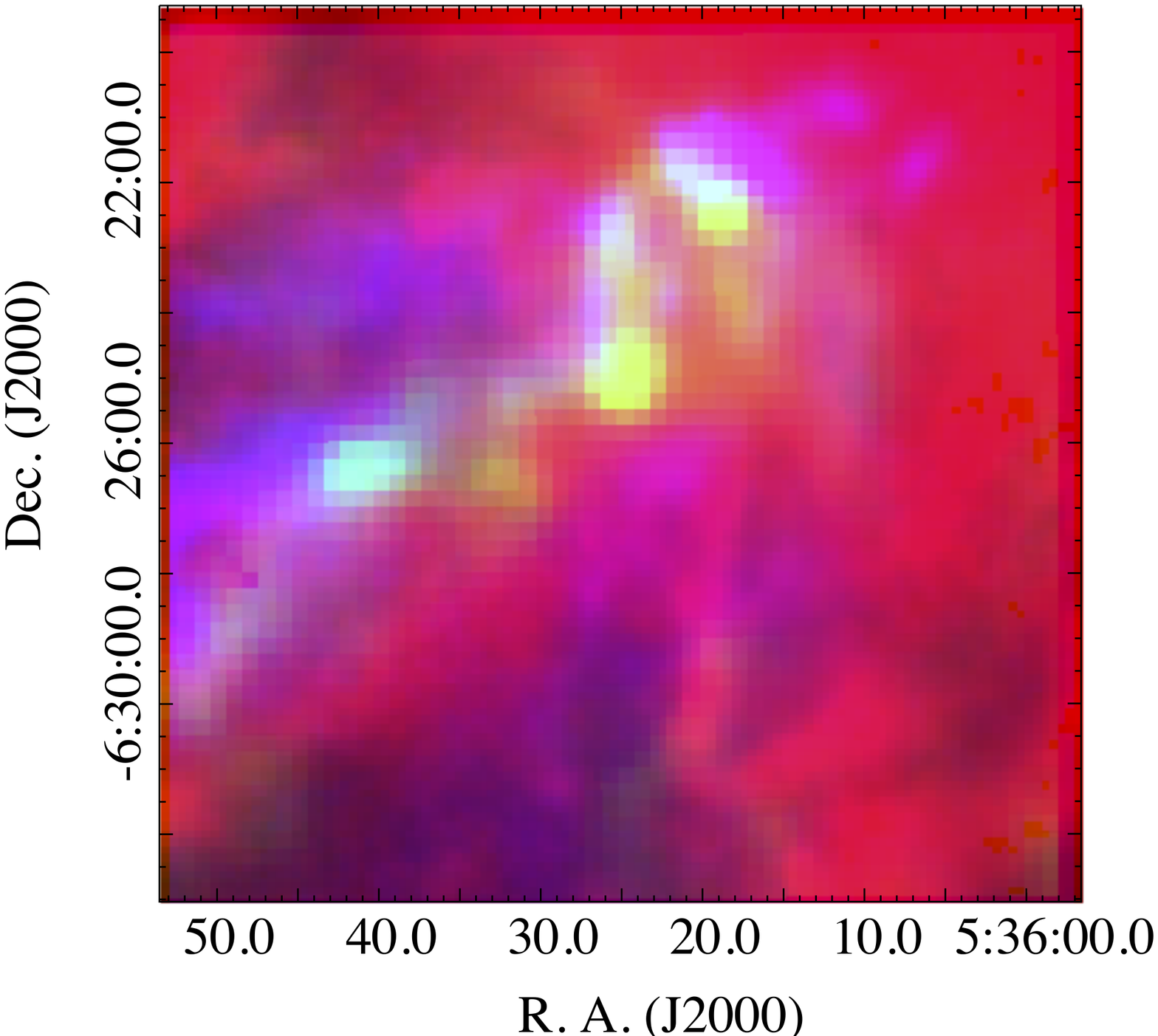}
 \end{center}
\caption{Three color image of the L1641N cluster-forming region with $^{12}$CO ($J=1-0$) intensity 
integrated from 10 km s$^{-1}$ to 16 km s$^{-1}$ (red),  $^{12}$CO ($J=1-0$) intensity 
integrated from 0 km s$^{-1}$ to 5 km s$^{-1}$ (blue), 
and {\it Herschel} column density (green).
\label{fig:L1641Noutflow}}
\end{figure}

In figure \ref{fig:L1641Noutflow}, we present the three-color image of the
L1641N region.
The red, blue and green images show the $^{12}$CO intensity image integrated from 11 km s$^{-1}$ to
15 km s$^{-1}$ (redshifted component), $^{12}$CO intensity image integrated from -20 km s$^{-1}$ to
0 km s$^{-1}$ (blueshifted components), and {\it Herschel} column density image, respectively.
In this region, \citet{stanke07} and \citet{nakamura12} identified molecular outflows in $^{12}$CO ($J=2-1$) 
and $^{12}$CO($J=1-0$), respectively.  The high-velocity components they previously identified are recognized
in our image.  For example, there are several dust cores in this region which are seen in green.
The redshifted collimated flow running from north to south blows out from the brightest dust core 
located at the position of (R.A., Dec.)=(5:36:19, $-$6:22:29). 

The result of accurate outflow identification is presented in Tanabe et al. (2018),
who found 44 CO outflows in Orion A, 17 out of which are new detections.
Based on the identified outflow physical parameters, they estimated the momentum injection rates due to the molecular outflows 
and found that the total momentum injection rate due to the outflows and the expanding shells identified in the $^{13}$CO data 
by \citet{feddersen18} is larger than the turbulence dissipation rate in Orion A. 
Thus, the stellar feedback such as the molecular outflows and expanding shells 
driven by stellar winds is an important mechanism to replenish the internal cloud turbulence.

\subsection{Aquila Rift (Serpens South)}

\begin{figure}
 \begin{center}
 \includegraphics[width=7cm, bb=-200 0 500 600]{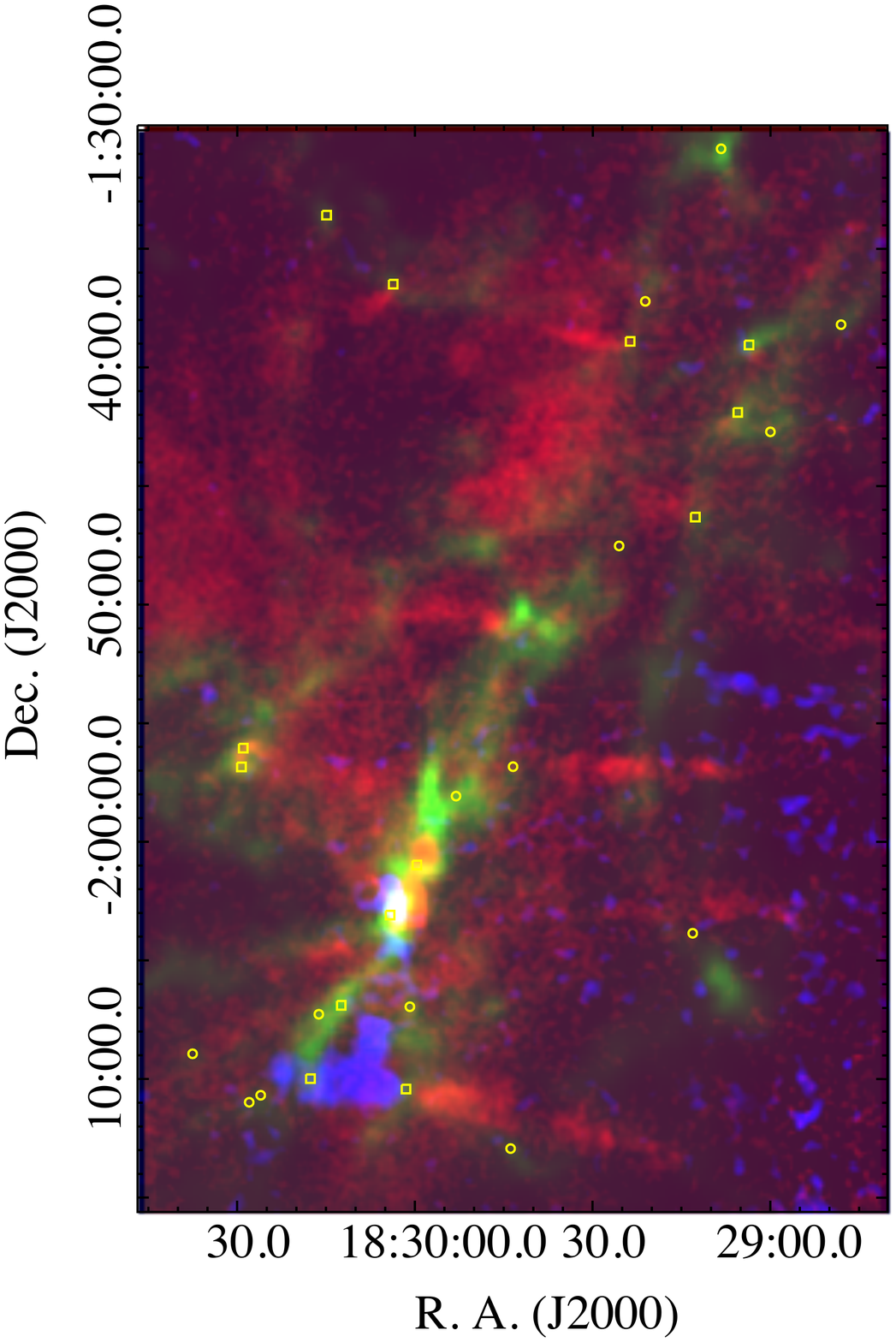}
 \end{center}
\caption{Three color image of the Serpens South region with $^{12}$CO ($J=1-0$) intensity 
integrated from 11 km s$^{-1}$ to 15 km s$^{-1}$ (red),  $^{12}$CO ($J=1-0$) intensity 
integrated from -20 km s$^{-1}$  to 0 km s$^{-1}$ (blue), and 
{\it Herschel} column density (green).
The squares and circles indicate protostellar core candidates with and without molecular outflows, respectively. The size of the circle is the same as the FWHM beam size of 21.7".
\label{fig:aquila_outflow}}
\end{figure}

In figure \ref{fig:aquila_outflow}, we present the three-color image of the Serpens South region.
\citet{nakamura11} conducted the molecular outflow survey toward the Serpens South cluster 
in $^{12}$CO ($J=3-2$).  The present paper is the first outflow survey using $^{12}$CO ($J=1-0$).
The coverage of the image shown in figure \ref{fig:aquila_outflow} is wider 
than that of \citet{nakamura11}.
The red, blue and green images indicate  the $^{12}$CO intensity image integrated from 11 km s$^{-1}$ to
15 km s$^{-1}$ (redshifted component), $^{12}$CO intensity image integrated from -20 km s$^{-1}$ to
0 km s$^{-1}$ (blueshifted components), and {\it Herschel} column density image, respectively.
The distribution of the high velocity components basically similar to that of $^{12}$CO ($J=3-2$).

By visual inspection, we attempt to identify the high-velocity components which are likely to 
originate from the molecular outflows toward {\it Herschel} protostellar cores in this region.
The result of the identification is summarized in table \ref{tab:outflow}.
We detected in the $^{12}$CO ($J=1-0$) emission 
almost all the outflow lobes identified by \citet{nakamura11}.
In total, we identified 13 outflow driving sources including the 3 tentative detections. 
From this survey, we identified 4 new outflow sources, all of which are located outside the
map of \citet{nakamura11}.

As discussed by \citet{shimoikura15}, the Aquila Rift region contains several
 cloud components with different line-of-sight velocities. The existence of such multiple 
 components sometimes precludes  clear identification of  high-velocity components 
 since high-velocity components tend to overlap with different cloud components 
 along the line of sight.
More careful inspection of the data cube is needed to fully identify the molecular outflows.
We expect that more outflows exist even in the area presented here.
We will present a complete outflow survey toward Aquila Rift in a separate paper.

The scientific results will be reported in more details 
 in separate papers
%We will conduct other sciences using the obtained data  in future.

\section{Summary}
\label{sec:summary}

In the present paper, we described the project overview of Nobeyama mapping project toward the three nearby molecular clouds,
Orion A, Aquila Rift, and M17.  
The main purpose of the present paper is to summarize complicated observational procedures and flux calibration methods.
%Individual sciences are presented in separate papers.  
We summarize the main results of the present paper as follows.

\begin{itemize}
\item[1.]  We conducted wide-field mapping observations toward three nearby molecular clouds, 
Orion A, Aquila Rift, and M17, in $^{12}$CO ($J=1-0$), $^{13}$CO ($J=1-0$), C$^{18}$O ($J=1-0$), 
N$_2$H$^+$ ($J=1-0$), and CCS  ($J_N=8_7-7_6$) using the Nobeyama 45-m telescope.
\item[2.] The map coverage is over 1$^\circ \times 1^\circ$.
We cover most of the molecular clouds seen in dust emission.
\item[3.]  We checked the absolute intensities obtained with the new 4-beam receiver, FOREST, 
by comparing the intensities obtained with the previous receiver, BEARS toward the same areas
for $^{12}$CO, $^{13}$CO, and C$^{18}$O.  
\item[4.] For N$_2$H$^+$, we compared the intensities of the Taurus molecular cloud
obtained with the IRAM 30-m telescope, and the fluxes taken with FOREST
coincide with those obtained with the IRAM 30-m telescope within an error of 5\%.
\item[5.] We obtained the column densities of $^{13}$CO ($J=1-0$) and C$^{18}$O ($J=1-0$)
and derived their fractional abundances toward Orion A.  Our maps indicate that 
the fractional abundances depends on the cloud environments, and varies from region to region by a factor of $\sim$10.
\item[6.] 
The cumulative column density distributions clearly show that only a fraction of the molecular gas
has column densities high enough to create high-mass stars for individual clouds. 
\item[7.]
Our maps have sufficient sensitivities to identify the molecular outflows.
In particular, in our $^{12}$CO ($J=1-0$) data, we confirmed all the outflows previously detected 
in $^{12}$CO ($J=3-2$) toward Serpens South, and identified 4 new outflows in the adjacent region. 
Using the catalog of the protostars, we identified the driving sources of these CO outflows.
\item[8.]  Finally we briefly describe results from our project published in separate papers.
We revealed hierarchical structure of Orion A, applying SCIMES and Dendrogram to $^{13}$CO cube data. 
In total, we identified about 80 clouds in Orion A. The abundance ratio of $^{13}$CO to C$^{18}$O
varies from region to region, affected by the far-UV radiation \citep{ishii19}. We identified 44 ouflows in Orion A. 
15 out of 44 are the new detections \citep{tanabe19}.  
We estimated a momentum injection rate of the identified outflows and found that they have significant injection momentum 
rate  in the surroundings. Using the data of the OMC-2 FIR 4 region, we characterize the spatial variation of the 
abundance ratios of several molecules and discussed the possible outflow-triggered star formation \citet{nakamura19}. 
The data was taken for the flux calibration of Orion A data.
For Aquila, \citet{shimoikura19} reveal evidence of the interaction of molecular cloud with the expanding H{\scriptsize II} 
region. 
\citet{kusune19} and \citet{sugitani19} carried out the near-infrared polarization observations toward Aquila and M17, 
respectively, and revealed that the global magnetic field tend to be perpendicular to the elongation of the molecular clouds.
\citet{quang19} revealed the global molecular gas distribution in M17, and found that in the IRDC region, the column densities 
are not dense enough to create high-mass stars, but ongoing cloud-cloud collisions are likely to be forming higher density 
regions. Thus, future high-mass star formation is expected.

\end{itemize}

\begin{ack}
This work was financially supported by JSPS KAKENHI Grant Numbers JP17H02863, JP17H01118, JP26287030, and JP17K00963.
This work was supported by NAOJ ALMA Scientific Research Grant Numbers 2017-04A.
This work was carried out as one of the large projects of the Nobeyama 
Radio Observatory (NRO), which is a branch of the National Astronomical 
Observatory of Japan, National Institute of Natural Sciences. 
We thank the NRO staff for both operating the 45 m and helping us with the data reduction.  
\end{ack}

%\bibliographystyle{aasjournal}
%\bibliography{reference}

\begin{table}
  \tbl{  Selected, recent wide-field survey toward Orion A, Aquila Rift, and M17}{%
  \begin{tabular}{llll}
  \hline
  Telescope/Survey & Line/continuum & Resolution & cloud/Key Reference\\
  &    & (arcsec/arcmin) &  \\ 
\hline
Osaka Pref. 1.85-m    & $^{12}$CO/$^{13}$CO/C$^{18}$O $J$=2--1   & 2\arcmin.7	&  Orion A/\citet{nishimura15} \\
Tsukuba 30-cm  & $^{12}$CO $J$=4--3   & 9\arcmin.4	&  Orion A/\citet{ishii16} \\
Harvard-CfA 1.2 m    & $^{12}$CO $J$=1--0   & 8\arcmin.4&  Orion A/\citet{wilson05} \\
ASTE, NRO 45-m    & 1.1m/$^{12}$CO $J$=1--0   & 36\arcsec/21\arcsec	&  Orion A/\citet{shimajiri11} \\
JCMT/GBS    & $^{13}$CO/C$^{18}$O $J$=3--2   & 17\arcsec	&  Orion A/\citet{buckle12} \\
FCRAO 14-m    & $^{12}$CO/$^{13}$CO $J$=1--0   & 46\arcsec	&  Orion A/\citet{ripple13} \\
IRAM 30-m    & $^{12}$CO/$^{13}$CO $J$=2--1   & 11\arcsec	&  Orion A/\citet{berne14} \\
ASTE 10-m    & $^{12}$CO $J$=3--2   & 30\arcsec	&  Orion A/\citet{takahashi08} \\
Herschel/HIFI IRAM 30-m   & CH$^+$/CO ($J=10-9$)/HCN/HCO$^+$ ($J=6-5)$, \dots   & 12\arcsec--27\arcsec	&  Orion A (OMC-1)/\citet{goicoechea18} \\
Herschel-Planck    & dust continuum   & 36\arcsec	&  Orion A/\citet{lombardi14} \\
Spitzer    & MIR 3--24 $\mu$ m   & 2\arcsec--5\arcsec	&  Orion A/\citet{megeath12} \\
VISTA/VISION    & NIR 0.85--2.4$\mu$ m   & 0.85\arcsec	&  Orion A/\citet{meingast16} \\
IN-SYNC    & NIR 1.5--1.6 $\mu$ m   & 1.6\arcsec	&  Orion A/\citet{dario16} \\
NRO 45-m    & N$_2$H$^+$ $J$=1--0   & 21\arcsec	&  Orion A/\citet{tatematsu08} \\ 
NRO 45-m    & H$^{13}$CO$^+$ $J$=1--0   & 21\arcsec	&  Orion A/\citet{ikeda07} \\ 
CARMA+NRO 45 m/CARMA-NRO Orion  & $^{12}$CO/$^{13}$CO/C$^{18}$O $J$=1--0   & 8\arcsec	&  Orion A/\citet{kong18} \\ 
NRO 45-m   & $^{12}$CO/$^{13}$CO/C$^{18}$O/N$_2$H$^+$ $J$=1--0 / CCS $J_N=7_6-6_5$   & 21\arcsec--24\arcsec	&  Orion A/this paper \\ 
\hline
Harvard-CfA 1.2 m    & $^{12}$CO $J$=1--0   & 8\arcmin.4&  Aquila/\citet{dame01} \\
Osaka Pref.  1.85-m  & $^{12}$CO/$^{13}$CO/C$^{18}$O $J$=2--1   & 2.\arcmin7	&  Aquila/ \citet{nakamura17} \\
  Herschel    & dust   & 36\arcsec	&  Aquila/\citet{andre10} \\
IRAM 30-m/MAMBO  & 1.2mm   & 11\arcsec	&  Aquila/\citet{maury11} \\
IRAM 30-m  & HCN/H$^{13}$CN/HCO$^+$/H$^{13}$CO$^+$ $J$=1--0   & 40\arcsec	&  Aquila/\citet{shimajiri17} \\
JCMT/GBS & 850$\mu$ m $^{12}$CO $J$=3--2   & 15\arcsec/22\arcsec	&  Aquila (W40)/\citet{rumble16} \\
ASTE 10-m    & $^{12}$CO $J=3-2$ / HCO$^+$ $J$=4--3   & 31\arcsec	&  Aquila (W40)/\citet{shimoikura15} \\
ASTE 10-m  & $^{12}$CO $J$=3--2 /HCO$^+$ $J=4-3$   & 24\arcsec	&  Aquila (Serpens South)/\citet{nakamura11b} \\
Spitzer    & IRAC   & 2\arcsec	&  Aquila (Serpens South)/\citet{gutermuth08} \\
NRO 45-m    & N$_2$H$^+$ $J$=1--0   & 24\arcsec	&  Aquila (Serpens South)/\citet{tanaka13} \\
MOPRA     & N$_2$H$^+$/H$^{13}$CN/HCN/HNC/HCO$^+$/H$^{13}$CO$^+$ $J$=1--0   & 40\arcsec	&  Aquila (Serpens South)/\citet{kirk13} \\
NRO 45-m    & CCS $J_N=4_3-3_2$/HC$_3$N $J=5-4$   & 37\arcsec	&  Aquila (Serpens South)/\citet{nakamura14} \\
NRO 45-m   & $^{12}$CO/$^{13}$CO/C$^{18}$O/N$_2$H$^+$ $J$=1--0 / CCS $J_N=7_6-6_5$   & 21\arcsec--24\arcsec	&  Aquila/this paper \\ 
\hline
Spitzer    &  MIR  & 2\arcsec	&  M17/\citet{povich10,povich16} \\
HHT 10-m   & $^{12}$CO/$^{13}$CO $J$=2--1   & 32\arcsec	&  M17/\citet{povich09} \\
 NRO 45-m  & $^{12}$CO/$^{13}$CO/C$^{18}$O $J$=1--0   & 20\arcsec	&  M17/\citet{nishimura18} \\
 NRO 45-m   & $^{12}$CO/$^{13}$CO/C$^{18}$O/N$_2$H$^+$ $J$=1--0 / CCS $J_N=7_6-6_5$   & 21\arcsec--24\arcsec	&  M17/this paper \\ 
\hline
  \end{tabular}}
  \label{tab:survey}
  \begin{tabnote}
  This is not a complete list of the recent wide-field survey.
See also table 1 of \citet{kong18} for Orion A.
  \end{tabnote}
\end{table}

\begin{table}
  \tbl{Spatial resolutions achieved for the Nobeyama 45-m, CARMA, and ALMA observations}{%
  \begin{tabular}{lllll}
  \hline
telescopes & angular resolution & nearest regions & intermediate regions & distant regions \\
  &  &  $\sim$140 pc & $\sim$400 pc& $\sim$3 kpc \\
  &  & (Taurus, $\rho$ Oph) & ({\bf Orion A, Aquila Rift}, California) & ({\bf M17}, several IRDCs) \\
\hline
NRO 45-m only & $\sim 20''$ & 2800 au (0.014 pc) & 8400 au (0.04 pc) & 40000 au (0.2 pc) \\
CARMA+NRO 45-m & $\sim 8''$ & $-$  &  3300 au& $-$ \\ 
ALMA	& $\sim 1''$ & 140 au & 420 au & 3000 au 	\\
\hline
 \end{tabular}}\label{tab:resolution}
 \begin{tabnote}
 The distances are adopted from the following references: 
 nearest regions, Taurus (137 pc, \cite{torres07} and \cite{loinard07}), 
 $\rho$ Oph (137 pc, \cite{ortez17a}), intermediate regions, Orion A (414 pc, \cite{menten07,kim08}),  Aquila Rift (436 pc, \cite{ortez17b}), and distant regions, M17 (2.0 kpc, \cite{xu11}).
 We note that for the M17 region, several clouds with different distances seem to be overlapped 
 along the line of sight \citep{povich16}.
Here we refer to the ALMA observations just for comparison 
of the spatial resolutions achieved. In the present paper, we present the data obtained with the Nobeyama 45-m telescope.
The detail of the CARMA+Nobeyama combined data is presented in \citet{kong18}.
\end{tabnote}
\end{table}

\begin{table}
  \tbl{Observed lines}{%
  \begin{tabular}{lllllll}
  \hline
  Molecule &Transition & Rest Frequency &Beam Size 
& $\Delta V$  & Main Beam Efficiency ($\eta$)\\
& &(GHz)  & (arcmin) & (km s$^{-1}$) & \\ 
\hline
$^{12}$CO	    & $J$=1--0    &	115.271204 	& 14.3 $\pm$ 0.4	& 0.1  & 0.416 $\pm$ 3\%\\
$^{13}$CO	    & $J$=1--0    & 	110.201354	& 14.9 $\pm$ 0.4	& 0.1  &  0.435 $\pm$ 3\% \\
C$^{18}$O	    & $J$=1--0    &	109.782176 	& 14.9 $\pm$ 0.4	& 0.1 & 0.437 $\pm$ 3\%\\
N$_{2}$H$^+$	& $J$=1--0    & 93.1737637	&17 $\pm$ 0.5	& 0.1   & 0.500  $\pm$ 5\%\\
CCS	& $J_N$=8$_7$--7$_6$ &	 93.870098 	&17 $\pm$ 0.5 & 0.1  & 0.497  $\pm$ 5\%\\
\hline
  \end{tabular}}\label{tab:obs}
  \begin{tabnote}
The rest frequency of the main hyperfine component of N$_2$H$^+$ is adopted from \citet{pagani09}. 
%In the sixth column, we present the archived rms noise levels in
%the brightness temperature scale.
The beam size and main beam efficiencies listed are those measured with FOREST.
The errors of the efficiencies are mainly due to the uncertainty of brightness temperature of the planets used for the measurements。See the NRO web page for more details.
  \end{tabnote}
\end{table}

\begin{table}
  \tbl{Sizes of Observation Boxes for the FOREST observations}{%
  \begin{tabular}{lllll}
  \hline
lines &  scan & $^{12}$CO/$^{13}$CO & C$^{18}$O/N$_2$H$^+$  & $^{13}$CO/C$^{18}$O/N$_2$H$^+$/CCS \\
\hline
season &  & 2014-2015/2015-2016/2016-2017 & 2015-2016  & 2016-2017 \\
\hline
Orion A   & $x$ & $10'\times 5'$  &	$20'\times 5'$ 	& $20'\times 5'$  \\
Orion A   & $y$ & $5'\times 10'$   &	$5'\times 20'$ 	& $5'\times 20'$ \\
Aquila Rift  & $x$ &   $20' \times 10'$  &  $-$  &  $20' \times 10'$ \\
Aquila Rift  & $y$ &   $10' \times 20'$  &  $-$  &  $-$ \\
M17  & $x$ &   $20' \times 10'$  &  $20' \times 10'$  &  $20' \times 10'$ \\
M17  & $y$ &   $10' \times 20'$  &  $10' \times 20'$  &  $10' \times 20'$ \\
\hline
  \end{tabular}}\label{tab:box}
  \begin{tabnote}
Taking into account the observation schedule of each season, we changed the observation box sizes.
For Aquila Rift, we could not obtain y-scan data of C$^{18}$O, N$_2$H$^+$, and CCS.
In addition, for M17, we significantly reduced the observation area for the 
C$^{18}$O, N$_2$H$^+$, and CCS observations. These incomplete observations 
are mainly due to the malfunction of the
master collimeter driving system happened in the 2016-2017 season.
  \end{tabnote}
\end{table}

\begin{table}
  \caption{Parameters of Observations with FOREST}\label{tab:otf}
  \begin{center}
    \begin{tabular}{llll}
     \hline
      Box size & $10' \times 5'$ & $20' \times 5'$ & $20' \times 10'$ \\ \hline
      Time for scan (s) & 15 & 30 & 26 \\
      Number of ONs per OFF & 4 & 2 & 3 \\
      Separation between scans  &  5$''$.17 & 5$''$.17 & 5$''$.17 \\
%      Map grid &  7$''$.5  &  7$''$.5 &  7$''$.5 \\
      Frequency resolution (kHz) & 15.26 & 15.26 & 15.26\\
%      Convolution function & spheroidal & spheroidal  &  spheroidal   \\
%      Observation Time & 20 min \\
      \hline
    \end{tabular}
  \end{center}
    \begin{tabnote}
 For all observations, scans of the OTF observations are separated 
 in the interval of \timeform{5''.17}.  Thus, individual scans 
 by 4 beams of FOREST are completely overlapped. This minimizes 
 the effort of flux calibration greatly. 
  \end{tabnote}
\end{table}

\begin{table}
  \tbl{Map coverage and positions of emission-free areas used for the observations 
  and SiO maser objects used for
  the pointing observations}{%
  \begin{tabular}{lllllll}
  \hline
  Regions & Map coverage & Emission-free areas&  & Pointing Objects&  (SiO maser line) & \\
& & R.A. (J2000.0) & Dec.(J2000.0) & R.A. (J2000.0)  & Dec.(J2000.0) & \\
\hline
Orion A   & $0.7^\circ \times 2^\circ$ & $\timeform{05h29m00.0s}$ &	$\timeform{-05D25'30.0''} $ & 
 $\timeform{05h35m14.16s}$ & \timeform{-05D22'21.5"} & Orion KL \\
Aquila Rift  & $1^\circ \times 1^\circ$ & $\timeform{18h41m19.09s}$ & \timeform{+04D12'00.0''} 
& $\timeform{18h37m19.26s}$ & \timeform{+10D25'42.2''}  & V1111-Oph \\
M17	  &  $2^\circ \times 0.5^\circ$ & $\timeform{18h37m19.26s}$  & \timeform{+10D25'42.2''} 
& $\timeform{18h37m19.26s}$ & \timeform{+10D25'42.2''} & V1111-Oph \\
\hline
  \end{tabular}}
   \label{tab:off}
 \begin{tabnote}
These areas and objects were used for both the BEARS and FOREST observations.
   \end{tabnote}
\end{table}

\begin{table}
  \tbl{Summary of the FOREST observations}{%
  \begin{tabular}{llllll}
   \hline
  Lines & Period & Observation Time & $T_{\rm sys}$ & vel. res. & noise level   \\
  \hline
  Orion A & & & & & \\
  \hline
$^{12}$CO & 2014 Dec $-$ 2016 Dec & 150 hrs & 350 $-$ 400 K & 0.1 km s$^{-1}$ & 0.50 $-$1.5 K \\
$^{13}$CO  & 2014 Dec $-$ 2017 March & 150 hrs &150 $-$ 200 K  & 0.1 km s$^{-1}$ & 0.20 $-$0.30 K  \\
C$^{18}$O  & 2017 Jan $-$ 2017 March & 150 hrs &150 $-$ 200 K & 0.1 km s$^{-1}$ & 0.26 $-$ 0.30 K  \\
CCS & 2017 Jan $-$ 2017 March & 150 hrs &150 $-$ 200 K & 0.1 km s$^{-1}$ & 0.30 $-$ 0.49 K   \\
N$_2$H$^+$ & 2017 Jan $-$ 2017 March & 150 hrs &150 $-$ 200 K & 0.1 km s$^{-1}$ & 0.26 $-$ 0.30  K   \\
  \hline
 Aquila Rift  & & & & & \\
  \hline
$^{12}$CO  & 2015 March $-$ 2017 March & 50 hrs & 300 $-$ 500 K & 0.1 km s$^{-1}$ & 0.38 $-$ 0.50 K \\
$^{13}$CO  & 2015 March $-$ 2017 March & 150 hrs & 150 $-$ 200 K  & 0.1 km s$^{-1}$ &0.38 $-$ 0.50 K   \\
C$^{18}$O & 2016 March $-$ 2017 March & 120 hrs & 150 $-$ 200 K & 0.1 km s$^{-1}$ & 0.20 $-$ 0.30 K \\
CCS  & 2016 March $-$ 2017 March & 120 hrs & 150 $-$ 200 K & 0.1 km s$^{-1}$ & 0.17 $-$ 0.20 K \\
N$_2$H$^+$ & 2016 March $-$ 2017 March & 120 hrs & 150 $-$ 200 K & 0.1 km s$^{-1}$ & 0.18 $-$ 0.22 K \\
\hline
M17 & & & & & \\
  \hline
$^{12}$CO  & 2015 March $-$ 2017 March & 25 hrs & 300 $-$ 500 K & 0.1 km s$^{-1}$ & 0.48 $-$1.7 K \\
$^{13}$CO  & 2015 March $-$ 2017 March & 150 hrs & 150 $-$ 200 K  & 0.1 km s$^{-1}$ & 0.16 $-$ 0.80 K   \\
C$^{18}$O & 2016 March $-$ 2017 March & 65 hrs & 150 $-$ 200 K & 0.1 km s$^{-1}$ & 0.17 $-$ 0.34 K  \\
CCS & 2016 March $-$ 2017 March & 65 hrs &150 $-$ 200 K & 0.1 km s$^{-1}$ & 0.20 $-$ 0.27 K  \\
N$_2$H$^+$ & 2016 March $-$ 2017 March & 65 hrs &150 $-$ 200 K & 0.1 km s$^{-1}$ & 0.15 $-$ 0.27 K \\
\hline
  \end{tabular}}
   \label{tab:forest}
 \begin{tabnote}
 The values of $T_{\rm sys}$ are given in the single side band.
    \end{tabnote}
\end{table}

\begin{table}
  \tbl{Summary of the BEARS observations}{%
  \begin{tabular}{llllllll}
   \hline
  Lines & Period & Observation Time & $T_{\rm sys}$ & vel. res. & noise level  & obs mode &  reference \\
  \hline
  Orion A & & & & & & & \\
  \hline
$^{12}$CO (north) & 2007 Dec $-$ 2008 May & $\sim$ 40 hrs & 250 $-$ 500 K & 0.2 km s$^{-1}$ & 0.4 K  & OTF &1 \\
$^{12}$CO (south) & 2009 Dec $-$ 2010 Jan & $\sim$ 20 hrs & 300 $-$ 600 K & 0.5 km s$^{-1}$&  0.52 K  & OTF & 2 \\
$^{13}$CO (north) & 2013 May & $\sim$ 50 hrs &270 $-$ 470 K  & 0.3 km s$^{-1}$ & 0.7 K  & OTF & 3 \\
$^{13}$CO (south) & 2012 Apr $-$ 2013 March & $\sim$ 60 hrs & 210 $-$ 400 K & 0.1 km s$^{-1}$ & 1.96 K  & OTF &  this paper \\
C$^{18}$O (north) & 2010 March $-$ 2013 May & $\sim$ 100 hrs &270 $-$ 470 K & 0.1 km s$^{-1}$ & 0.3 K  & OTF & 3 \\
  \hline
 Aquila Rift (Serpens South) & & & & & & & \\
  \hline
$^{12}$CO  & 2011 Apr $-$ 2011 May & $\sim$15 hrs & 250 $-$ 500 K & 0.5 km s$^{-1}$ & 1.3 K  & OTF&this paper \\
$^{13}$CO  & 2011 Apr $-$ 2011 May & $\sim$ 30 hrs &210 $-$ 400 K  & 0.5 km s$^{-1}$ & 0.88 K  & OTF& this paper \\
C$^{18}$O & 2011 Apr $-$ 2014 Apr & $\sim$ 30 hrs &200 $-$ 400 K & 0.1 km s$^{-1}$ & 0.9 K  & OTF& this paper \\
\hline
  \end{tabular}}
   \label{tab:bears}
 \begin{tabnote}
The values of $T_{\rm sys}$ are given in  the double side band.
References --- 1: \citet{shimajiri11}, 2: \citet{nakamura12}, 3:\citet{shimajiri14}
   \end{tabnote}
\end{table}

\begin{table}
  \caption{Molecular Gas Mass Estimated from $^{13}$CO toward Orion A, Aquila Rift, and M17}\label{tab:mass}
  \begin{center}
    \begin{tabular}{lll}
     \hline
      Name & Region & Mass  \\ 
       &  & $M_\odot$ \\ \hline
      Orion A  &  total  &  $3.86\times 10^{4}$  \\
      Orion A North  &  Dec. $\ge$ --05:30:26.0  &  $1.56\times 10^{4}$  \\
      Orion A South  &  Dec. $\le$ --0.5:16:03.5 &  $2.30\times 10^{4}$  \\
      Orion A OMC-1  &  --05:30:26.0 $\le$ Dec. $\le$ --0.5:16:03.5 &  $7.4\times 10^{3}$\\ \hline
           Aquila Rift  &  total  &  $2.67\times 10^{4}$  \\
      Aquila East (W40)  &  R.A. $\le$ 18:30:39.6  &  $1.30\times 10^{4}$  \\
      Aquila West (Serpens South)   &  R.A. $\ge$ 18:30:39.6 &  $1.37\times 10^{4}$  \\ \hline
       M17 &  total  &  $8.1\times 10^{5}$  \\
     M17  &  $ l \ge$ 14:26:22.5  &  $3.6\times 10^{5}$  \\
     M17 SWex &  $l \le$ 14:26:22.5 &  $4.5\times 10^{5}$  \\
            \hline
    \end{tabular}
  \end{center}
    \begin{tabnote}
 OMC-1 is a part of the north area of Orion A.  The distances of 414 pc, 436 pc, and 2.0  kpc are adopted for 
 Orion A, Aquila Rift, and M17, respectively. 
 The excitation temperature of $^{13}$CO is estimated from the peak intensity of $^{12}$CO at each pixel.
  \end{tabnote}
\end{table}

\begin{table}
  \tbl{Molecular Outflow Candidates in Serpens South}{%
  \begin{tabular}{llllll}
   \hline
  No. & R.A. (J2000) & Dec. (J2000) & Outflows & classification &   \citet{nakamura11}\\
  \hline
135 & 18:28:48.09  & -01:38:11.6 & N &   & \\                                                                                                 
147 & 18:29:00.01  & -01:42:42.8 & N &  & \\
151 & 18:29:03.62  & -01:39:03.0 & BR & C & B12, R7 \\                                                                                                         
155 & 18:29:05.52 & -01:41:53.6 & R & C & B11, R6 \\                                                                                                       
163 & 18:29:08.34 & -01:30:46.8 & N & & \\                                                                                                       
171 & 18:29:12.68 & -01:46:18.4 & R & M & new\\
174 & 18:29:13.10 & -02:03:51.2& N & & \\                                                    
196 & 18:29:21.13 & -01:37:12.8 & N & & \\
202 & 18:29:23.69 & -01:38:54.0 & R & C & B10, R5 \\                                                                                                              
209 & 18:29:25.54 & -01:47:31.5 & N & & \\                                                                                                             
250 & 18:29:43.44 & -01:56:49.9 & N & & \\
251 &18:29:43.84 & -02:12:56.0 & N & & \\                                                                                                              
271 &18:29:53.06 & -01:58:04.5 & N & & \\
285 & 18:29:59.67, & -02:00:58.7 & R & C & R2 \\                                                                                                           
289 & 18:30:00.85 & -02:06:57.3 & N & & \\                                                                                                           
290 & 18:30:01.20 & -02:06:09.8 & N & & \\
292 &18:30:01.50 & -02:10:25.5& BR & C & B15, R8 \\                                                                                                        
297 & 18:30:03.68 & -01:36:29.4 & R & C & new \\                                                                                                              
299 & 18:30:04.19 & -02:03:05.5 & BR & C & B1,B2, B3,B6, R1, R3,  R4? \\                                                                                                             
315 & 18:30:12.44 & -02:06:53.6 & B & C & B7 \\                                                                                                              
321 & 18:30:14.93 & -01:33:34.9 & BR & M & new \\                                                                                                              
323 & 18:30:16.22 & -02:07:16.3 & N & & \\                                                                                                                                                                                                                                                                                                                                                                                                                                                                                          326 & 18:30:17.64 & -02:09:59.3 & B & C  & B14 \\
342 & 18:30:26.03 & -02:10:41.2 & & & \\                                                                                                              
347 & 18:30:27.97 & -02:10:59.0 & & & \\
349 & 18:30:28.98 & -01:56:03.2& R & Y & R9 \\                                                                                                        
351 & 18:30:29.28 & -01:56:50.6 & R & M & new \\     
362 & 18:30:37.53 & -02:08:56.3 & N & N & \\                                                                                                    
\hline
  \end{tabular}}
   \label{tab:outflow}
 \begin{tabnote}
 1st column: the number of {\it Herschel} protostellar core catalog  by \citet{konyves15}.
 4th column: B = blueshifted component, R = redshifted component.
 5th column:  C = clear, M = marginal.
 6th column: comparison with the identification by \citet{nakamura11},
 new = new detection (this paper).
In the densest part of the Serpens South cluster, outflow components from several different 
are observed. We assigned all such lobes to No. 299.
    \end{tabnote}
\end{table}

%%%%%%%%%%%%%%%%
%%% Figures
%%%%%%%%%%%%%%%%
\clearpage

%\begin{figure}
% \begin{center}
%% \includegraphics[width=8cm]{newfig/M1713cocolumndensitymap.eps}
% \end{center}
%\caption{The opacity-corrected $^{13}$CO column density map of M17.
%\label{fig:M1713cocolumndensity}}
%\end{figure}

\begin{figure}
 \begin{center}
\includegraphics[angle=90,width=7cm, bb=-200 -200 500 600]{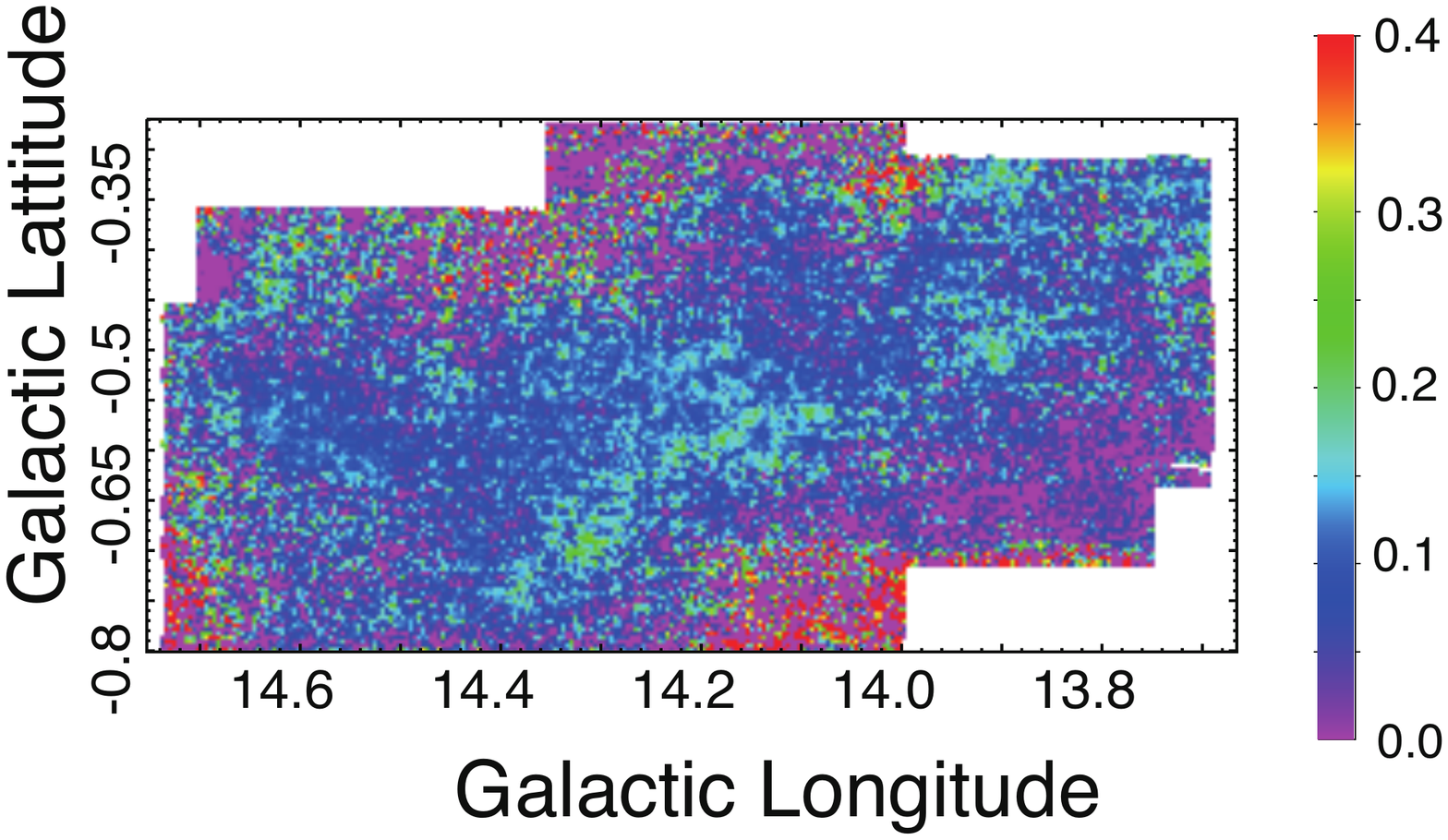}
 \end{center}
\caption{The C$^{18}$O-to-$^{13}$CO fractional abundance ratio of M17.
\label{fig:m17_13co-c18o}}
\end{figure}

\end{document}